\documentclass[12pt]{article}
\usepackage{amsfonts,amssymb,epsfig,amsmath,mathtools}
\usepackage{color}

\usepackage{mathrsfs}
\usepackage{graphicx}
\usepackage{subcaption}

\renewcommand{\baselinestretch}{1.2}

\def\Tr{{\rm Tr}}

\def\det{{\rm det}}

\newcommand{\be}{\begin{eqnarray}}
\newcommand{\ee}{\end{eqnarray}}

\newcommand{\bn}{\begin{enumerate}}
\newcommand{\en}{\end{enumerate}}

\renewcommand{\d}[1]{\mathinner{d#1}} 

\begin{document}

\makeatletter \@addtoreset{equation}{section} \makeatother
\renewcommand{\theequation}{\thesection.\arabic{equation}}
\renewcommand{\thefootnote}{\alph{footnote}}

\begin{titlepage}

\begin{center}
\hfill {\tt SNUTP21-001}\\

\vspace{2cm}

{\Large\bf The Yang-Mills duals of small AdS black holes}

\vspace{2cm}

\renewcommand{\thefootnote}{\alph{footnote}}

{\large Sunjin Choi$^{1,2}$, Saebyeok Jeong$^{3,4}$ and Seok Kim$^1$ }

\vspace{0.3cm}

\textit{$^1$Department of Physics and Astronomy \& Center for
Theoretical Physics,\\
Seoul National University, 1 Gwanak-ro, Gwanak-gu, Seoul 08826, Republic of Korea}\\

\vspace{0.2cm}

\textit{$^2$School of Physics, Korea Institute for Advanced Study,\\
85 Hoegi-ro, Dongdaemun-gu, Seoul 02455, Republic of Korea}\\

\vspace{0.2cm}

\textit{$^3$New High Energy Theory Center, Rutgers University,\\
136 Frelinghuysen Road, Piscataway, New Jersey 08854-8019, USA}\\

\vspace{0.2cm}

\textit{$^4$Department of Theoretical Physics, CERN, 1211 Geneva 23, Switzerland}\\

\vspace{0.7cm}

E-mails: {\tt sunjinchoi@kias.re.kr,
saebyeok.jeong@physics.rutgers.edu, seokkimseok@gmail.com}

\end{center}

\vspace{1cm}

\begin{abstract}

We study the large $N$ matrix model for the index of 4d $\mathcal{N}=4$
Yang-Mills theory and its truncations to understand the dual AdS$_5$ black holes.
Numerical studies of the truncated models provide insights on the black hole
physics, some of which we investigate analytically with the full Yang-Mills matrix
model. In particular, we find many branches of saddle points
which describe the known black hole solutions. We analytically construct the
saddle points dual to the small black holes whose sizes are much smaller than the
AdS radius. They include the asymptotically flat
BMPV black holes embedded in large AdS with novel thermodynamic instabilities.

\end{abstract}

\end{titlepage}

\renewcommand{\thefootnote}{\arabic{footnote}}

\setcounter{footnote}{0}

\renewcommand{\baselinestretch}{1}

\tableofcontents

\renewcommand{\baselinestretch}{1.2}

\section{Introduction}

In this paper, we study the matrix model
\cite{Romelsberger:2005eg,Kinney:2005ej} for the index of
the 4d maximal super-Yang-Mills theory with $U(N)$ gauge group at large $N$.
Our goal is to better understand the black holes in
AdS$_5\times S^5$ \cite{Gutowski:2004ez,Gutowski:2004yv,Chong:2005da,Kunduri:2006ek}.
There have been some recent works on this subject: see
\cite{Cabo-Bizet:2018ehj,Choi:2018hmj,Benini:2018ywd} and references thereof.
Many of these works are not directly based on the matrix model of
\cite{Romelsberger:2005eg,Kinney:2005ej}, except those on large black holes \cite{Choi:2018hmj}, because this matrix model is difficult to study.

Recently, \cite{Copetti:2020dil} discussed the truncated versions of
this matrix model. The truncations keep finite numbers of terms
(among infinitely many) appearing in the potential of
the matrix model. The simplest truncation keeps only one term,
closely related to the Gross-Witten-Wadia model \cite{Gross:1980he,Wadia:1980cp}
at complex coupling. Improved truncations include more terms in the
potential, forming an infinite sequence of truncated matrix models.
Strictly speaking, the truncations are
justified only at low temperature. In practice, \cite{Copetti:2020dil} showed that
the truncations provide fairly good descriptions of some physics even at
finite temperature. \cite{Copetti:2020dil} demonstrated it by studying the
confinement-deconfinement transition
\cite{Witten:1998zw,Sundborg:1999ue,Aharony:2003sx}
of the truncated model which keeps
two terms. This model `approximates' the Hawking-Page transition
\cite{Hawking:1982dh} of black holes fairly well.

Higher truncated models will describe the physics with more precision.
Also, one can get valuable
qualitative insights from the truncated models with relatively
simpler computations. In this paper, we explain some progress
along these lines.

We start by explaining the truncated matrix model analysis in a
streamlined fashion. We employ the analysis of the truncated models of
\cite{Aharony:2003sx}, slightly generalized to cover our problem. This is
equivalent to the setup explored in \cite{Copetti:2020dil}. We numerically study
the physics of the first three truncated models. We first
study the deconfinement transitions, repeating \cite{Copetti:2020dil},
and semi-quantitatively explain how they describe the Hawking-Page transition.
We also study the black hole saddle points at fixed charge, by numerically performing the Legendre transformation. For higher models, there appear
multiple branches of saddle points whose physics is close to the known
AdS black holes.  The fine structures revealed by our studies
are delicate. See section 3 for the details, and also sections 4 and 5
for partial analytic accounts.

With insights from the numerical studies, we find the exact analytic
saddle points for small black holes, whose sizes are much smaller than
the AdS radius. Our primary interest is the 2-parameter free
energy as a function of the inverse-temperature $\beta$ and
an angular chemical potential $\gamma$. (See section 5.3 for a 4-parameter
generalization.) Small black holes are reached by taking $\beta\ll 1$.
The 1-parameter free energy at $\gamma=0$ will account for
the black holes of \cite{Gutowski:2004ez}. In the truncation
keeping $p$ terms in the matrix model potential, we find the exact
leading saddle point solutions. From this, we can take the
$p\rightarrow\infty$ limit to study the full
Yang-Mills matrix model without truncations. The leading free energy
at $p\rightarrow\infty$ is given by
\begin{equation}\label{small-free}
  \log Z\approx -\frac{4N^2}{\pi^2}\beta^3\ .
\end{equation}
Its Legendre transformation yields the following entropy at fixed charge $q$:
\begin{equation}\label{small-entropy}
  S(q)=\log Z(\beta)+ q\beta\rightarrow
  \pi\left[\frac{q^3}{27N^2}\right]^{\frac{1}{2}}\ .
\end{equation}
This accounts for the Bekenstein-Hawking entropy of the small BPS black holes in
AdS$_5\times S^5$ \cite{Gutowski:2004ez} from the canonical saddle point analysis.
See section 5.2 for the generalization to $\gamma\neq 0$
with an extra spin, with $\log Z\approx-\frac{4N^2\beta^3}{\pi^2+\gamma^2}$ and
$S(q,j)=\pi\sqrt{\frac{q^3}{27N^2}-j^2}$.

Our results on small black holes are interesting for many reasons. First of all,
the black holes much smaller than the AdS radius
locally behave like asymptotically flat black hole solutions.
The black holes explained above reduce to the 5d asymptotically
flat black holes of Strominger, Vafa \cite{Strominger:1996sh}
and BMPV \cite{Breckenridge:1996is}.
The microscopic details of the small AdS black holes also exhibit
similarities with \cite{Strominger:1996sh,Breckenridge:1996is}. Namely,
\cite{Kinney:2005ej} made a heuristic account for the entropy (\ref{small-entropy})
with D3-branes in AdS$_5\times S^5$, in close parallel to the counting of
\cite{Strominger:1996sh}. Our work can be regarded as
counting asymptotically flat black holes of \cite{Strominger:1996sh}
from first principles, with an IR regularization by AdS, without
ad hoc assumptions like D-branes in quantum gravity.

The details of the small black hole saddles are also interesting.
The saddle point is given by a distribution function $\rho(\theta)$ of
matrix eigenvalues. (See section 2 for the precise definition.)
For large black holes, it was found
\cite{Choi:2018hmj,Honda:2019cio,ArabiArdehali:2019tdm,Kim:2019yrz,Cabo-Bizet:2019osg} that $\rho(\theta)\sim \delta(\theta)$ asymptotically. Its physics
is understood from the deconfinement of the gauge theory:
at high temperature for large black holes,
the system wants to maximally liberate the deconfined gluons.
This is realized by putting all eigenvalues asymptotically on top of each
other, $\rho(\theta)\sim\delta(\theta)$ \cite{Aharony:2003sx}.
On the other hand, it is a priori unclear
what $\rho(\theta)$ should be for small black holes.
While large black holes represent the exotic high temperature phase of gravity,
small black holes represent microstates of gravity in the conventional phase
such as D-branes \cite{Kinney:2005ej}. For the 1-parameter small black holes of
\cite{Gutowski:2004ez}, we find (at $p\rightarrow\infty$)
the following triangular eigenvalue distribution,
\begin{equation}
  \rho(\theta)=\frac{1}{\pi^2}(\pi-|\theta|)\ \ \ \textrm{for}\
  -\pi<\theta<\pi\ ,\ \ \rho(\theta)=\rho(\theta+2\pi)
  \ .
\end{equation}
This is different from the small black hole
limit of the Bethe ansatz \cite{Benini:2018ywd}.

We further study the QFT duals of the small spinning black holes at
$\gamma\neq 0$, related to the BMPV black holes
\cite{Breckenridge:1996is}. We emphasize the appearance of a curious
thermodynamic instability in AdS which extends that of \cite{Bena:2011zw}.
It is analogous to the well-known instability of the Kerr-AdS black holes \cite{Hawking:1999dp}. We microscopically
study these unstable saddle points. We expect this to be a ubiquitous
aspect of asymptotically flat spinning black holes embedded in large AdS.

Combining our numerical and analytic insights, we expect the
small AdS black holes to provide a novel and powerful route to study
the challenging problems of asymptotically flat BPS black holes.
Studying different parameter regimes and the saddle point ansatz,
hopefully more interesting asymptotically flat black holes can be
identified and studied.

The remaining part of this paper is organized as follows.
In section 2, we explain the matrix model, its truncations and
the saddle point analysis. In section 3, we study the deconfinement and the
Hawking-Page transition. We then study the black hole saddle points at fixed charges.
Section 4 provides a short analytic explanation on the large black hole limit,
clarifying its universality. In section 5, we construct the exact saddle points for small black holes and explore the physics, also discussing the thermodynamic
instabilities. Section 6 concludes with discussions.
Appendix A explains the saddle point analysis related to section 2.
Appendix B explains the BPS black hole
solutions and the small BMPV black hole limit.

\section{The matrix models and the saddle points}

The index of the $\mathcal{N}=4$ Yang-Mills theory on $S^3\times\mathbb{R}$ \cite{Romelsberger:2005eg,Kinney:2005ej} is defined by
\begin{equation}
  Z(\Delta_I,\omega_i)={\rm Tr}\left[(-1)^Fe^{-\sum_{I=1}^3 R_I\Delta_I
  -\sum_{i=1}^2 J_i\omega_i}\right]\ ,
\end{equation}
with chemical potentials constrained by
$\Delta_1+\Delta_2+\Delta_3-\omega_1-\omega_2=0$. The three charges $R_I$
are for the $U(1)^3\subset SO(6)$ R-symmetry, and the two angular momenta
$J_i$ are for the $U(1)^2\subset SO(4)$ which rotates the spatial $S^3$.
With the constraint satisfied by the chemical potentials, the exponential
measure inside the trace commutes with two of the $32$ supercharges of
this system. Calling the commuting Poincare supercharge as $Q$, it satisfies
$[R_I,Q]=+\frac{1}{2}Q$, $[J_i,Q]=-\frac{1}{2}Q$. Its conjugate conformal
supercharge $S\equiv Q^\dag$ carries $R_I=-\frac{1}{2}$ and $J_i=+\frac{1}{2}$.
The index counts BPS states which are annihilated by $Q$ and $S$. The BPS
states saturate the bound coming from the following algebra:
\begin{equation}
  r\{Q,Q^\dag\}\sim rE-(R_1+R_2+R_3+J_1+J_2)\geq 0\ ,
\end{equation}
where $E$ is the energy and $r$ is the radius of $S^3$.
We are also interested in the following 1- and 2-parameter unrefined versions
of this index. After the 1-parameter unrefinement $e^{-\Delta_1}=e^{-\Delta_2}=e^{-\Delta_3}
\equiv x^2$ and $e^{-\omega_1}=e^{-\omega_2}\equiv x^3$, the index is given by
\begin{equation}\label{index-1-para}
  Z(x)={\rm Tr}\left[(-1)^F x^{6\left(\frac{R_1+R_2+R_3}{3}
  +\frac{J_1+J_2}{2}\right)}\right]\equiv
  {\rm Tr}\left[(-1)^F x^{6\left(R+J_+\right)}\right]\ ,
\end{equation}
where $R\equiv\frac{R_1+R_2+R_3}{3}$, $J_+\equiv\frac{J_1+J_2}{2}$.
$q\equiv 6(R+J_+)$ is quantized to be integers. This index will be
used in sections 3, 4 and 5.1 to study the 1-parameter BPS black holes of
\cite{Gutowski:2004ez} and the related asymptotically flat black holes
\cite{Strominger:1996sh}.
The 2-parameter index with equal $e^{-\Delta_I}\equiv x^2$
and $e^{-\omega_1}=x^3y$, $e^{-\omega_2}=x^3y^{-1}$ is given by
\begin{equation}\label{index-2-para}
  Z(x,y)={\rm Tr}\left[(-1)^F x^{6(R+J_+)}y^{2J_-}\right]\ ,
\end{equation}
where $J_-\equiv\frac{J_1-J_2}{2}$. This index will be used in section 5.2
to study the 2-parameter black holes of \cite{Chong:2005da}
(also \cite{Kunduri:2006ek} at equal $\mu_I$'s) and the related BMPV
black holes \cite{Breckenridge:1996is}.

For $\mathcal{N}=4$ Yang-Mills theories with weak coupling limits, the index admits
a unitary matrix integral representation. For the $U(N)$ gauge group, one obtains
\cite{Romelsberger:2005eg,Kinney:2005ej}
\begin{eqnarray}\label{index-matrix-model}
  Z(\Delta_I,\omega_i)&=&
  \frac{1}{N!}\prod_{a=1}^N\int_0^{2\pi}\frac{d\alpha_a}{2\pi}
  \cdot\prod_{a<b}\left(2\sin\frac{\alpha_{ab}}{2}\right)^2\cdot
  \exp\left[\sum_{a,b=1}^N\sum_{n=1}^\infty\frac{a_n(\Delta_I,\omega_i)}{n}
  e^{in\alpha_{ab}}\right]\nonumber\\
  a_n&\equiv& 1-\frac{\prod_{I=1}^3(1-e^{-n\Delta_I})}
  {(1-e^{-n\omega_1})(1-e^{-n\omega_2})}
\end{eqnarray}
where $\alpha_{ab}\equiv\alpha_a-\alpha_b$.
The variables $e^{i\alpha_a}$ are eigenvalues of a $U(N)$ unitary matrix.
This integral can also be written as
\begin{equation}\label{index-potential}
  Z(\Delta_I,\omega_i)=
  \frac{1}{N!}\prod_{a=1}^N\int_0^{2\pi}\frac{d\alpha_a}{2\pi}
  \exp\left[N\sum_{n=1}^\infty\frac{a_n}{n}\right]
  \exp\left[-\sum_{a<b}V(\alpha_a-\alpha_b)\right]\ ,
\end{equation}
with the 2-body eigenvalue potential $V(\theta)$ given by
\begin{equation}\label{2-body-potential}
  V(\theta)\equiv-\log\left[4\sin^2\frac{\theta}{2}\right]
  -\sum_{n=1}^\infty\frac{a_n}{n}(e^{in\theta}+e^{-in\theta})\ .
\end{equation}
The first factor in the integrand of (\ref{index-potential})
comes from the $N$ Cartans of the adjoint fields, and will be irrelevant
for studying the large $N$ free energy proportional to $N^2$.

Since its discovery, this index used to be studied at real
chemical potentials $\Delta_I$, $\omega_i$ until recently. The recent studies
\cite{Cabo-Bizet:2018ehj,Choi:2018hmj,Benini:2018ywd} demand studying
this index at complex chemical potentials. The physical reasonings are explained in
\cite{Choi:2018hmj,Choi:2018vbz,Choi:2019miv,Choi:2019zpz}, with gradual upgrades,
until \cite{Agarwal:2020zwm} provided very concrete interpretations
and evidences. Here we summarize the interpretation comprehensively.
The discussions can be made from the microcanonical
viewpoint or the grand canonical viewpoint. Although the two are closely related,
we provide both interpretations for the sake of completeness.
For simplicity, we discuss the 1-parameter unrefined index (\ref{index-1-para}).

In the microcanonical
viewpoint, the index (\ref{index-1-para}) is a generating function for the
degeneracies $\Omega_q$ at fixed integral charges $q\equiv 6(R+J_+)$:
\begin{equation}\label{Z-series}
  Z(x)=\sum_{q=0}^\infty \Omega_q x^q\ .
\end{equation}
We consider $\Omega_q$ at large charge $q$.
Large charge could mean either $q\gg 1$ at finite $N$, or
$q\sim N^2$ at large $N$. $|\Omega_q|$ grows macroscopically at
large $q$. For instance, one can be confident
about its macroscopic growth by computing (\ref{index-1-para})
numerically in a series expansion \cite{Murthy:2020rbd,Agarwal:2020zwm} at
various values of $N$.
Here, note that $Z(x)$ is an index which grades fermionic states with
$-1$. So $\Omega_q$ may, and actually does, make sign oscillations as
a function of discrete $q$ \cite{Agarwal:2020zwm}.

$\Omega_q$ can be obtained from $Z(x)$ by the following integral,
\begin{equation}
  \Omega_q=\oint \frac{dx}{2\pi ix} x^{-q} Z(x)\ ,
\end{equation}
where the integral contour may be taken to be a small circle at constant
$|x|$. One can compute the integral at large $q$ by the saddle point
approximation, which is the Legendre transformation.
How can this approximation compute macroscopic numbers $\Omega_q$ with
sign oscillations? The answer is by having a pair of
mutually complex conjugate saddles \cite{Agarwal:2020zwm}. Namely, if
a complex saddle $x=x_\ast$ and its conjugate $\bar{x}_\ast$ are
equally dominant, the approximation yields
\begin{equation}\label{Omega-macro}
  \Omega_q\sim e^{S(q)+\cdots}+e^{\overline{S(q)}+\cdots}\sim
  e^{{\rm Re}[S(q)]+\cdots}\cos\left[{\rm Im}[S(q)]+\cdots\right]\ .
\end{equation}
Here $S(q)$ is the complex saddle point action, and $\cdots$ are possible
subleading terms at large $q$. ${\rm Re}(S(q))$ at the
complex saddles measures the leading entropy of the index, while
${\rm Im}(S(q))$ measures the sign oscillations of $\Omega_q$.
So it is crucial to know $Z(x)$ at complex
fugacity $x$, at least in the region where we expect the saddles to be.
This is the microcanonical reason to consider complex
chemical potentials: to extract sign-oscillating macroscopic degeneracies
at large charges from the Legendre transformation.

It is also illustrating to understand the role of complex fugacities in
the grand canonical ensemble. The key ideas are already presented in
\cite{Choi:2018vbz}. In the summation over $q$ in (\ref{Z-series})
at fixed $x$, $\Omega_q$ at nearby $q$'s may undergo partial cancelations
if $\Omega_q$ sign-oscillates fast enough. Therefore, although
each $\Omega_q$ may be macroscopic, the sum over $q$ in (\ref{Z-series})
may apparently look much smaller due to this cancelation.
With the macroscopic $\Omega_q$ taking the form of (\ref{Omega-macro}),
one can estimate when to expect more/less cancelations.
For each term appearing in (\ref{Omega-macro}), the corresponding contribution
to (\ref{Z-series}) will take the form of
\begin{equation}
  Z(x)\leftarrow\sum_{q}e^{{\rm Re}(S(q))}
  e^{\pm i {\rm Im}(S(q))-\mu q}\ ,
\end{equation}
where $x\equiv e^{-\mu}$. The oscillating phase
$e^{\pm i{\rm Im}(S(q))}$ can cause destructive interference of nearby
terms. Note here that, if $\mu$ has nonzero imaginary part, and if
the corresponding term $-{\rm Im}(\mu)q$ combines with either
$\pm {\rm Im}(S(q))$ to yield a slower phase oscillation, the cancelation
can be obstructed to certain extent. In fact \cite{Choi:2018vbz} found that,
by turning on ${\rm Im}(\mu)$ to various values, $Z$ with less cancelation
can show apparent phase transitions at lower temperatures. (See
\cite{Copetti:2020dil} for a more precise statement about the transition,
which we shall review in section 3.1.)

One can ask the optimal value of ${\rm Im}(\mu)$ which maximally
obstructs the cancelation. In general the optimal choice of ${\rm Im}(\mu)$
can be made only locally, since it depends on the region of $q$
one wishes to study. Namely, for a destructive interference near a given
$q$ to be maximally obstructed, the phase $e^{\pm i{\rm Im}(S(q))-i{\rm Im}(\mu)q}$
should locally remain constant near the chosen $q$. This
amounts to demanding ${\rm Im}(\mu)$ to satisfy the stationary phase condition
in $q$:
\begin{equation}
  \pm\frac{d}{dq}{\rm Im}(S(q))={\rm Im}(\mu)\ .
\end{equation}
This is nothing but the imaginary part of the Legendre transformation,
which is either $\frac{dS(q)}{dq}=\mu$ or $\frac{d\overline{S(q)}}{dq}=\mu$
depending on the saddle point.
Therefore, to summarize, the grand canonical viewpoint is more general
than the microcanonical one. Imaginary chemical potential is introduced to obstruct
the cancelation of the nearby terms of (\ref{Z-series}),
which is related to the microcanonical picture only when ${\rm Im}(\mu)$ is
chosen optimally to maximally obstruct the cancelation.

With these understood on the complex parameters, let us now review the
procedures of \cite{Aharony:2003sx} to construct the large $N$
saddle point solutions, somewhat generalized to accommodate our setup.
This is also completely equivalent to the procedures of \cite{Copetti:2020dil}.
In the large $N$ limit, the integral (\ref{index-potential})
can be evaluated using the saddle point approximation. The saddle point
consists of $N$ eigenvalues forming distributions along certain `cuts,'
which are intervals in the complex $\alpha_a$ plane. The distribution is
complex since our chemical potentials are, which complexify the potential $V$
in (\ref{index-potential}).
In this paper, we only consider the eigenvalue distributions forming a single cut.
The single cut saddle points will include the known BPS black hole solutions in AdS
\cite{Gutowski:2004ez,Gutowski:2004yv,Kunduri:2006ek,Chong:2005da}.
Using the translation symmetry of $\alpha_a$'s, and
also the symmetry of flipping all $\alpha_a\rightarrow-\alpha_a$, we seek
for a single cut distribution which is symmetric under the reflection
$\alpha\rightarrow-\alpha$. We take the two endpoints of the cut $I$ to be
$\theta_0$ and $-\theta_0$. where $\theta_0$ is a $\mathcal{O}(N^0)$ complex number.

The saddle point equation for $\alpha_a$ is given by
\begin{equation}\label{saddle-eqn-discrete}
  \sum_{b(\neq a)}V^\prime(\alpha_a-\alpha_b)=0\ .
\end{equation}
We introduce the eigenvalue density function $\rho(\theta)$ defined by
\begin{equation}\label{rho-definition}
  \rho(\alpha_a)\equiv\frac{1}{N}\frac{\Delta a}{\Delta\alpha_a}\ ,
\end{equation}
with $a=1,\cdots,N$. $\Delta\alpha_a\equiv \alpha_{a+\Delta a}-\alpha_a$ is
the difference between the two complex eigenvalues in the infinitesimal neighbor.
$\rho(\theta)$ defined in this way for $\theta\in I$ is in general complex
since $\alpha_a$'s are. Using this $\rho(\theta)$, the saddle point equation
(\ref{saddle-eqn-discrete}) can be written as
\begin{equation}
  \int_{-\theta_0}^{\theta_0}d\theta \rho(\theta) V^\prime(\alpha-\theta)=0
\end{equation}
where the $\theta$ integral is over the contour $I$ that ends on
$-\theta_0$ and $\theta_0$. Inserting (\ref{2-body-potential}),
one obtains
\begin{equation}\label{saddle-eqn}
  -\sum_{a}V^\prime(\alpha-\alpha_a)=
  \int_{-\theta_0}^{\theta_0}\cot\left(\frac{\alpha-\theta}{2}\right)
  \rho(\theta)d\theta-2\sum_{n=1}^\infty a_n\rho_n\sin(n\alpha)=0\ ,
\end{equation}
where $\rho_n$ are the moments of the distribution $\rho(\theta)$
defined by
\begin{equation}\label{moment}
  \rho_n\equiv\int_{-\theta_0}^{\theta_0} d\theta \rho(\theta)
  \cos(n\theta)\ \ \ ,\ \ n=1,2,\cdots\ .
\end{equation}
Here we used the fact that $\rho(\theta)$ is an even function so that the $\sin(n\theta)$ moments are zero,
\begin{equation}
  \int_{-\theta_0}^{\theta_0}d\theta \rho(\theta)\sin(n\theta)=0\ .
\end{equation}
At this point, we comment on some generalization that we made for complex
$\rho(\theta)$ defined on a complex cut $I$. In \cite{Aharony:2003sx},
real $\rho(\theta)$ for real $\theta$ was considered. There one
could Fourier expand
\begin{equation}
  \rho(\theta)\stackrel{\textrm{real}}{\longrightarrow}\frac{1}{2\pi}
  +\frac{1}{\pi}\sum_{n=1}^\infty\rho_n\cos(n\theta)\ .
\end{equation}
Namely, $\rho(\theta)$ was extended beyond the real interval $(-\theta_0,\theta_0)$
as $\rho(\theta)=0$, and the Fourier expansion of an even function was made on
the whole circle $\theta\in[-\pi,\pi]$.
Unlike this, we abstractly define $\rho_n$ as
the moments (\ref{moment}) of the complex $\rho(\theta)$ defined only
on the cut $I$.

Before proceeding, we explain the invertibility of the map between
$\rho(\theta)$ and the distribution of $\alpha_a$ in the complex case.
Obtaining $\rho(\theta)$ at a given complex distribution of $\alpha_a$ is
obvious, as given by (\ref{rho-definition}). Conversely, suppose that
a complex $\rho(\theta)$ is given on the complex plane of $\theta$. (This
is what we shall obtain by the saddle point analysis, after following the
procedures analogous to \cite{Aharony:2003sx}.) Then defining
$s\equiv\frac{a}{N}-\frac{1}{2}$ which satisfies $-\frac{1}{2}<s<\frac{1}{2}$,
$\alpha_a$ in the continuum limit can be written as a complex function
$\alpha(s)$ for real $s\in(-\frac{1}{2},\frac{1}{2})$. This can be obtained
by integrating (\ref{rho-definition}). Namely,
$\rho(\alpha_a)=\frac{1}{N}\frac{da}{d\alpha_a}$
can be integrated to yield
\begin{equation}\label{curve-s}
  s(\alpha)=\int \rho(\alpha) d\alpha+{\rm constant}\ .
\end{equation}
One can obtain the curve $\alpha(s)$  on the complex $\alpha$ plane
by demanding the last integral to be real. Therefore, one can locally
construct the $\alpha_a$ distribution from complex $\rho(\theta)$.
$s(\alpha)$ has to be either an increasing or a decreasing function
along the cut, and should terminate on $s=\pm\frac{1}{2}$.
Whether such a cut exists, starting from $\theta=-\theta_0$, passing
through $\theta=0$ and ending on $\theta=\theta_0$, is a nontrivial question.
There will appear parameter regions
in which the desired cuts do not exist. Once such a cut exists,
it will be guaranteed in our solution below that
$s(\pm\theta_0)=\pm\frac{1}{2}$, corresponding to
$\int_{-\theta_0}^{\theta_0} d\theta \rho(\theta)=1$.

Now we explain the single cut saddle point solutions,
following \cite{Aharony:2003sx}. As in \cite{Aharony:2003sx}, we
treat $\rho_n$ as independent variables and solve for (\ref{saddle-eqn})
and (\ref{moment}). We first solve (\ref{saddle-eqn}) for $\rho(\theta)$ at
fixed independent $\rho_n$'s.\footnote{In the context of \cite{Copetti:2020dil},
making $\rho_n$ independent and fixed for a while corresponds to introducing
the Hubbard-Stratonovich parameters and integrating over $\alpha_a$'s first.}
For the single cut distribution,
the solution to this linear equation is \cite{Jurkiewicz:1982iz,Aharony:2003sx}
\begin{equation}\label{gap-solution}
  \rho(\theta)=\frac{1}{\pi}
  \sqrt{\sin^2\frac{\theta_0}{2}-\sin^2\frac{\theta}{2}}~
  \sum_{n=1}^\infty Q_n\cos\left[\textstyle{\left(n-\frac{1}{2}\right)}\theta\right]
\end{equation}
where
\begin{equation}\label{Q-definition}
  Q_n\equiv2\sum_{l=0}^\infty a_{n+l}\rho_{n+l}P_l(\cos\theta_0)\ ,
\end{equation}
and $P_l$ are the Legendre polynomials given by
\begin{equation}
  \sum_{l=0}^\infty P_l(x) z^l=(1-2xz+z^2)^{-\frac{1}{2}}\ .
\end{equation}
An extra condition, as part of the solution to (\ref{saddle-eqn}), is given by
\begin{equation}\label{normalization}
  Q_1=Q_0+2\ ,
\end{equation}
where $Q_0$ is defined from (\ref{Q-definition}) by starting the
summation from $l=1$. Strictly speaking, the solution
(\ref{gap-solution}), (\ref{Q-definition}), (\ref{normalization}) is derived
in \cite{Jurkiewicz:1982iz} at real $a_n\rho_n$'s. In Appendix A,
we explain that the results can be extended to the case with complex couplings.

The remaining equation (\ref{moment})
to be solved becomes a linear equation of $\rho_n$'s, after inserting
the solution (\ref{gap-solution}) to (\ref{saddle-eqn}).
This linear equation and (\ref{normalization}) take the form of \cite{Aharony:2003sx}
\begin{equation}\label{linear-rho-n}
  R\vec{\rho}=\vec{\rho}\ \ ,\ \ \vec{A}\cdot\vec{\rho}=1\ ,
\end{equation}
where $\vec{\rho}=(\rho_1,\rho_2,\cdots)$ is an $\infty$ dimensional column vector,
and the $\infty$ dimensional matrix $R$ and vector $\vec{A}$ are defined by
\begin{eqnarray}\label{R}
  R_{ml}&=&a_l\sum_{k=1}^l\left[B^{m+k-\frac{1}{2}}(s^2)+
  B^{\left|m-k+\frac{1}{2}\right|}(s^2)\right]P_{l-k}(1-2s^2)\nonumber\\
  A_m&=&a_m\left[P_{m-1}(1-2s^2)-P_m(1-2s^2)\right]\ .
\end{eqnarray}
Here $s$ and $B^{n-\frac{1}{2}}$ are
defined by $s^2\equiv\sin^2\frac{\theta_0}{2}$ and
\begin{eqnarray}
  B^{n-\frac{1}{2}}(s^2)&=&\frac{1}{\pi}\int_{-\theta_0}^{\theta_0}
  d\theta\sqrt{s^2-\sin^2\frac{\theta}{2}}~\cos\left[
  \textstyle{\left(n-\frac{1}{2}\right)}\theta\right]\nonumber\\
  \sum_{n=0}^\infty B^{n+\frac{1}{2}}(x)z^n&=&
  \frac{\sqrt{(1-z)^2+4zx}+z-1}{2z}\ .
\end{eqnarray}
The variables to be determined by these equations are
$\rho_n$ and $\theta_0$. The first equation of
(\ref{linear-rho-n}) admits an eigenvector with eigenvalue $1$ if
the matrix $R-{\bf 1}$ is degenerate,
\begin{equation}\label{det-gap}
  \det(R-{\bf 1})=0\ .
\end{equation}
This equation determines $\theta_0$ in terms of
$a_n$'s (which in turn depend on chemical potentials).
After $\theta_0$ is determined this way, $\vec{\rho}$ is given by
\cite{Aharony:2003sx}
\begin{equation}\label{rho-formal-sol}
  \vec{\rho}=M^{-1}\vec{e}_1\ ,
\end{equation}
where $M$ is a matrix obtained by replacing the first row of
the matrix $({\bf 1}-R)$ by the vector $\vec{A}$, and $\vec{e}_1$ is
a column vector given by $(1,0,0,\cdots)$.

As explained in \cite{Aharony:2003sx}, the above procedures
become more tractable in truncated
matrix models, in which all $a_{n>p}$ with an integer cutoff $p$ are taken
to be zero by hand. This defines the $p$'th entry in the sequence of the
truncated matrix models, in which only $p$ out of infinite terms are
kept in the second term of the potential given by (\ref{2-body-potential}).
In this case, the matrix $M$ takes the form of
\begin{equation}
  M=\left(
    \begin{array}{cc}
      M_{p\times p}&{\bf 0}_{p\times\infty}\\
      L_{\infty\times p}&{\bf 1}_{\infty\times\infty}
    \end{array}
  \right)\ .
\end{equation}
Therefore, $\vec{\rho}$ is given by
\begin{equation}
  \vec{\rho}=
  \left(
    \begin{array}{cc}
      M_{p\times p}^{-1}&{\bf 0}_{p\times\infty}\\
      -L_{\infty\times p}M_{p\times p}^{-1}&{\bf 1}_{\infty\times \infty}
    \end{array}
  \right)\vec{e}_1\ .
\end{equation}
The first $p$ lowest moments $\rho_{n\leq p}$ can be determined by
just knowing the inverse of the finite matrix $M_{p\times p}$. This suffices
to determine the eigenvalue distribution in this truncated model with
$a_{n\geq p}=0$, since in this case (\ref{gap-solution}) and nonzero
$Q_{n\leq p}$'s depend only on $\rho_{n\leq p}$. Similarly, the
equation (\ref{det-gap}) determining the cut size $\theta_0$ also
effectively reduces to a $p\times p$ matrix determinant equation.
It turns out that this is a degree $p^2+p$ polynomial equation in
$s^2$. Choosing one of these $p^2+p$ solutions for $s^2$, one can
in principle solve for $\rho(\theta)$ by a linear procedure in the
truncated matrix models. So in the $p$'th truncated model, there are at
most $p^2+p$ distinct saddle points in the single-cut setup. (We shall
see in section 3 with the examples of $p=1,2,3$ that not all these branches
are physical.) In particular, the finite $p$ models will
make the numerical analysis of the saddle points easier, as we shall
study in section 3. Also, in section 5, we shall obtain exact analytic solutions
for arbitrary $p$'th truncated models in the so-called small black hole limit.
This will allow us to eliminate the truncation by taking the limit
$p\rightarrow\infty$ and obtain the analytic saddle point solutions for the
full matrix model.

Once the saddle point values of $\rho_{n\leq p}$ and $\rho(\theta)$ are known,
the saddle point effective action $\log Z$ can be computed from
\begin{equation}\label{finite-p-free}
  \log Z=\frac{N^2}{2}\int_{-\theta_0}^{\theta_0} d\theta_1
  d\theta_2 \log \left(4\sin^2\frac{\theta_1-\theta_2}{2}\right)
  \rho(\theta_1)\rho(\theta_2)
  +N^2\sum_{n=1}^p\frac{a_n}{n}|\rho_n|^2\ ,
\end{equation}
where we used the moment formula (\ref{moment}), and one should
insert (\ref{gap-solution}) for $\rho(\theta)$. In practice, this can be
computed more easily in the following way. One first integrates the saddle point
equation (\ref{saddle-eqn}) in $\alpha$ to obtain
\begin{equation}\label{potential-eom}
  \mu=\int_{-\theta_0}^{\theta_0}d\theta \rho(\theta)
  \log\left[4\sin^2\frac{\alpha-\theta}{2}\right]
  +\sum_{n=1}^p\frac{2a_n}{n}\rho_n\cos\left(n\alpha)\right)\ ,
\end{equation}
where $\mu$ is the integration constant. Then
one can insert the solution $\rho(\theta)$ and $\rho_{n\leq p}$ to
compute $\mu$ for the solution of one's interest.
This can be computed by evaluating the right hand side at any value
of $\alpha$, say at $\alpha=0$. With $\mu$ computed this way, one integrates
both sides of (\ref{potential-eom}) with
$\frac{N^2}{2}\int_{-\theta_0}^{\theta_0}d\alpha
\rho(\alpha)$, which yields
\begin{equation}\label{free-constant}
  \log Z=\frac{N^2\mu}{2}\ .
\end{equation}
Therefore, one can compute the free energy $\log Z$ by
knowing the constant $\mu$ from (\ref{potential-eom}).

\begin{figure}[!t]
\centering
\begin{subfigure} [b]{0.32\textwidth}
\includegraphics[width=\textwidth]{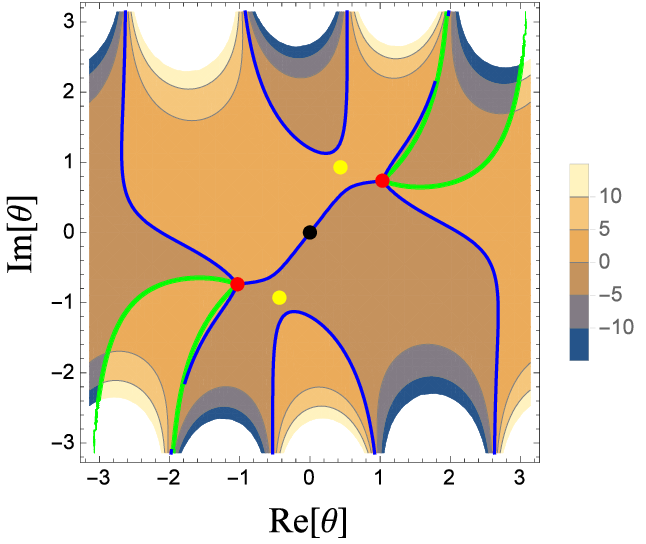}
\subcaption{$x\approx .721 e^{2.296 i}$: $\exists$cut}
\end{subfigure}
\hspace{0cm}
\begin{subfigure} [b]{0.32\textwidth}
\includegraphics[width=\textwidth]{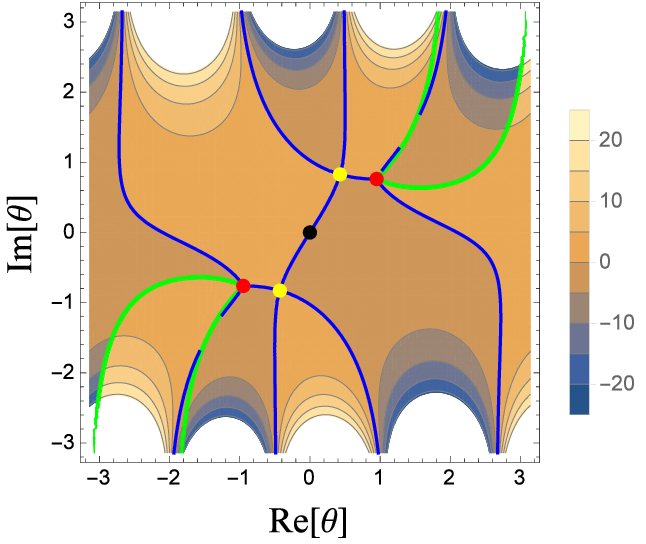}
\subcaption{$x\approx .744 e^{2.252 i}$: marginal}
\end{subfigure}
\hspace{0cm}
\begin{subfigure} [b]{0.32\textwidth}
\includegraphics[width=\textwidth]{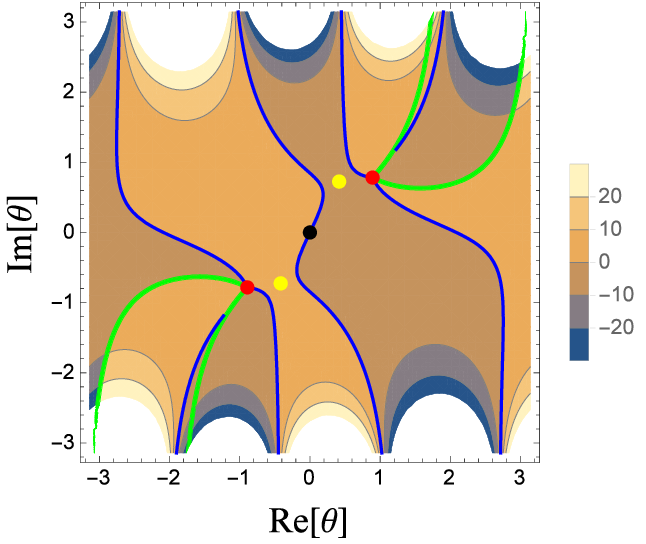}
\subcaption{$x\approx .769 e^{2.214 i}$: $\nexists$cut}
\end{subfigure}
\caption{Illustrating how the critical point $\theta_\ast$ (yellow dots)
satisfying $\rho(\theta_\ast)=0$ can destroy the cut as $x$ crosses a wall.
The blue lines are ${\rm Im}[s(\theta)]=0$ lines, red dots the branch points
$\theta=\pm\theta_0$, and the black dots the origin $\theta=0$.
The background contour plots are those for ${\rm Im}[s(\theta)]$.
(Green lines are the branch cuts of
$s(\theta)$ chosen by mathematica's convention.)}
\label{cut-change}
\end{figure}

The function $\rho(\theta)$ can be integrated to obtain
$s(\theta)$, which determines the cut by the condition ${\rm Im}(s(\theta))=0$.
As we explained above, our ansatz seeks for a single cut solution which is
invariant under $\theta\rightarrow-\theta$ flip: it passes through $\theta=0$,
and ends on the two branch points $\theta=\pm\theta_0$.
Whether such a cut exists is a nontrivial question, whose answer
depends on the parameters
of the matrix model such as $x$. Even if the single cut solution exists for
certain $x$'s, it may cease to exist after $x$ passes through a wall. This means
that the part of line ${\rm Im}(s)=0$ which connects $\pm\theta_0$ will suddenly
change, so that the ${\rm Im}(s)=0$ line starting from one end $\theta_0$ will
escape to infinity rather than ending on $-\theta_0$.
This can happen when two ${\rm Im}[s(\theta)]=0$ lines
meet. This is illustrated in Fig. \ref{cut-change}, where the cut connecting
$\theta=\pm\theta_0$ (red dots) suddenly disappears after meeting other
parts of the ${\rm Im}(s)=0$ lines.
These are for a particular branch of saddle points
in the $p=2$ model with one parameter $x$. See section 3.2 for more details.
Fig. \ref{cut-change}(b) shows the cut when $x$ is on the wall.
If $x$ crosses the wall, the single-cut saddle point disappears as illustrated
in Fig. \ref{cut-change}(c).

If two ${\rm Im}(s)=0$ lines meet at a point $\theta=\theta_\ast$, it means that
${\rm Im}[s(\theta)]$ remains constant along two independent directions
at $\theta=\theta_\ast$. This can happen only if $s(\theta)$
is extremal at $\theta=\theta_\ast$. This is because,
making a Taylor expansion
\begin{equation}
  s(\theta)\approx s(\theta_\ast)+s^\prime(\theta_\ast)(\theta-\theta_\ast)
  +\frac{1}{2}s^{\prime\prime}(\theta_\ast)(\theta-\theta_\ast)^2+\cdots\ ,
\end{equation}
the presence of the second term $s^\prime(\theta_\ast)(\theta-\theta_\ast)$
would give a unique direction along which ${\rm Im}(s(\theta))$
remains constant if $s^\prime(\theta_\ast)\neq 0$. Therefore, if two such lines
meet, this implies $\rho(\theta_\ast)=s^\prime(\theta_\ast)=0$.
Therefore, a necessary condition for the wall in $x$ space is
the critical point satisfying $\rho(\theta_\ast)=0$ meeting the cut.
The yellow dots of Fig. \ref{cut-change} are the points satisfying
$\rho(\theta_\ast)=0$. The critical point $\theta_\ast$ meeting the cut
is only a necessary condition for the cut to disappear,
since the cut may continue to exist
after the critical point crosses the cut. Such an example can be found
in the $p=1$ model, although we shall not illustrate it here.

\begin{figure}[!t]
\centering
\begin{subfigure} [b]{0.32\textwidth}
\includegraphics[width=\textwidth]{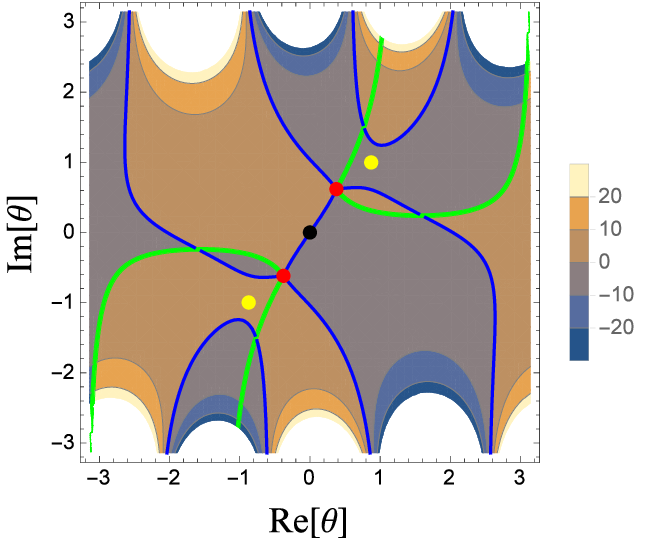}
\subcaption{$x\approx .766 e^{2.215 i}$}
\end{subfigure}
\hspace{0cm}
\begin{subfigure} [b]{0.32\textwidth}
\includegraphics[width=\textwidth]{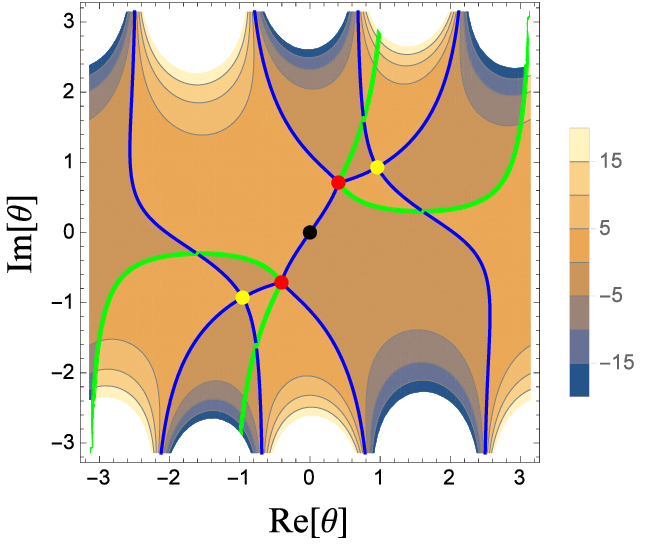}
\subcaption{$x\approx .742 e^{2.255 i}$}
\end{subfigure}
\hspace{0cm}
\begin{subfigure} [b]{0.32\textwidth}
\includegraphics[width=\textwidth]{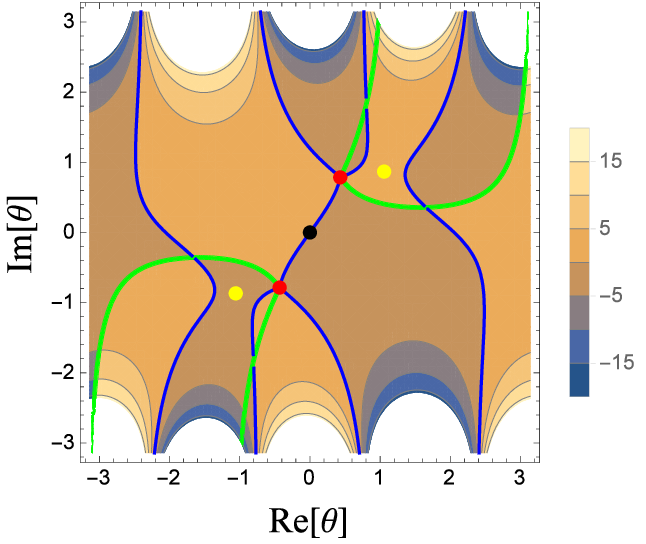}
\subcaption{$x\approx .721 e^{2.304 i}$}
\end{subfigure}
\caption{Illustrating why (\ref{cut-wall}) is only a necessary condition
for the wall, by the critical point meeting the ${\rm Im}[s(\theta)]=0$
line not through the cut. Color conventions are all same
as Fig. \ref{cut-change}.}
\label{cut-no-change}
\end{figure}

The critical point $\theta_\ast(x)$ satisfying
$\rho(\theta_\ast)=0$ is a function of $x$.
A further necessary condition
for $\theta_\ast(x)$ to meet the cut is
\begin{equation}\label{cut-wall}
  {\rm Im}[s(\theta_\ast(x))]=0\ .
\end{equation}
This is just a necessary condition because $\theta_\ast(x)$ may cross
the line ${\rm Im}[s(\theta)]=0$ either through the finite segment
$[-\theta_0,\theta_0]$ or through the other part of this line.
Fig. \ref{cut-no-change} shows an example of $\theta_\ast$ meeting an irrelevant
part of the line ${\rm Im}[s(\theta)]=0$, thus not destroying the cut.
This figure describes another branch of saddle points in the
$p=2$ model: see section 3.2. Therefore, in concrete models
of section 3, one should first draw the lines in the $x$ space defined by
(\ref{cut-wall}). Then one should investigate the behaviors of the cuts near
these lines (by studying the configurations like Figs. \ref{cut-change} and
\ref{cut-no-change}) to determine which part of (\ref{cut-wall}) are the boundaries
of a region admitting the saddle points.

For $p=1$, one finds
\begin{equation}\label{s-p1}
  s(\theta)=\frac{1}{\pi}{\rm arcsin}\left[\frac{\sin\frac{\theta}{2}}
  {t^{\frac{1}{2}}}\right]
  +\frac{1}{\pi t}\sin\frac{\theta}{2}\sqrt{t-\sin^2\frac{\theta}{2}}\ .
\end{equation}
The only critical point of this model is $\theta_\ast=\pi$.
After solving for $t(x)$ and investigating ${\rm Im}[s(\theta)]$ by
changing $x$, one finds that the cut is never destroyed by
$\theta_\ast$. See section 3.1 for the results.

For $p=2$, one finds
\begin{equation}\label{s-p2}
  s(\theta)=\frac{1}{\pi}{\rm arcsin}\left[\frac{\sin\frac{\theta}{2}}
  {t^{\frac{1}{2}}}\right]
  +\frac{1}{\pi}\sin\frac{\theta}{2}\sqrt{t-\sin^2\frac{\theta}{2}}
  \left[\frac{1}{t}-a_2(2-8t+14t^2-7t^3)+2a_2(1-t)^2\cos\theta\right]
\end{equation}
which satisfies $s(0)=0$, $s(\pm\theta_0)=\pm\frac{1}{2}$.
The critical points satisfying $\rho(\theta_\ast)=0$ are given by
\begin{equation}
  \theta_\ast=\pi,\theta_1,-\theta_1\ \ ,\ \ \textrm{where}\ \
  \cos\theta_1\equiv 1-\frac{Q_1}{Q_2}=-\frac{1}{2a_2t(1-t)^2}
  +\frac{4-10t+12t^2-5t^3}{2(1-t)^2}\ .
\end{equation}
The critical point $\theta_\ast=\pi$ can destroy the cut in certain branches
of saddle points. However, this will happen in a region in the $x$ space
that we are not interested in. Only the critical points $\theta=\pm \theta_1$
will play important roles in this paper. At $\theta=\theta_1$, one finds
\begin{eqnarray}
  s(\theta_1)&=&\frac{1}{\pi}{\rm arcsin}
  \left[\frac{\sqrt{a_2^{-1}-2t+6t^2-10t^3+5t^4}}{2t(1-t)}\right]\\
  &&+\frac{1+t}{2\pi t}\sqrt{-1+2a_2(2t\!-\!4t^2\!+\!6t^3\!-\!3t^4)
  -a_2^2(4t^2\!-\!16t^3\!+\!36t^4\!-\!44t^5\!+\!36t^6\!-\!20t^7\!+\!5t^8)}\ .\nonumber
\end{eqnarray}
Here $t=t(x)$ is one of the six solutions solving the equation (\ref{det-gap}).
The two critical points $\pm\theta_1$ meet the cut at the same time, from the
symmetry of the cut. As illustrated in Fig. \ref{cut-change}, these critical points
create nontrivial walls, beyond which the saddle point solutions cease to exist
within our single-cut ansatz. See section 3.2 for the details.
On the wall, such as $x$ of Fig. \ref{cut-change}(b),
$\rho(\theta)$ becomes zero at the two points $\pm\theta_1$ on the wall
(except the cut boundaries
$\pm\theta_0$). This means that the single-cut distribution is making a
phase transition to a triple cut distribution on the wall. We shall
not study the triple cut distributions beyond the walls.
However, our findings predict the existences of large $N$ saddle points
beyond the single cut ansatz.

Similar phenomena also happen at $p=3$, but in a more complicated manner.
There are five possible values of critical points $\theta_\ast(x)$ at
each $x$. We shall not explicitly show the formulae here, and just show
the final numerical results in section 3.3.

\section{Numerical studies}

The key objective of this section is to study deconfinement
and the black hole like saddles in the truncated models. To this end,
we start by reviewing the ideas of \cite{Choi:2018vbz,Copetti:2020dil}
about the confinement-deconfinement
phase transition of this system. For conceptual discussions here, it is
helpful to change the real integral variables $\alpha_a$'s
of (\ref{index-potential}) to the eigenvalue distribution $\rho(\theta)$
on a circle $\theta\sim\theta+2\pi$ \cite{Aharony:2003sx,Kinney:2005ej}.
The effective action of this matrix integral can be written as
\begin{equation}
  S_{\rm eff}=\frac{N^2}{2}
  \int d\theta_1 d\theta_2 V(\theta_1-\theta_2)\rho(\theta_1)\rho(\theta_2)\ .
\end{equation}
$\rho(\theta)$ is a real function,
constrained to be (1) periodic:
 $\rho(\theta)=\rho(\theta+2\pi)$, (2) normalized:
$\int_0^{2\pi}d\theta \rho(\theta)=1$ and (3) non-negative:
$\rho(\theta)\geq 0$. In particular,
condition (3) demands the allowed domain for $\rho(\theta)$
to have a boundary. $\rho(\theta)$ can be written in terms of
its Fourier modes $\rho_n$, $\rho(\theta)=
\frac{1}{2\pi}+\frac{1}{2\pi}\sum_{n\neq 0}\rho_ne^{in\theta}$. One
imposes $\rho_{-n}=\overline{\rho_n}$ for the reality of $\rho(\theta)$.
The conditions (1), (2) above are also met.
In terms of $\rho_n$, The effective action is given by
\begin{equation}\label{Z-rho-Fourier}
  S_{\rm eff}=N^2\sum_{n=1}^\infty\frac{1-a_n}{n}|\rho_n|^2\ .
\end{equation}
The condition (3) introduces a boundary of the
allowed domain for $\{\rho_n\}$. This boundary has a complicated
shape, as one can easily check from finite dimensional
subspaces of $\{\rho_n\}$.

An important question is whether the large $N$ partition function
confines or deconfines, and when the confinement-deconfinement phase
transition happens. This
phase transition is dual to the Hawking-Page transition of the AdS quantum gravity
\cite{Witten:1998zw}, which happens due to the thermal competition
of large black holes and thermal gravitons. An order parameter of
this transition is the Polyakov loop operator, which is the Wilson loop along
the thermal circle \cite{Polyakov:1978vu}.
It is particularly important in our context to consider the Polyakov loop
in the fundamental representation \cite{Aharony:2003sx}
\begin{equation}\label{polyakov-loop}
  \frac{1}{N}{\rm Tr}_{\rm fund}\left[P\exp\left(i\oint d\tau A_\tau\right)
  \right]\ .
\end{equation}
This quantity is zero/nonzero
when the system confines/deconfines, respectively. $-\log$ of its
normalized expectation value is the extra free energy cost for inserting an
external quark loop. So vanishing Polyakov loop implies that the system abhors
this insertion. In our matrix variables or those of \cite{Aharony:2003sx},
this operator is given by
\begin{equation}\label{polyakov-weak}
  \frac{1}{N}\sum_{a=1}^N e^{i\alpha_a}=\rho_1\ ,
\end{equation}
which is nothing but the first Fourier coefficient $\rho_1$ \cite{Aharony:2003sx}.
See section 5.7 of \cite{Aharony:2003sx} for a more careful definition of
this order parameter. Strictly speaking, if one wishes to compute its
strong coupling expectation value using SUSY, one has to supersymmetrize
(\ref{polyakov-loop}) and insert it in the path integral. Since (\ref{polyakov-loop})
is not supersymmetric, inserting (\ref{polyakov-weak}) into our matrix model
integrand yields the expectation value of (\ref{polyakov-loop}) at weak coupling
only, unprotected by SUSY non-renormalization.
The weak-coupling behavior of (\ref{polyakov-loop})
will still provide useful guidance along the spirit of \cite{Aharony:2003sx}.
In particular, it is natural to expect deconfinement when
$\rho_1$ (or $|\rho_1|^2$) wants to condense
at weak coupling. This is because $S_{\rm eff}$ of
(\ref{Z-rho-Fourier}) will then acquire a nonzero contribution
$N^2(1-a_1)|\rho_1|^2$ proportional to $N^2$, which implies deconfinement
unless this term precisely cancels with others.
This is also true in the setup of section 2, from the formula (\ref{finite-p-free}).

Integrals with the effective action (\ref{Z-rho-Fourier}) is subtler
than it naively looks. Although the integrand is Gaussian in $\rho_n$'s,
the integral domain would have a boundary
which is a nontrivial hypersurface. Inspired by \cite{Aharony:2003sx} in
which the role of the fundamental Polyakov loop $\rho_1$ was crucial,
consider integrating over $\rho_1$ first,
\begin{equation}\label{rho-integral-many}
  Z\sim \int \prod_{n=2}^\infty d\rho_n d\rho_{-n}
  \exp\left[N^2\sum_{n=2}^\infty\frac{a_n(x)-1}{n}|\rho_n|^2\right]
  \int_{f_-(\rho_n)}^{f_+(\rho_n)}d\rho_1\exp\left[N^2(a_1(x)-1)\rho_1^2\right]\ .
\end{equation}
Here we took $\rho_1$ to be real using the translation symmetry of $\theta$
\cite{Aharony:2003sx}. Due to the presence of the boundary of the
integral domain, $\rho_1$ is constrained in a range
$f_-(\rho_n)\leq \rho_1\leq f_+(\rho_n)$ which depends on other variables
$\rho_{n\geq 2}$. In particular,
$f_+(\rho_n=0)=\frac{1}{2}$, $f_-(\rho_n=0)=-\frac{1}{2}$ when all the
other variables are at the confining saddle $\rho_{n\geq 2}=0$.
We consider whether the first integral
\begin{equation}\label{rho-integral-1}
  \int_{f_-(\rho_n)}^{f_+(\rho_n)}d\rho_1\exp\left[N^2(a_1(x)-1)\rho_1^2\right]
\end{equation}
can exhibit a condensing behavior to $\rho_1\neq 0$.
As this is simply expressible as the error functions
at complex coefficient $a_1(x)-1$, it is easy to derive that the dominant
contribution to this integral comes from either
$\rho_1=f_\pm$ when ${\rm Re}(a_1(x)-1)>0$.
So supposing that we consider the 1 dimensional integral
(\ref{rho-integral-1}) rather than (\ref{rho-integral-many}),
the integral can be approximated as
\begin{equation}\label{rho-1-result}
  \sim \exp\left[N^2(a_1(x)-1)(\max |f_\pm|)^2\right]
\end{equation}
when ${\rm Re}(a_1(x)-1)>0$. The complex number (\ref{rho-1-result}) has large
absolute value at large $N$. So the 1 dimensional integral (\ref{rho-integral-1})
exhibits a deconfining behavior.

Let us call the region ${\rm Re}(a_1(x)-1)>0$ the tachyonic region
of $\rho_1$, for an obvious reason. As explained in \cite{Choi:2018vbz},
the Hawking-Page `temperature' of known BPS black holes in AdS$_5\times S^5$
is higher than the tachyon threshold ${\rm Re}(a_1(x)-1)=0$.
This led \cite{Choi:2018vbz} to conjecture new
hypothetical black holes with a lower Hawking-Page temperature.
However, \cite{Copetti:2020dil} suggested a much simpler resolution of
this discrepancy, basically by showing that the integral
(\ref{rho-integral-many}) may not acquire dominant contribution at
$\rho_1=f_\pm$ due to the integral of other $\rho_{n\geq 2}$'s.
Namely, although (\ref{rho-1-result}) has a large absolute value, it also
has a fast-oscillating phase factor depending on $\rho_{n\geq 2}$'s at complex $x$.
This phase factor can render extra cancelations during the
$\rho_{n\geq 2}$ integrals, which may invalidate the dominance of the region
$\rho_1=f_\pm$ for the full integral. This way,
the deconfinement transition can be delayed relative to the tachyon threshold
\cite{Copetti:2020dil}. So the tachyon threshold need not agree with the
deconfinement point. Rather, it is a lower bound of deconfinement.
This will be a useful guidance of where to seek for black holes and
deconfinement.

The viewpoint of the previous $4$ paragraphs, directly regarding
$\rho(\theta)$ or $\{\rho_n\}$'s as the integral variables, is cumbersome in
practice since the integral domain has boundaries.
Note that in section 2, we introduced $\rho(\theta)$ and its complexification
rather conservatively, only for the purpose of estimating saddle point quantities.
From now we go back to the setup of section 2 and investigate deconfinement and
black holes in the truncated models.
In this setting, all procedures are linear except solving
the degree $p^2+p$ polynomial equation $\det(R-{\bf 1})=0$ for
the gap $t\equiv\sin^2\frac{\theta_0}{2}$. For the lowest truncation at $p=1$,
one can exactly solve the quadratic polynomial equation. For the higher models at
$p\geq 2$, the polynomial equations are solved numerically in general.
At $p=1,2,3$, we find solutions which describe the known black holes.
For $p\geq 2$, there appear multiple solutions which
combine to describe the known black holes. The interpretations of
these multiple branches are discussed in section 5 and 6.

\subsection{The $p=1$ model}

\begin{figure}[!t]
\centering
\begin{subfigure} [b]{0.45\textwidth}
\includegraphics[width=\textwidth]{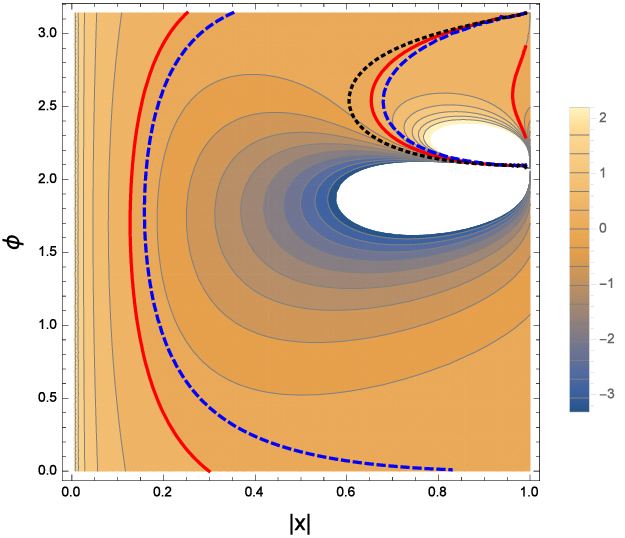}
\subcaption{$\frac{1}{N^2}{\rm Re(\log Z_+)}$ for the $p=1$ model}
\end{subfigure}
\hspace{0.7cm}
\begin{subfigure} [b]{0.45\textwidth}
\includegraphics[width=\textwidth]{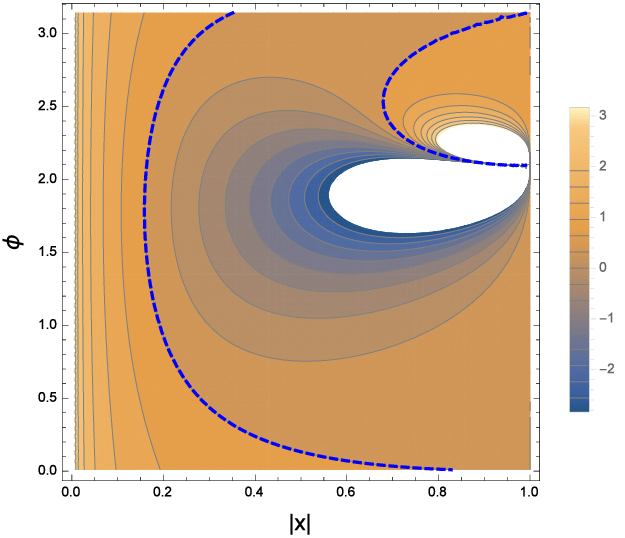}
\subcaption{$\frac{1}{N^2}{\rm Re}(\log Z_{\rm BH})$}
\end{subfigure}
\caption{The contour plots of: (a) $\frac{1}{N^2}{\rm Re}(\log Z_+)$,
(b) $\frac{1}{N^2}{\rm Re}(\log Z_{\rm BH})$. The red curves
denote ${\rm Re}(\log Z_+)=0$ lines. The black dotted curve is
the tachyon threshold ${\rm Re}(a_1-1)=0$, right of which is the tachyonic
region. The dashed blue curves denote ${\rm Re}(\log Z_{\rm BH})=0$ lines.}
\label{p1-free-1}
\end{figure}

\begin{figure}[!t]
\centering
\includegraphics[width=0.45\textwidth]{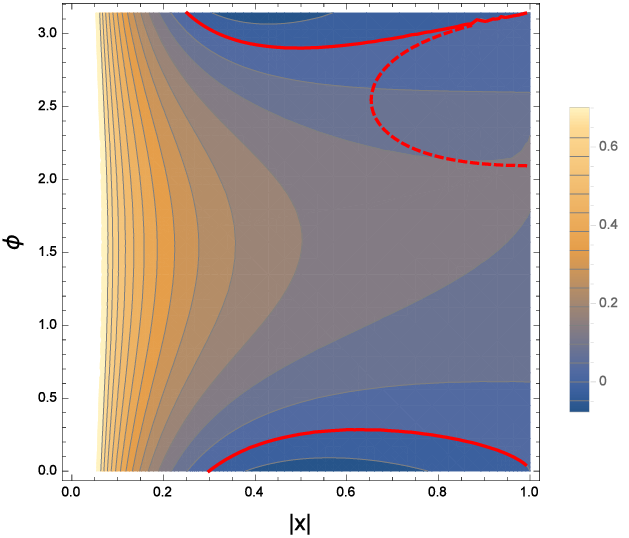}
\caption{The contour plots of $\frac{1}{N^2}{\rm Re}(\log Z_-)$ for the saddle point
$g_-$. The solid red curves denote ${\rm Re}(\log Z_-)=0$ lines.
The dashed red curve is the deconfinement line ${\rm Re}(\log Z_+)=0$.}
\label{p1-free-2}
\end{figure}

This model is related to the complex Gross-Witten-Wadia (GWW) model.
Namely, the intermediate model of section 2 which keeps
$\rho_1$ independent and fixed is the complex GWW model.
In the setup of section 2, $R$ and $\vec{A}$ are numbers at $p=1$,
given by
\begin{equation}
  R=a_1(2t-t^2)\ ,\ \ A=2a_1 t\ ,
\end{equation}
where $t\equiv s^2=\sin^2\frac{\theta_0}{2}$ is the gap parameter.
The degree $p^2+p$ equation $\det(R-{\bf 1})=0$ of $t$
and the solutions are given by
\begin{equation}\label{p=1-quadratic}
  t^2-2t+\frac{1}{a_1}=0\ \rightarrow\ \
  t\left(=\sin^2\textstyle{\frac{\theta_0}{2}}\right)
  =t_\pm\equiv 1\mp\sqrt{1-\frac{1}{a_1}}\ .
\end{equation}
The two solutions with upper/lower signs are called the saddles
$g_\pm$ in \cite{Copetti:2020dil}, respectively. $\rho_1$ is given by
$\rho_1=\frac{1}{2a_1t}$. The function $\rho(\theta)$
is given by
\begin{equation}
  \rho(\theta)=\frac{\cos\frac{\theta}{2}}{\pi t}
  \sqrt{t-\sin^2\frac{\theta}{2}}\ .
\end{equation}
Integrating this, one obtains $s(\theta)=\int d\theta \rho(\theta)$
given by (\ref{s-p1}). Demanding ${\rm Im}[s(\theta)]=0$ determines the cut.
For all $x$ in the range $|x|<1$, the cut connecting $\pm\theta_0$ through
$\theta=0$ exists. The `free energy' at these
saddles can be computed from (\ref{potential-eom}) and (\ref{free-constant}).
The result is
\begin{equation}
  \log Z_\pm=\frac{N^2\mu_\pm}{2}=
  \frac{N^2}{2}\left[-1+\log t_\pm+\frac{1}{t_\pm}\right]
\end{equation}
at the two saddle points $g_\pm$ given by (\ref{p=1-quadratic}).
Throughout this section, we study the 1-parameter index (\ref{index-1-para}),
for which $a_1=1-\frac{(1-x^2)^3}{(1-x^3)^2}$ with a complex $x$. To study
the grand canonical phases, we study ${\rm Re}\left[\log Z\right]$ which determines
the dominant saddle point. We also compare it with the free energy
${\rm Re}(\log Z_{\rm BH})$ of the BPS black holes of
\cite{Gutowski:2004ez}, in the form
presented in \cite{Hosseini:2017mds,Agarwal:2020zwm}:
\begin{equation}\label{BH-free-equal}
  \log Z_{\rm BH}=\frac{N^2}{2}\frac{\Delta^3}{\omega^2}\ \ ,
  \ \ \ x=e^{-\frac{\omega}{3}+\frac{2\pi i}{3}}=-e^{-\frac{\Delta}{2}}\ .
\end{equation}
Fig. \ref{p1-free-1}(a) shows the contour plots of
${\rm Re}(\log Z_+)$ for the saddle $g_+$, as a function of
$x=|x|e^{i\phi}$. We have only shown the plots in the region
$0<\phi<\pi$, since the remaining region $-\pi<\phi<0$ is the complex
conjugate region related by the map $\phi\rightarrow-\phi$.
The free energy ${\rm Re}(\log Z_{\rm BH})$ of the AdS black hole is
plotted in Fig. \ref{p1-free-1}(b). Fig. \ref{p1-free-2} shows a similar
plot for the saddle $g_-$.

We start by mentioning that, in both Yang-Mills matrix model and
the truncated models,
there are confining saddle points in which $\rho(\theta)$ is constant
along the real $\theta$ circle. This is an ungapped distribution, not captured
by the ansatz of section 2. Its free energy is $\log Z=0$ at $N^2$ order.
This is dual to the thermal graviton saddle. We physically
believe this is the dominant saddle at sufficiently
low $|x|$ \cite{Aharony:2003sx,Kinney:2005ej}. Note here that the truncation
to $p=1$ is a very good approximation at small $|x|$,
since $|a_1|\gg |a_2|\gg |a_3|\gg \cdots$. So the confining saddle
should be dominant also for the $p=1$ model at low $|x|$.
With these understood, let us discuss Fig. \ref{p1-free-1}(a).
There are four regions separated by three red lines for
${\rm Re}(\log Z_+)=0$. In the low temperature region
bounded by the leftmost red line, one finds $N^{-2}{\rm Re}(\log Z_+)>0$.
Had this saddle $g_+$ been physical there, it would have been
more dominant than the graviton saddle. This should not be correct.
So the saddle $g_+$ should be irrelevant
for small enough $|x|$, not being on the matrix integration contour.

On the other hand, consider the region in Fig. \ref{p1-free-1}(a) near
the middle red line. If this saddle point is relevant in this region, the red curve
is the deconfinement transition point. Comparing this curve with the
blue dashed curve representing the Hawking-Page transition, one finds that
they exhibit fairly good qualitative agreement.
So we empirically learn that the saddle point $g_+$ near this region
should be on the integration contour. Combining
this with our observation in the previous paragraph, we conclude that there
should be a Stokes' phenomenon of this saddle point at certain intermediate value
of $|x|$. Namely, as we increase $|x|$ at given $\phi$, we expect
the steepest descent contour to pass through $g_+$ beyond certain
threshold. Checking this is beyond our scope, so we shall leave it as
a conjecture.

Both deconfinement and the Hawking-Page transition
happen within the tachyonic region of $\rho_1$, enclosed by the black dotted
curve of Fig. \ref{p1-free-1}(a). The transition of the $p=1$ model is delayed
relative to the tachyon threshold \cite{Copetti:2020dil}, but still lower
than the Hawking-Page transition. The gap between the two transition points
will decrease in the higher $p$ models.
As we change $\phi$, the apparent transition temperatures $\sim -\log |x|$ change
as shown by the middle red line or the blue dashed line.
As explained in section 2, this is just an apparent delay of the
transition caused by the cancelations of the nearby
$\Omega_q$'s at non-optimal $\phi$.
The transition temperature is minimal at certain optimally
tuned $\phi$. These minima are the actual transition points of the $p=1$
model and the black holes as seen by the index.

Related to the apparent delay at different $\phi$'s,
one also finds a strange region in
Fig. \ref{p1-free-1}(a) at high temperature, on the right side
of the rightmost red curve. Since ${\rm Re}(\log Z)_+<0$
in this region, the system looks
apparently confining. We also interpret this as coming from the non-optimal
choice of $\phi$. The optimal choice
is $\phi=\frac{2\pi}{3}$ when $|x|\rightarrow 1^-$,
which is the large black hole region \cite{Choi:2018hmj,Choi:2018vbz}.
Similar non-optimal region at high temperature with ${\rm Re}(\log Z)<0$ will
continue to appear in the higher $p\geq 2$ models, which we shall interpret
similarly. One may be unconfident about this because
a similar high temperature region does not exist in Fig. \ref{p1-free-1}(b),
so that the qualitative agreement between
$g_+$ and the black hole saddle seems to break down here.
This apparent mismatch is an artifact of the $p=1$ truncation.
We shall study the higher $p\geq 2$ model in detail, with
$p^2+p$ branches of saddle points. There the branches analogous to $g_+$
have no apparently confining high temperature region and behave like Fig.
\ref{p1-free-1}(b).
We find that the exotic high temperature confining region shows up in
a different branch.

We also discuss the saddle $g_-$. The contour plot of
${\rm Re}(\log Z_-)$ is shown in Fig. \ref{p1-free-2}.
We find that there are no reasons to trust that this
saddle point is relevant for the large $N$ physics in any temperature
range. Firstly, ${\rm Re}(\log Z_-)$ is positive at very low temperature,
meaning that $g_-$ should not be on the integration contour at low $|x|$.
As we increase the temperature, ${\rm Re}(\log Z_-)$ just remains positive
all the way to infinite temperature, except in some small corners of the
parameter space which will never be important.
In particular, one finds ${\rm Re}(\log Z_-)>0$ on the deconfinement
curve ${\rm Re}(\log Z_+)=0$ of this model. See the dashed red line of Fig.
\ref{p1-free-2}. So the presence of $g_-$ on the integration contour in the intermediate temperature region would spoil the deconfinement physics of
the $g_+$ saddle. So we conjecture that the saddle $g_-$ will have
no relevance to the large $N$ physics at any temperature region. In the higher
$p\geq 2$ models, many of the $p^2+p$ solutions partly behave like $g_-$.

We now discuss the Legendre transformation of  $\log Z$
at real positive charge $q\equiv 6(R+J_+)$. $|x|$ and $\phi$ are
determined in terms of $q$.
One can understand this calculus in two different ways.
Firstly, this can simply be regarded as considering the microcanonical ensemble.
Secondly, one can interpret the results in the grand canonical ensemble at fixed
$|x|$. Holding $|x|$ fixed and letting $q$ to vary, phase transitions can happen
by absorbing latent heat. In this picture, $\phi(q)$ is viewed as a function
of $|x|$. $\phi(|x|)$ is optimally tuned to
minimize the cancelations of nearby $\Omega_q$'s at fixed $|x|$.
As explained in section 2, this freezing of $\phi$ allows one to extract
the proper information of $|\Omega_q|$'s without the phase factors obscuring
the physics.

\begin{figure}[!t]
\centering
\includegraphics[width=0.45\textwidth]{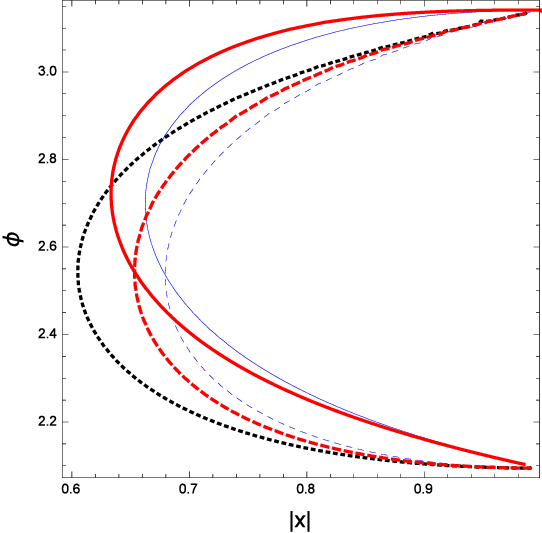}
\caption{Saddles of Legendre transformation at real charges $0<q<\infty$
for the $g_+$ saddle (solid red) and the black holes (solid blue).
We also show the red/blue dashed lines for the deconfinement and
the Hawking-Page transitions, respectively. Plots shown only near the tachyonic
region of $\rho_1$ ($\frac{2\pi}{3}<\phi<\pi$),
whose boundary is the black dotted line.}
\label{p1-legendre}
\end{figure}

We consider the $g_+$ saddle only, inside or near the
tachyonic region of $\rho_1$. We extremize
\begin{equation}
  \log Z_+(x)-q\log x
\end{equation}
in $x$. The resulting $x(q)$ is a curve in the $|x|$-$\phi$ space.
This is shown as the red solid line of Fig. \ref{p1-legendre}.
For comparison, we also show $x(q)$ obtained by
Legendre transforming the black hole free energy $\log Z_{\rm BH}$
by the blue solid line.
We have also shown the deconfinement and the Hawking-Page transition points
by the red/blue dashed lines, respectively.
Both solid curves start from $|x|=1$, $\phi=\pi$ at $q=0$, and ends at
$|x|=1$, $\phi=\frac{2\pi}{3}$ at $q=\infty$. For black holes, they are
the small/large black hole limits, respectively.
As the charge $q$ increases,
on both curves the temperature decreases for certain while until $|x|$ reaches
its minimum. After passing the minimum, the temperature increases. On the two
branches, the specific heat (or more precisely the susceptibility) of the
system is negative/positive, respectively.
One finds that the saddle points of the $p=1$ model shows similar behaviors
as the black holes. When the solid curve crosses the dashed line with same color,
a phase transition happens in the grand canonical ensemble
which holds $|x|$ fixed and $\phi(|x|)$ tuned.
For black holes, this defines the Hawking-Page transition. For the
$p=1$ model, this is the deconfinement transition.
For both black holes and the matrix model, the
transitions happen precisely at the minimal transition temperature on
the dashed lines.

Now in the microcanonical viewpoint, the saddles $x(q)$ of the $p=1$
model end precisely on the large and small
black hole limits. The large black hole limit is well understood
analytically \cite{Choi:2018hmj,Choi:2018vbz} from QFT.
The small black hole limit is not well understood so far.
So we expect the truncated models to provide useful insights,
on which we shall elaborate in section 5.
Defining $x\equiv -e^{-\beta}$, the small black hole limit
is given by $\beta\rightarrow 0$. At small $\beta$, one finds
\begin{equation}
  (t-1)^2\approx -2\beta^{3}\ \ ,
  \ \ \log Z\approx -\frac{N^2}{2}\beta^3\ .
\end{equation}
We call the branch with these scalings in $\beta$ as the `standard' branch
for the small black holes, as there will always exist such a branch at
arbitrary $p$. (The coefficients will depend on $p$.)

\subsection{The $p=2$ model}

In this case, the matrix $R$ and the vector $A$ are given by
\begin{equation}
  R=\left(\begin{array}{cc}
    a_1(2t-t^2)&4a_2t(1-t)^2\\
    2a_1t(1-t)^2&a_2 t(4-14t+20t^2-9t^3)
  \end{array}\right)\ ,\ \
  \vec{A}=\left(\frac{}{}\!2a_1t~,~2a_2(2t-3t^2)\right)\ .
\end{equation}
The degree $p^2+p=6$ polynomial equation for $t$ is given by
\begin{equation}\label{polynomial-p=2}
  1-2(a_1+2a_2)t+(a_1+14a_2)t^2-20a_2t^3+3a_2(3+2a_1)t^4
  -6a_1a_2 t^5+a_1a_2 t^6=0\ .
\end{equation}
One finds six distinct one-cut saddle points
for the six solutions $t_a(x)$, $a=1,\cdots,6$. We shall study them numerically
below. At each saddle point with given $t=t_a(x)$,
one finds
\begin{equation}
  \rho_1=\frac{1-4a_2t+14a_2t^2-20a_2t^3+9a_2t^4}{2a_1t(1-4a_2t^3+3a_2t^4)}\ ,
  \ \ \rho_2=\frac{(1-t)^2}{1-4a_2t^3+3a_2t^4}
\end{equation}
and
\begin{equation}
  Q_1=2a_1\rho_1+2a_2\rho_2(1-2t^2)\ ,\ \
  Q_2=2a_2\rho_2\ .
\end{equation}
The free energy $\log Z$
is given by
\begin{equation}
  \log Z=\frac{N^2}{2}\left[\left(tQ_1+(t-t^2)Q_2\right)\log t
  -\left(tQ_1+(t+{\textstyle \frac{t^2}{2}})Q_2\right)
  +2a_1\rho_1+a_2\rho_2\right]\ .
\end{equation}

Before proceeding, we comment on labeling the
six solutions $t_a(x)$. Numerically solving
(\ref{polynomial-p=2}) at various $x$, Wolfram Mathematica labels
the six roots in the order of increasing real part, which
causes discontinuities in $x$. We want to label the six branches so that
$t_a(x)$ are all continuous functions of $x$. To do so,
we discretize the $|x|$-$\phi$ space into small grids and solve the
polynomial equation to get $t_a(x)$ in each grid. (We use
$1001\times 1001$ grids for $p=2$ plots in this subsection, and
less refined $40\times 34$ grids for more demanding $p=3$ plots in the
next subsection.) Then we reorder
them if necessary to make $t_a(x)$ to behave `continuously' within our
discretized setup.
This strategy exhibits ambiguities in some regions,
because branch cuts may develop from the degenerate roots.
This problem did not arise at $p=1$ since the branch points were all
at $|x|=1$, so that we can choose the branch cuts in
the unphysical region $|x|>1$ and ignore them.
For the internal branch cuts, quantities are continuous only
after branch mixings. At $p=2$, degenerate roots can be found by solving
(\ref{polynomial-p=2}) together with the equation obtained by taking
$t$ derivative of this polynomial to vanish,
\begin{equation}
  -(a_1+2a_2)+(a_1+14a_2)t-30a_2t^2+6(3+2a_1)t^3-15a_1a_2 t^4+3a_1a_2 t^5=0\ .
\end{equation}
Solving these equations, one finds two
internal branch points for the triple roots of $t$:
\begin{eqnarray}\label{p=2-branch-point}
  t\approx.2727+.1198i&:&x\approx .8503\exp\left[.3793\pi i\right]\equiv x_1\\
  t\approx .0016+.2655i&:&x\approx .8003\exp\left[.7256\pi i\right]\equiv x_2\ .
  \nonumber
\end{eqnarray}
Three of the six branches mix around each branch point.

\begin{figure}[!t]
\centering
\includegraphics[width=.6\textwidth]{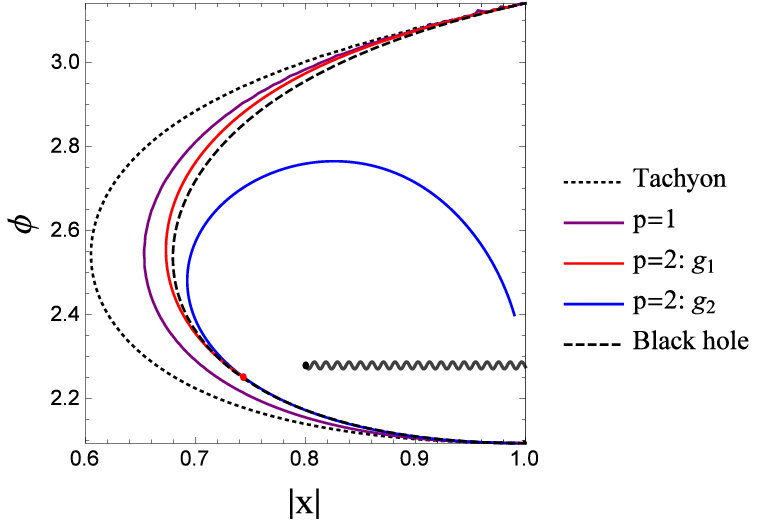}
\caption{`Local' transition points ${\rm Re}(\log Z_a)=0$
for the saddles $g_1,g_2$.
We also display the points of: Hawking-Page transition (black dashed),
transition of the $p=1$ model (purple), $\rho_1$ tachyon
threshold (black dotted). Branch cut which mixes
$g_1,g_2$ are also shown.}
\label{p2-free}
\end{figure}

As explained at the end of section 2, the eigenvalue distribution
within our ansatz should form a cut ending on $\theta=\pm\theta_0$,
passing through $\theta=0$. Depending on the choice of the branch $g_a$
(where $a=1,\cdots,6$), such a cut does not exist in some region of $x$.
We shall only show the two branches, which we label $a=1,2$, which
exhibit nontrivial physics near the $\rho_1$ tachyon region (which
we take to be $0.6<|x|<1$ and $\frac{2\pi}{3}<\phi<\pi$). Some branches
do not exist in this region, and other branches do not exhibit
proper physics (like the $g_-$ saddle of the $p=1$ model).
For simplicity, in Fig. \ref{p2-free} we only display
the ${\rm Re}(\log Z_a)=0$ lines for the $g_1$ and $g_2$ saddles around
the $\rho_1$ tachyon region. These are the lines above which the
saddle $g_a$ locally becomes more dominant than the thermal graviton saddle.
At each $\phi$, the curve with lower $|x_a|(\phi)$ would determine the
deconfinement transition temperature. One finds that the minimum curve
$\min(|x_1|(\phi),|x_2|(\phi))$ is closer to the Hawking-Page
temperature (dashed line) than the deconfinement temperature of the
$p=1$ model.

We note that the saddle $g_1$ does not exist in the lower-right region
of the figure, since the eigenvalue cut connecting $\pm\theta_0$ does
not exist. Along the line ${\rm Re}(\log Z_1)=0$ (solid red
line for $g_1$), the cut does not
exist beyond the red point of Fig \ref{p2-free}. The shapes of the cuts
along this line are illustrated by Fig. \ref{cut-change}.
In particular, Fig. \ref{cut-change}(b) shows the cut when
$x$ is on the red point of Fig. \ref{p2-free}.
The cut is just about to disappear at this point.
As explained in section 2, this does not mean that this saddle point
suddenly disappears. It rather implies that the single cut distribution
should undergo a phase transition to a triple cut distribution beyond
the red point. Beyond this point, we find that the $g_2$ branch
describes the Hawking-Page transition (black dashed) fairly well. Also, before
$g_1$ disappears, the two transition temperatures for $g_1,g_2$ are
very close. ($g_2$ is slightly more dominant.) We therefore do not
attempt to construct the triple cut solution after $g_1$ disappears.
To conclude, we find that multiple branches are patched to describe
the deconfinement transition of this model. This feature will be
more important below, when we study the saddle points of the
Legendre transformation.

\begin{figure}[!t]
\centering
\includegraphics[width=.6\textwidth]{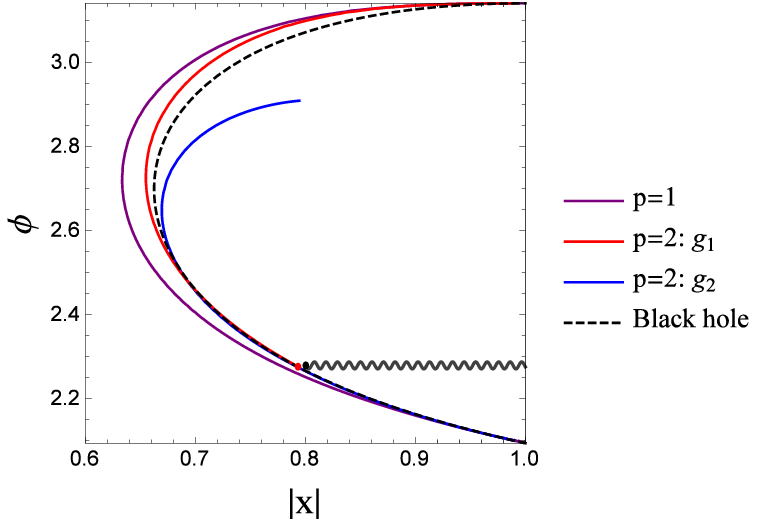}
\caption{Legendre transformation lines with macroscopic entropies.}
\label{p2-legendre}
\end{figure}

Now we study the Legendre transformation,
extremizing $S_a(q;x)=\log Z_a(x)-q\log x$ at
$q\equiv 3(2R+J_1+J_2)>0$. Again, we show the results for $a=1,2$
around the tachyonic region of $\rho_1$.
The results are shown in Fig. \ref{p2-legendre}, when the corresponding
macroscopic entropies $S_a(q)$ are positive.
Let us first explain the aspects of two branches $g_1$ (red), $g_2$ (blue) in turn.

The red curve denotes the Legendre transformation curve in the
$g_1$ branch. The curve starts from $|x|=1$, $\phi=\pi$ on the
upper right corner at small charge $q=0$. As we move along the curve
from this point, $q$ increases until it ends on the red point.
Just like Fig. \ref{p2-free}, the eigenvalue cut does not to exist
beyond this point. This happens at a finite nonzero charge $q$.
Section 4 will analytically explain why this saddle cannot exist
all the way up to the large charge limit $q\rightarrow\infty$.
The gap and the free energy in the small charge limit
are given by
\begin{equation}
  (t-1)^2\approx-2\beta^{3}\ \ ,\ \
  \log Z_1\approx-\frac{N^2}{2}\beta^3\ .
\end{equation}
The coefficients are accidentally same as the $p=1$ model,
which will not be true for $p\geq 3$.

The blue curve denotes the Legendre transformation curve in the
$g_2$ branch. We start to consider this curve from its end
$|x|=1$, $\phi=\frac{2\pi}{3}$ on the lower right corner at large
charge $q\rightarrow\infty$. As we move along the curve from
this point, $q$ decreases until we stop displaying the curve
at a finite nonzero charge (also at $|x|<1$). The saddle point
continues to exist beyond this endpoint, but the entropy
$S_2(q)$ becomes negative beyond the part shown in Fig. \ref{p2-legendre}.
The saddle point with negative entropy may still play some role to
describe the subleading corrections to the large $N$ free energy,
but will not describe any black holes.

At small $q$, only one saddle $g_1$ exists. This qualitatively describes
the black hole (black dashed line) better than the $g_+$ saddle of
the $p=1$ model. As $q$ is increased, the red and blue curves approach
very close to each other before the red curve disappears.
At the charge of the red point, the entropies of the two saddles are
very close to each other. The combination of the red curve (when it exists)
and the blue curve (when the red one does not exist) describes the black
hole (black dashed) better than the purple curve of the $p=1$ model.
It is again very crucial that multiple branches have to be combined to
describe the known black holes. We will show that this will continue to
be true, perhaps in a more dramatic manner, in the $p=3$ model
(section 3.3) and the $p=\infty$ model (section 5.1, small charge limit).

Although we do not explicitly show the results here, we have also
found the saddle points of the Legendre transformation in
the region outside Fig. \ref{p2-legendre}. In particular,
we find saddle points in the small charge limit
$|x|=1$, $\phi=\frac{\pi}{2}$ around the tachyonic region of $\rho_2$.
The solutions we report here all have one cut. We think one also
has to consider the two cut saddle points to fully understand the structures
of possible black hole like saddles in the $\rho_2$ tachyon region.
Although it is likely that the $\rho_1$ tachyon region plays
the most important role in the AdS thermodynamics,
$\rho_2$ tachyon region may also host
interesting black holes. We hope to come back to this subject
in the near future, with more quantitative and analytic understandings.

\subsection{The $p=3$ model}

The matrix $R$ and the vector $A$ are given by
\begin{eqnarray}
  R\!&\!=\!&\!\left(\!\begin{array}{ccc}
    a_1(2t-t^2)&4a_2t(1-t)^2&3a_3t(1-t)^2(2-5t)\\
    2a_1t(1\!-\!t)^2&a_2 t(4\!-\!14t\!+\!20t^2\!-\!9t^3)
    &6a_3t(1\!-\!t)^2(1\!-\!4t\!+\!6t^2)\\
    a_1t(1\!-\!t)^2(2\!-\!5t)&4a_2t(1\!-\!t)^2(1\!-\!4t\!+\!6t^2)&
    a_3t(6\!-\!51t\!+\!200t^2\!-\!366t^3\!+\!312t^4 \!-\!100t^5)
  \end{array}\!\right)\nonumber\\
  \vec{A}\!&\!=\!&\!\left(\frac{}{}\!2a_1t~,~2a_2(2t-3t^2)~,~
  2a_3 t(3-12t+10t^2)\right)\ .
\end{eqnarray}
The degree $p^2+p=12$ polynomial equation for $t$ is given by
\begin{eqnarray}
  0&=&1-2(a_1+2a_2+3a_3)t+(a_1+14a_2+51a_3)t^2-20(a_2+10a_3)t^3\\
  &&+(9a_2+6a_1a_2+366a_3+64a_1a_3+50a_2a_3)t^4
  -(6a_1a_2+312 a_3 + 224 a_1a_3 + 250 a_2a_3)t^5\nonumber\\
  &&+(a_1a_2 +100a_3 + 288a_1a_3 + 535 a_2a_3)t^6
  -a_3(152 a_1 + 640 a_2) t^7+ a_3(25 a_1 +470 a_2) t^8
  \nonumber\\
  &&  -20a_2a_3(10 + a_1) t^9 + a_2a_3(36 + 30 a_1) t^{10}
  - 12 a_1 a_2 a_3 t^{11} + a_1 a_2 a_3 t^{12}\ .\nonumber
\end{eqnarray}
$\rho_{n}$ and $Q_n$ are given by
(\ref{rho-formal-sol}) and (\ref{Q-definition}), respectively.
$\log Z$ is given from (\ref{potential-eom}) and
(\ref{free-constant}) by
\begin{eqnarray}
  \log Z&=&\frac{N^2}{2}\left[
    \left\{tQ_1 + (t-t^2) (Q_2 + (1-2t)Q_3)\right\} \log t
    -t ({\textstyle Q_1 + (1+\frac{t}{2})Q_2 +
    (1+ \frac{3}{2} t- \frac{5}{3} t^2)Q_3})\right.\nonumber\\
    &&\hspace{.7cm}
    \left.+  2a_1\rho_1+a_2\rho_2+{\textstyle \frac{2}{3}}a_3\rho_3\right]\ .
\end{eqnarray}
Using these formulae, we computed the $12$ branches of $t_a(x)$
and $\log Z_a$ numerically. Although all not explicitly shown below,
we carefully chose the directions of various branch cuts.

\begin{figure}[!t]
\centering
\begin{subfigure} [b]{0.48\textwidth}
\includegraphics[width=\textwidth]{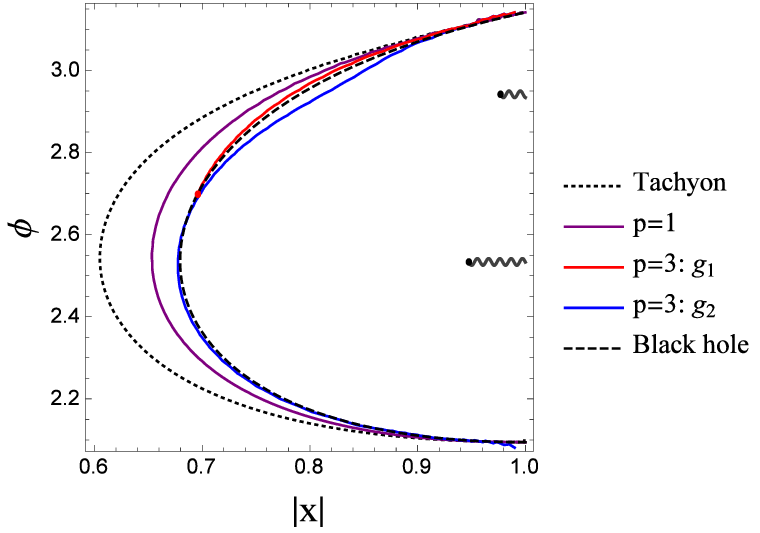}
\subcaption{${\rm Re}(\log Z_a)=0$ lines}
\end{subfigure}
\hspace{0.3cm}
\begin{subfigure} [b]{0.48\textwidth}
\includegraphics[width=\textwidth]{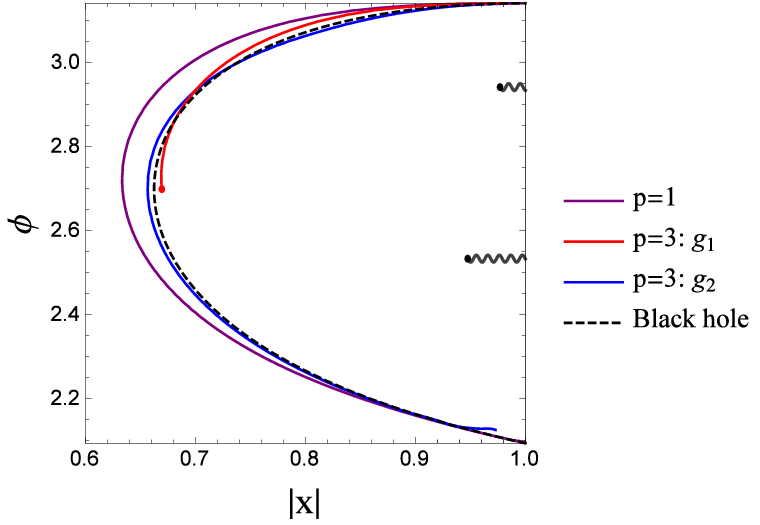}
\subcaption{Legendre transformations}
\end{subfigure}
\caption{Two branches $g_1,g_2$ around the tachyonic region of $\rho_1$.
Blue curves are imprecise near the large charge limit
due to the coarse resolution. Two branch cuts are also shown.
The upper cut mixes $g_2$ and other branches, while the lower cut mixes
$g_1,g_2$ and other branches.}
\label{p3-deconfine-legendre}
\end{figure}

Around the tachyonic region of $\rho_1$, we find $2$ branches which
exhibit nontrivial black hole like behaviors. See Fig. \ref{p3-deconfine-legendre}.
Other branches are all irrelevant either in the sense of the $g_-$ saddle
of the $p=1$ model, or because the cut does not exist in this region.
Again we name the two  branches $g_1$ (red), $g_2$ (blue).
We first take a look at the local deconfinement points
${\rm Re}(\log Z_a)=0$. The lower $|x|$ between the two at each $\phi$
describes the known black hole's Hawking-Page transition (black dashed)
much better than the $p=1$ model as shown in Fig. \ref{p3-deconfine-legendre}(a).
Although we did not display the $p=2$ curves together,
one can notice an improvement by comparing with Fig. \ref{p2-free},
especially in the upper region.

Fig. \ref{p3-deconfine-legendre}(b) shows the Legendre transformation
curves of both branches. Both curves start from $|x|=1$, $\phi=\pi$ at
$q=0$. As will be explained in more detail in section 5.1, the
small $q$ or $\beta$ behaviors of these two saddles are somewhat different.
The saddle $g_1$ exhibits the small $\beta$ scaling rather similar
to the $g_+$ saddle of the $p=1$ model and the $g_1$ saddle of
the $p=2$ model:
\begin{equation}
  (t-1)^2\approx-\beta^{3}\ ,\ \
  \log Z\approx-\frac{9N^2}{20}\beta^3\ .
\end{equation}
On the other hand, the $g_2$ saddle will exhibit a gap
with $(t-1)^4\propto \beta^{3}$, while $\log Z$ is still proportional
to $\beta^3$ with a different coefficient. We shall analytically study both
types of small charge limit at general $p$ in section 5.1.

As we move along the Legendre transformation curves, $q$ increases.
For $g_1$, the eigenvalue cut does not exist beyond the red point of
Fig. \ref{p3-deconfine-legendre}(b). On the other hand, $g_2$
continues to exist all the way to the large charge limit $q\rightarrow\infty$.
We find that both branches are rather close to the black hole curve
(black dashed) when they exist.

\section{Large black holes}

The large charge limit has been analytically
studied in the literature, at $p=\infty$. We reconsider
this problem in the setup of section 2, at general finite $p$ (which admits
the limit $p\rightarrow\infty$). Among others, we shall gain better
insights on the numerical results of section 3.

In the grand canonical viewpoint, the large charge limit
amounts to taking temperature large $|x|\rightarrow 1^-$ while tuning the
phase to $\phi=\frac{2\pi}{3}$ \cite{Choi:2018hmj,Choi:2018vbz}. We shall
study the degree $p^2+p$ polynomial equation
(\ref{det-gap}), ${\rm det}(R-{\bf 1})=0$, approximately at
small $\beta$, defined by $x=e^{\frac{2\pi i}{3}}e^{-\beta}$.\footnote{
$\beta$ in this section is different from $\beta$
in sections 3 and 5 for small black holes, $x=-e^{-\beta}$.} We first summarize
the small $\beta$ scalings of the roots
$t=\sin^2\frac{\theta_0}{2}$, which we learn by studying sufficiently many
cases of $p$. We are interested in the cases in which the cut length
asymptotically shrink to zero, $\theta_0\rightarrow 0$. This is natural
since the system would want to maximally deconfine
in the high temperature limit \cite{Aharony:2003sx,Choi:2018hmj}.
To make a general classification of the roots with this behavior, let us call
$p=3l+m$, where $l\in\mathbb{Z}_{\geq 0}$ and $m=0,1,2$. At the leading
order in the $\beta\rightarrow 0^+$ limit, one finds that the polynomial behaves like
\begin{equation}
  \det(R-{\bf 1})\sim t^{(2l+m)^2}(1-t)^{l^2+l} \times
  \left[\textrm{degree }(2l+1)(2l+m)\textrm{ polynomial}\right]\ .
\end{equation}
Namely, there is a $(2l+m)^2$-fold degeneracy at $t=0$.
We study how this degeneracy at $t=0$ splits at
small nonzero $\beta$. One finds that the $(2l+m)^2$ roots of
$t^{(2l+m)^2}=0$ are split into
\begin{eqnarray}
  1\textrm{ root}&:&t\sim\beta^2\\
  3\textrm{ roots}&:&t^{3}\sim\beta^2\nonumber\\
  &&\vdots\nonumber\\
  2(2l+m)-3\textrm{ roots}&:&t^{2(2l+m)-3}\sim\beta^2\nonumber\\
  2(2l+m)-1\textrm{ roots}&:&t^{2(2l+m)-1}\sim\beta^2\ .\nonumber
\end{eqnarray}
For instance, at $p=1$ (i.e. $l=0$, $m=1$), there is only one root having
small $t$ at $t\sim\beta^2$. This is the $g_+$ branch of section 3.1.
At $p=2$ ($l=0$, $m=2$), there are $4$ roots with small $t$.
One root at $t\sim \beta^2$ is the $g_2$ branch of section 3.2.
Also, there are three roots at $t\sim\beta^{\frac{2}{3}}$, one of which
being the $g_1$ branch of section 3.2.
At $p=3$ ($l=1$, $m=0$), again there are $4$ roots with small $t$.
One root at $t\sim \beta^2$ is the $g_2$ branch of section 3.3, while
there are three roots at $t\sim\beta^{\frac{2}{3}}$. One of these three
is the $g_1$ of section 3.3.

For all $p$'s, there is a unique root at $t\sim\beta^2$, or $\theta_0\sim\beta^1$.
This is the natural splitting that one expects from the eigenvalue dynamics.
To explain this, we first note the divergent behaviors of
$a_n(x)$ in the small $\beta$ limit. One finds
\begin{equation}
  a_n(x)=1-\frac{(1-x^{2n})^3}{(1-x^{3n})^2}=
  1-\frac{(1-e^{\frac{4\pi in}{3}}e^{-2n\beta})^3}{(1-e^{-3n\beta})^2}\sim
  \left\{\begin{array}{ll}
    \frac{(1-e^{\frac{4\pi in}{3}})^3}{9n^2\beta^2}&\textrm{if }n\notin 3\mathbb{Z}\\
    1&\textrm{if }n\in 3\mathbb{Z}
  \end{array}
  \right.\ .
\end{equation}
With these in mind, we study the potential (\ref{2-body-potential})
between two eigenvalues separated at distance $\theta$,
and also its force $V^\prime(\theta)$ given by
\begin{equation}
  V^\prime(\theta)=-\cot\frac{\theta}{2}+2\sum_{n=1}^p
  a_n(x)\sin(n\theta)\ .
\end{equation}
The first term, coming from the Haar measure of the matrix model, diverges
if the eigenvalues get close to each other. On the other hand, the second
term becomes large for $n\notin 3\mathbb{Z}$ in the large black hole limit
from $a_n\sim\frac{1}{\beta^2}$. The large $N$ saddle point equation
in the small $\beta$ limit thus demands
\begin{equation}\label{force-large}
  0=\sum_{b(\neq a)}V^{\prime}(\alpha_{ab})=
  \sum_{b(\neq a)}\left[-\cos\frac{\alpha_{ab}}{2}
  +2\sum_{n=1}^pa_n(x)\sin(n\alpha_{ab})\right]\sim
  \sum_{b(\neq a)}\left[-\frac{2}{\alpha_{ab}}
  +\frac{\#_p\alpha_{ab}}{\beta^2}\right]
\end{equation}
with certain $p$-dependent coefficient $\#_p$.
We used the fact that $\alpha_a$'s want to coalesce in the
small $\beta$ limit, and made small $\alpha_{ab}$ expansions.
The balance of the two terms in (\ref{force-large}) naturally
demands $\alpha_{ab}\sim\beta$ for most of the pairs. This is
why the $\theta_0\sim\beta$ scaling is natural. The $p\rightarrow\infty$
version of this analysis exhibits slightly different intermediate steps, but
still leads to the $\theta_0\sim\beta$ scaling
in the large black hole limit \cite{GonzalezLezcano:2020yeb}.
This unique root with $t\sim\beta^2$ scaling yields the Cardy
limit of this partition function. For $p=1,2,3$, we have explicitly
seen that the cut exists in section 3 ($g_+$ or $g_2$ saddles).
We expect this to be true for general $p$.

For all other roots with small $t$, one finds that
$\theta_0$ approaches zero at much slower rates than $\theta_0\sim\beta^1$.
If one investigates the structure of the force balancing
equation (\ref{force-large}), such a slow coalescence appears to be
impossible. This is compatible with the numerical results of section 3.
Namely, for the $g_1$ branches of the $p=2,3$ models with
$t\sim\beta^{\frac{2}{3}}$ scalings, we have found that the cut does
not exist in a region which contains the large black hole limit.
We expect that this phenomenon will continue to be true for higher $p$.

It appears that this illustrates the universality of the Cardy limit of
\cite{Choi:2018hmj} near the large black hole point $|x|=1$, $\phi=\frac{2\pi}{3}$.
Namely, although the matrix model can have multiple branches of saddle points
at given $x$ in general, their structures tend to be simpler in the large
black hole limit. This morally sounds somewhat similar to the universality of
the 2d Cardy limit. This is in sharp contrast to the situations away from
the large black hole limit. In section 3, we already saw that
multiple black hole like saddle points may exist at fixed $x$ or fixed $q$.
In section 5.1, this will be more concretely illustrated in the small black hole
limit.

However, we should comment that there are possible
caveats of the Cardy universality in 4d gauge theories. Firstly, the universal
behavior we explained above (by having all but one saddle points forbidden)
is strictly within the single cut ansatz. Although
the single cut saddle point provides the dominant physics in some
region of $x$ (probably the most interesting region),
multi-cut saddles may be more dominant in
other regions. Curiously, this possibility exists near the Cardy
point $|x|=1$, $\phi=\frac{2\pi}{3}$. Note that in section 3 we studied the physics
of the single cut saddles in the $\rho_1$ tachyon region. This region
exists for $\phi>\frac{2\pi}{3}$. However, if one approaches the
Cardy limit $|x|\rightarrow 1$, $\phi\rightarrow\frac{2\pi}{3}$ from the
$\phi<\frac{2\pi}{3}$ side, there is a reason to believe that physics is
richer. In terms of the chemical potential $\omega$ appearing in
(\ref{BH-free-equal}), $\phi<\frac{2\pi}{3}$ corresponds to
${\rm Im}(\omega^2)>0$. It has been first noticed in \cite{ArabiArdehali:2019tdm}
that there can be nontrivial holonomy saddle issues in this region,
where the eigenvalues $\alpha_a$ do not necessarily want to coalesce.
In fact such issues exist in a wide class of 4d $\mathcal{N}=1$ SCFTs
studied in \cite{Kim:2019yrz}. Although the main focus of
\cite{Kim:2019yrz} was the region ${\rm Im}(\omega^2)<0$ in which the
eigenvalues all want to coalesce, the sign-flipped matrix model potentials
of \cite{Kim:2019yrz} can be studied to conclude that there are issues
of nontrivial holonomies when ${\rm Im}(\omega^2)>0$.
At least for the maximal super-Yang-Mills theory, we think that a
natural class of large $N$ solutions in this region is the two-cut saddles.
Note also that the tachyonic region of $\rho_2$ is within $\phi<\frac{2\pi}{3}$.

To summarize, we found that there is a certain sense of Cardy universality
in our matrix models, but with possible caveats which could make the large
charge physics richer. We wish to study this problem in more detail
as a separate project.

\section{Small black holes}

\subsection{1-parameter solutions}

We study the small black hole limit of the
1-parameter index (\ref{index-1-para}) analytically.
Defining $x\equiv -e^{-\beta}$, this limit is defined by
$\beta\rightarrow 0$ with ${\rm Re}(\beta)\rightarrow 0^+$.
In both large and small black hole limits, the `inverse temperature'
asymptotically vanishes, ${\rm Re}(\log x^{-1})\rightarrow 0^+$. This means
that both limits are
the BPS analogues of the high temperature limit. Large black holes
represent the new deconfined high temperature phase of the full quantum
gravity in AdS. Small black holes, whose sizes are much smaller than
the AdS radius, locally behave as asymptotically flat black holes
in many ways. Like Schwarzschild black holes in flat spacetime, they have
negative specific heat (susceptibility). This is why the temperature
diverges in the small black hole limit.

We again start by expanding the polynomial equation (\ref{det-gap})
in small $\beta\rightarrow 0$ for the small black hole. At the leading $\beta^0$
order, we find that all $p^2+p$ roots are degenerate at $t=1$.
Namely, the polynomial reduces to
${\rm det}(R-{\bf 1})\sim (t-1)^{p^2+p}$.
Investigating how this degeneracy is split at small but nonzero $\beta$,
we find the following patterns at generic $p$, which are organized into two classes (A) and (B):
\begin{eqnarray}\label{small-roots}
\left. \begin{array}{lll}
  2\textrm{ roots}&:&(t-1)^2\sim\beta^3\\
  4\textrm{ roots}&:&(t-1)^4\sim\beta^3\nonumber\\
  &&\vdots\nonumber\\
  2\left\lceil \textstyle{\frac{p}{2}} \right\rceil \textrm{ roots}&:&(t-1)^{2\lceil \frac{p}{2} \rceil}\sim\beta^3\nonumber\\
\end{array} \right\} \  \textrm{(A)} \\
\\
\left. \begin{array}{lll}
  2 \left\lceil \textstyle{\frac{p}{2}} \right\rceil + 2 \textrm{ roots}&:&(t-1)^{2 \lceil \frac{p}{2} \rceil + 2 }\sim\beta^1\nonumber\\
  2 \left\lceil \textstyle{\frac{p}{2}} \right\rceil + 4 \textrm{ roots}&:&(t-1)^{2 \lceil \frac{p}{2} \rceil + 4}\sim\beta^1\nonumber\\
  &&\vdots\nonumber\\
  2p\textrm{ roots}&:&(t-1)^{2p}\sim\beta^1\ \nonumber
\end{array} \right\} \  \textrm{(B)}
\end{eqnarray}
where $\lceil x\rceil$ is the ceiling function. (For instance,
$\lceil 1\rceil=1$, $\lceil 1.5\rceil=2$, etc.)
At $p=1$, there are only $2$ roots
at $(t-1)^2\sim\beta^3$, corresponding to the first line of
(\ref{small-roots}). One of these roots describe the small black holes
in the region $\phi<\pi$, while another describes the mirror branch
at $\phi>\pi$ related by $\phi\rightarrow 2\pi-\phi$. From $p\geq 2$,
both classes (A) and (B) shown in (\ref{small-roots}) appear.
For both $p=2,3$, we empirically observe from our numerical studies of
section 3 that the roots of the class (B) at $(t-1)^{2p}\sim\beta^1$ do
not exhibit interesting black hole like behaviors. We shall disregard the roots of the class (B) and study only the roots of the class (A) throughout this paper. For both $p=2,3$,
the saddle $g_1$ exhibits the scaling $(t-1)^2\sim\beta^3$ in
the small charge limit. For $p=3$, $g_2$ also
reaches the small charge limit, with the scaling $(t-1)^4\sim\beta^3$.
Just for the technical reason, we call the first branch at
$(t-1)^2\sim\beta^3$ the `standard' small black hole branch. However,
as far as we can see, there is no fundamental reason to believe that this
branch is more important. In fact, in the
$p\rightarrow\infty$ limit, we shall explain that infinitely many branches of the class (A)
degenerately describe the physics of the known black hole
solutions. (Its possible interpretation will be discussed
at the end of this subsection.)

We expand the functions $P_l$ and $B^{l+\frac{1}{2}}$
appearing in the analysis of section 2. We learned in the previous paragraph
that $u\equiv t-1$ is small at small $\beta$. The functions
can be written as
\begin{equation}
  P_l(1-2t)=(-1)^l\sum_{n=0}^l \frac{(l+n)!}{(n!)^2(l-n)!}u^n
  \ \ ,\ \
  B^{l+\frac{1}{2}}(t)=
  \delta_{l,0}+(-1)^l\sum_{n=0}^{l}
  \frac{(l+n)!}{(n+1)!n!(l-n)!}u^{n+1}\ .
\end{equation}
To study various branches of small black holes,
we make double expansions of the matrix elements $R_{ml}$ in
small $u$ and $\beta$. We first expand
\begin{eqnarray}\label{expand-functions}
  r_{ml}(u)&\equiv&\sum_{k=1}^l\left[B^{m+k-\frac{1}{2}}+B^{|m-k+\frac{1}{2}|}\right]
  P_{l-k}\\
  &=&\delta_{m,l}-(-1)^{l+m}ml^2
  \left[u^2+{\textstyle \frac{m^2+l^2-2}{3}}u^3+
  {\textstyle \frac{11-8(m^2+l^2)+m^4+l^4+3m^2l^2}{24}}u^4
  +\mathcal{O}(u^5)\right]\ .\nonumber
\end{eqnarray}
The terms shown above
will be sufficient for concrete illustrations. All one needs to know about
the general $u^n$ order at $n\geq 3$ is that its coefficient is given by
$ml^2$ times a degree $n-2$ polynomial of $m^2$ and $l^2$.
The equation $(R-{\bf 1})\cdot \rho=0$ which determines
$\rho_n$ can be written up to $u^4$ order as
\begin{eqnarray}\label{eigen-eqn}
  0&=&R_{ml}\rho_l-\rho_m=\left[r_{ml}(u)-a_m^{-1}\delta_{ml}\right](a_l\rho_l)\\
  &=&m\left[
  \frac{1-a_m^{-1}}{m^3}\delta_{m,l}-
  (-1)^{l+m}
  \left(u^2+{\textstyle \frac{m^2+l^2-2}{3}}u^3+
  {\textstyle \frac{11-8(m^2+l^2)+m^4+l^4+3m^2l^2}{24}}u^4
  +\mathcal{O}(u^5)\right)\right]l^2a_l\rho_l\ .
  \nonumber
\end{eqnarray}
Now we use the expansions $a_m^{-1}-1\approx 2m^3\beta^3$ for odd $m$
and $a_m^{-1}-1\approx\frac{8m\beta}{9}$ for even $m$, and rephrase the
above zero eigenvector equation in terms of the even and odd blocks.
$(-1)^ll^2a_l\rho_l\equiv(v_{\rm odd},v_{\rm even})$ has to satisfy the following
equations:
\begin{eqnarray}\label{eigen-expand}
  0&\approx&2\beta^3 (v_{\rm odd})_m+u^2 n_m(n\cdot v)
  +u^3 (M_1\cdot v)_m+u^4 (M_2\cdot v)_m+\mathcal{O}(u^5)\ ,\\
  0&\approx&\frac{8\beta}{9m^2}(v_{\rm even})_m+u^2n_m(n\cdot v)+u^3
  (M_1\cdot v)_m+u^4(M_2\cdot v)_m+\mathcal{O}(u^5)\nonumber\ ,
\end{eqnarray}
where
\begin{equation}
  n_m=1\ ,\ \ (M_1)_{ml}=\frac{m^2+l^2-2}{3}\ ,\ \
  (M_2)_{ml}=\frac{11-8(m^2+l^2)+m^4+l^4+3m^2l^2}{24}\ .
\end{equation}
From this equation, one can construct various leading order solutions
at small $\beta$. They will exhibit various scalings of (\ref{small-roots}),
as we shall explain shortly. Or more generally, starting from
(\ref{eigen-eqn}), one can iteratively construct the small $\beta$ expansions
of $\rho_m$ and other physical quantities.

Before getting into the details, let us first comment on
the nature of the expansion that one can make in this setup.
There is no particular subtlety at finite and fixed $p$. However,
we are ultimately interested in the large $p$ limit to reach the full
Yang-Mills matrix model. So we consider the double expansion of
physical quantities in small $\frac{1}{p}$ and $\beta$. Physically, we
want to take $\frac{1}{p}$ to approach zero first, and then take $\beta$
to be small. In practice, we fix $p$ and make a small $\beta$ calculus first.
Changing the order of the two limits may cause a subtle structure, which we want to
clarify first. In particular, making the double expansion,
we find that one obtains a series in small $\frac{1}{p}$ and $p\beta$.
In other words, the radius of convergence for $\beta$ appears to be at order
$\frac{1}{p}$ at any given $p$, so that the double series expansion makes
good sense in the rather unphysical setting $\beta\lesssim\frac{1}{p}\ll 1$.
Let us briefly explain why this structure appears, and how one can make
physically meaningful approximations in this situation.

The radius of convergence in the $p$'th matrix model can be understood
as follows. The matrix integral contains the measure
given by the truncated 2-body potential,
\begin{equation}
  V(\alpha)=-\log\left[4\sin^2\frac{\alpha}{2}\right]
  -2\sum_{n=1}^p\frac{a_n(x)}{n}\cos(n\alpha)
\end{equation}
with $x=-e^{-\beta}$. $V$ diverges when any
$a_n(-e^{-\beta})$ does. Note that
\begin{equation}
  a_n(-e^{-\beta})=\left\{\begin{array}{ll}
    1-\frac{2\sinh^3(n\beta)}{\sinh^2\frac{3n\beta}{2}}&\textrm{for even }n\\
    1-\frac{2\sinh^3(n\beta)}{\cosh^2\frac{3n\beta}{2}}&\textrm{for odd }n
  \end{array}
  \right.\ ,
\end{equation}
so that only $a_n$'s for even $n$ can diverge near $\beta=0$ if $n$ is
large enough. The closest
pole to $\beta=0$ for even $a_n$ is $\beta=\pm\frac{2\pi i}{3n}$.
Among them, the closest poles are located at
$\beta=\pm \frac{\pi i}{3\lfloor\frac{p}{2}\rfloor}$. This explains
why the radius of convergence of the $\beta$ expansion is proportional
to $\frac{1}{p}$ for large $p$.

So in the framework of this subsection, we
shall make double expansions of physical quantities in $\frac{1}{p}$ and $p\beta$,
\begin{equation}\label{taylor-small}
  f(p,\beta)=\sum_{a,b}f_{a,b}
  \left({\textstyle \frac{1}{p}}\right)^a(p\beta)^b\ ,
\end{equation}
where $a$ and $b$ label infinite towers of terms.
At given $a$ with fixed small $\frac{1}{p}$, the sum over $b$
should be a Taylor series with its radius of convergence for $\beta$ at
order $\frac{1}{p}$. In the Yang-Mills matrix model at $p\rightarrow\infty$,
which is our ultimate interest, we can find poles arbitrarily close to
$\beta=0$. So the small $\beta$ expansion of physical quantities should
be an asymptotic series at zero radius of convergence. The last asymptotic
series is related to the summation of $b$ above in a nontrivial manner.
Namely, since the series (\ref{taylor-small}) in $p\beta$ involves positive powers
$p^{b-a}$ at large enough order in $b$, it does not make sense in the
strict $p\rightarrow\infty$ limit. To relate it to the asymptotic series at
infinite $p$, one has to resum over $b$ and take the $p\rightarrow\infty$ limit.
This implies that the series (\ref{taylor-small}) before the resummation
is useless in general for studying the matrix model at $p\rightarrow\infty$ of
our interest. In particular, the series is useless for studying certain
subleading corrections in the small $\beta$ expansion, by having explicit
positive powers in $p$.

However, the series (\ref{taylor-small}) is still useful for computing
certain leading small $\beta$ contributions at $p=\infty$. This is the case
if the physical quantity $f$ has a smooth $\frac{1}{p}=0$ and $p\beta=0$ limit.
In our discussions below, the observable $f$ having the smooth limit
will be the eigenvalue distribution $\rho(\theta)$ and its coefficients
$\rho_n$. If the series (\ref{taylor-small}) has its lowest order term at
$a=b=0$, then $f_{0,0}$ provides the strict $p=\infty$, $\beta=0$ value of
that observable. This will be the case for $\rho(\theta)$. Knowing the
strict $\beta=0$ limit of $\rho(\theta)$ at $p=\infty$, we will
derive below other important quantities such as $\log Z$
at strict $p=\infty$, at its leading order in small $\beta$.\footnote{The series
(\ref{taylor-small}) that can be computed using our framework here
will not be directly useful for computing the subleading corrections
in $\beta$. In fact the situation is similar for the calculus of \cite{Choi:2018hmj}
in the large black hole limit. The calculus of \cite{Choi:2018hmj} is reliable
only for the leading Cardy limit, while for subleading corrections one should use
a more elaborate approach \cite{GonzalezLezcano:2020yeb}. }

With these understood, let us first study the `standard' small charge branch at
the leading order in $\beta$. We shall then study other `non-standard'
small charge branches of the class (A).

To get the standard solution, we set $u=u_0\beta^{\frac{3}{2}}+\cdots$,
where $\cdots$ are higher order terms in small $\beta$.
From (\ref{eigen-expand}), this scaling of $u$ admits a solution
by balancing the first two terms,
\begin{equation}
  0\approx 2\beta^3(v_{\rm odd})_m+u^2n_m(n\cdot v_{\rm odd})\ \ ,\ \
  0\approx \frac{8\beta}{9m^2}(v_{\rm even})_m+u^2n_m(n\cdot v_{\rm odd})
\end{equation}
and taking the leading odd/even moments to be
\begin{equation}
  v_{\rm odd}=v_0\beta^0+\cdots\ ,\ \
  v_{\rm even}=w_{0}\beta^2+\cdots\ .
\end{equation}
In this scaling, all the ignored terms of
(\ref{eigen-expand}) are subleading. Inserting
$u\approx u_0\beta^{\frac{3}{2}}$, $w_0$ will be determined in terms of
$v_0$, which should meet the following eigenvector equation:
\begin{equation}
  2v_0+u_0^2 n(n\cdot v_0)=0\ .
\end{equation}
One therefore finds that $v_0$ has to be proportional to
$n=(1,\cdots,1)$. In particular, the eigenvalue of this equation should be
\begin{equation}
  u_0^2=-\frac{2}{\lceil\frac{p}{2}\rceil}\ .
\end{equation}

We have thus determined the leading order values of the moments
$\rho_m$ up to an overall scaling, by computing
$v_m=(-1)^mm^2 a_m\rho_m\approx (-1)^m m^2\rho_m$ at the leading order.
We found that $v_{\rm even}$ is at order $\beta^2$, so can be ignored
at the leading $\beta^0$ contributions to $\rho(\theta)$.
$v_m$ for odd $m$ are required to be $m$ independent, being proportional to
$n_m$. So $\rho_m$ should be proportional to $\frac{1}{m^2}$.
The overall coefficient can be determined by the second equation of
(\ref{linear-rho-n}), which at the leading order is
given by
\begin{equation}
  2\sum_{l=1}^{\lceil\frac{p}{2}\rceil}\rho_{2l-1}\approx 1\ .
\end{equation}
Therefore, one finally obtains the leading order moments to be
\begin{equation}
  \rho_{2n-1}=\frac{\frac{1}{2(2n-1)^2}}
  {\sum_{l=1}^{\lceil\frac{p}{2}\rceil}\frac{1}{(2l-1)^2}}
\end{equation}
at fixed $p$. We are interested in the limit
$p\rightarrow\infty$, which yields
\begin{equation}\label{rho-exact-final}
  \rho_{2n-1}=\frac{\frac{1}{2(2n-1)^2}}
  {\sum_{l=1}^{\infty}\frac{1}{(2l-1)^2}}=
  \frac{4}{\pi^2}\frac{1}{(2n-1)^2}\ .
\end{equation}
In the rest of this subsection, we shall only consider the
full matrix model at $p\rightarrow\infty$, with the understanding that
only the leading order calculus is reliable. As for $\rho_{2n}$,
it will suffice to remember $\rho_{2n}\sim\mathcal{O}(\beta^2)$.

It is easy to compute the eigenvalue distribution $\rho(\theta)$ for
(\ref{rho-exact-final}). The quickest way to find it is to note that
(\ref{rho-exact-final}) defines a real positive function
for real $\theta\in(-\pi,\pi)$, so that the interpretation of $\rho_n$
as the Fourier transformation on a circle applies. (Note also that the
gap $t=\sin^2\frac{\theta_0}{2}\rightarrow 1$ closes in this limit.)
So $\rho(\theta)$ is given by
\begin{equation}\label{rho-triangular}
  \rho(\theta)=\frac{1}{2\pi}+\frac{1}{\pi}\sum_{n=1}^\infty\rho_n\cos(n\theta)
  =\frac{1}{2\pi}\left[1+\frac{8}{\pi^2}\sum_{n=1}^\infty
  \frac{\cos(2n-1)\theta}{(2n-1)^2}\right]=
  \frac{1}{\pi^2}(\pi-|\theta|)
\end{equation}
for $-\pi<\theta<\pi$. This is a triangular distribution centered
around $\theta=0$. One can obtain the same result by starting from the more
abstract definition of $\rho(\theta)$ in terms of $\rho_n$ as explained in
section 2, based on $\rho_n$ defined as
the moments on the complex interval. In particular, the eigenvalue cut
at the leading order is given by $(-\pi,\pi)$ on the real axis.
Note that this triangular distribution is different from
the so-called `Bethe root' distribution \cite{Benini:2018ywd}
in the small black hole limit, which is given by
\begin{equation}
  \rho(\theta)_{{\rm Bethe}}\stackrel{\beta\rightarrow 0}{\longrightarrow}
  \left\{
    \begin{array}{ll}
      \frac{1}{\pi}&\textrm{for }|\theta|<\frac{\pi}{2}\\
      0&\textrm{for }\frac{\pi}{2}<|\theta|<\pi
    \end{array}
  \right.
\end{equation}
on the unit circle. This is a rectangular distribution
which fills half of the circle. There is no contradiction here,
because \cite{Benini:2018ywd} does not use our matrix model for this problem.

We next compute the free energy of our saddle point, which will allow us
to count the dual black hole microstates. Again, we only consider the
full Yang-Mills partition function $Z=Z_{\infty}$ at $p\rightarrow\infty$.
The general large $N$ free energy $\log Z$ is given by
\begin{eqnarray}\label{free-fourier}
  \log Z&=&-\frac{N^2}{2}\int_{-\theta_0}^{\theta_0}
  d\theta_1 d\theta_2 V(\theta_1-\theta_2)\rho(\theta_1)\rho(\theta_2)\\
  &=&N^2\sum_{n=1}^\infty\frac{1}{n}\int_{-\theta_0}^{\theta_0}
  d\theta_1 d\theta_2 \left[a_n-1\right]e^{in(\theta_1-\theta_2)}
  \rho(\theta_1)\rho(\theta_2)=N^2\sum_{n=1}^\infty
  \frac{a_n-1}{n}(\rho_n)^2\ .\nonumber
\end{eqnarray}
The last expression is an exact formula at $p=\infty$, supposing that
the infinite sum converges. (And it does converge in our problem.)
Here recall that at the leading order,
$\rho_{\rm odd}\sim\mathcal{O}(\beta^0)$ and
$\rho_{\rm even}\sim\mathcal{O}(\beta^2)$. Also, $a_n-1$ at
small $\beta$ are given by
\begin{equation}
  a_n(\beta)-1\approx\left\{\begin{array}{ll}
    -2n^3\beta^3&\textrm{for odd }n\\
    -\frac{8n\beta}{9}&\textrm{for even }n
  \end{array}\right.\ .
\end{equation}
Of course these expansions are invalid at very large $n$, but the fast
damping of $(\rho_n)^2\propto n^{-4}$ allows the calculation of the leading
term at $\beta\ll 1$ using (\ref{free-fourier}).
The last expression of (\ref{free-fourier}) acquires leading
$\mathcal{O}(\beta^3)$ contribution from odd $n$'s, while the terms with even
$n$ are at the subleading order $\mathcal{O}(\beta^5)$. One obtains
\begin{equation}\label{free-small-final}
  \log Z\approx-2N^2\beta^3\sum_{n=1}^\infty(2n-1)^2\cdot
  \left(\frac{4}{\pi^2}\frac{1}{(2n-1)^2}\right)^2
  =-\frac{4N^2\beta^3}{\pi^2}\ .
\end{equation}
This precisely agrees with the free energy of the small black holes in
AdS$_5\times S^5$ \cite{Hosseini:2017mds}. To see this, the general
free energy of the 1-parameter black holes of \cite{Gutowski:2004ez}
in our convention is given by
\begin{equation}
  \log Z\sim \frac{N^2}{2}\frac{\left(2\beta\right)^3}
  {\left(-\pi i+3\beta\right)^2}\ .
\end{equation}
See \cite{Agarwal:2020zwm} for converting the result of
\cite{Hosseini:2017mds} to the convention we use here.
In this setting, $\beta\rightarrow 0$ is the small black hole limit,
which precisely yields (\ref{free-small-final}).

$\log Z\approx-\frac{4N^2\beta^3}{\pi^2}$ is negative at real positive
$\beta$. This means that the small black hole saddle will never be more
dominant than the graviton saddle. Anyway, small black hole saddles
are unstable in the grand canonical ensemble, with negative specific heat.
We should consider this saddle point in the microcanonical ensemble.
We Legendre transform $\log Z$ at fixed charge $q$,
which is $N^2$ times an independently small number which does not scale in $N$.
The Legendre transformation of this free energy at fixed
$q=6(R+J_+)\ll N^2$ yields the entropy
\begin{equation}\label{SV-entropy}
  -\frac{4N^2}{\pi^2}\beta^3+\beta q
  \ \stackrel{\textrm{extremize}}{\longrightarrow}\
  S(q)=\pi\sqrt{\frac{q^3}{27N^2}}=\pi\sqrt{\frac{8(R+J_+)^3}{N^2}}\ ,
\end{equation}
which precisely agrees with the Bekenstein-Hawking entropy of small
BPS black holes in AdS$_5\times S^5$. See appendix B for the details of taking
the limit. Generalizing this, the saddle points with three independent $R_I$ is
derived in section 5.3. At $J_1=J_2$, the entropy is given by
\begin{equation}
  S=\pi\sqrt{\frac{8(R_1+J_+)(R_2+J_+)(R_3+J_+)}{N^2}}\ .
\end{equation}
This again completely agrees with the Bekenstein-Hawking entropy
of the small black holes of \cite{Gutowski:2004yv}.
On the known black hole solutions at $J_1=J_2$, the angular momentum is much
smaller than the electric charges, $\frac{J_+}{N^2}\sim \left(\frac{R}{N^2}\right)^2\ll\frac{R}{N^2}$, in the small black hole limit.
With this extra input, the entropy can be written as
\begin{equation}
  S\approx\pi\sqrt{\frac{8R_1R_2R_3}{N^2}}\ .
\end{equation}

In the local region of spacetime including the black hole
whose size is much smaller than the AdS radius $\ell$,
the small black hole solution is precisely the same as
the asymptotically flat 5d BPS black holes of Strominger and Vafa
\cite{Strominger:1996sh}. There, the embedding into the 10d
string theory is different from ours. We embed the small black hole into
large AdS$_5$, also keeping the black holes rather `uniform' in the
large internal $S^5$ at the same radius $\ell$.
In this picture, the quantized charges $R_I$
are realized as momenta along the large $S^5$. On the other hand,
the black holes of \cite{Strominger:1996sh} are traditionally
embedded into type IIB string theory compactified on $K3\times S^1$ or
$T^5=T^4\times S^1$. We compare our studies with the $T^4\times S^1$ embedding.
The size of the internal $T^5$ can be much smaller than the size of
the black holes. The three charges carried by the same 5d gravity solution
are quantized differently. The first two charges may be
$Q_1$ D1-branes wrapping $S^1$ and $Q_2$ D5-branes wrapping $T^5$.
Alternatively, they can be $Q_1$ D3-branes wrapping a $T^3$ and $Q_2$
D3-branes wrapping a different $T^3$. The third charge is
the quantized momentum $p$ along $S^1$. In this realization,
the same black hole entropy is written as $S=2\pi\sqrt{pQ_1Q_2}$.
The different prefactor in front of $(R_1R_2R_3)^{\frac{3}{2}}$ or $q^{\frac{3}{2}}$
in our formulae is due to the different charge quantizations.
With different realizations of charges, different size of
the internal manifold and also the presence/absence of the AdS gravitational wall,
various aspects of the black holes are different in the two setups.
See our section 5.2 for one such example, for the BMPV black holes
embedded in AdS. However, as for explaining
near-horizon properties of a given black hole such as the area law,
we are studying precisely the same object as \cite{Strominger:1996sh}.
We emphasize that we made a first-principle counting of the same
black hole solutions of \cite{Strominger:1996sh}, without extra
ad hoc assumptions like D-branes.

Ironically, precisely because of this abstract nature of our approach,
it is not even clear whether the notion of D-branes is relevant at all
for the microstates of small black holes.
We believe that D-branes will be the relevant degrees of freedom,
from an interesting D-brane-based argument \cite{Kinney:2005ej}
for the entropy (\ref{SV-entropy}). See the section 5.4 of \cite{Kinney:2005ej}.
The idea is to use D3-brane giant gravitons in $S^5$,
and to distribute the charges $R_I$ suitably to these branes
and the momentum on their worldvolumes. This approach
has a technical limitation, in that it uses an unjustified 2d QFT approach.
However, we feel that their results illustrate an essential nature
of small black holes. Namely, as far as we are aware of, the small black holes
are not expected to be described by the fully deconfined plasma of gluons.
For instance from Fig. \ref{p1-legendre}, the Legendre transformation
line at small charge is always outside the
deconfining region. Rather, it is natural to expect their
microstates to consist of more conventional objects of gravity in the
traditional low temperature phase. Quantum gravity at low temperature phase
shows rich towers of states, which are the `confining spectrum' from the gauge
theory point of view. In this sense, D-branes ($\sim$ baryons) are the most natural
objects which make it possible for the entropy to see
$N$ in the high energy confining spectrum. It will be nice to clarify
how one can concretely see these D3-branes within our abstract approach.
To this end, perhaps studying the Polyakov loop \cite{Polyakov:1978vu} operators
at higher rank symmetric representations may be useful, since they are related
to D3-branes. They could be studied rather intuitively from our triangular
distribution (\ref{rho-triangular}), or perhaps more rigorously by
inserting the BPS Polyakov loop operators in $S^3\times S^1$ \cite{Gang:2012yr}.

As the final subject of this subsection, we study the `non-standard'
small black hole branches, defined by the scalings of the gap
parameter $u\equiv t-1$ in (\ref{small-roots}) other than $u^2\propto \beta^3$ in the class (A).
We discuss the branches on the second line of (\ref{small-roots}) in
some detail, at $u\approx u_0\beta^{\frac{3}{4}}$,
after which the other cases can be understood more easily.

After carefully inspecting various terms appearing in (\ref{eigen-expand}),
one finds that the first/fourth terms of the first equation can be balanced
by making these terms to be at the leading $\beta^3$ order.
Also, the first/fourth terms of the second equation can be balanced at
the leading $\beta^{3}$ order as well.
This is achieved by taking $v_{\rm odd}\sim\mathcal{O}(\beta^0)$
and $v_{\rm even}\sim\mathcal{O}(\beta^{2})$. There are apparently
more leading terms than $\beta^3$, from the second and third terms
containing $v_{\rm odd}$. These terms have
to cancel for our non-standard ansatz to work. The last requirement
will impose further constraints on $v_{\rm odd}$.
Let us explain this with the first equation of (\ref{eigen-expand}),
since the second equation can be understood in exactly the same manner.
Consider the following expansions:
\begin{equation}
  v_{\rm odd}=v_0+v_1\beta^{\frac{3}{4}}+v_2\beta^{\frac{3}{2}}+\cdots\ \ ,
  \ \ \ u=\beta^{\frac{3}{4}}\left(u_0+u_1\beta^{\frac{3}{4}}+u_2\beta^{\frac{3}{2}}
  +\cdots\right)\ .
\end{equation}
Then in the first equation of (\ref{eigen-expand}), the terms which are
apparently more leading than or at $\beta^3$ order are given by
\begin{eqnarray}
  \hspace*{-.7cm}&&2\beta^3v_0+u_0^2\beta^{\frac{3}{2}}
  n\left(n\cdot (v_0+\beta^{\frac{3}{4}}v_1+\beta^{\frac{3}{2}}v_2)\right)
  +2u_0u_1\beta^{\frac{9}{4}}n\left(n\cdot(v_0+\beta^{\frac{3}{4}}v_1)\right)
  +(2u_0u_2\!+\!u_1^2)\beta^3n(n\cdot v_0)\nonumber\\
  \hspace*{-.7cm}&&+u_0^3\beta^{\frac{9}{4}}M_1\cdot(v_0+\beta^{\frac{3}{4}}v_1)
  +3u_0^2u_1\beta^3 M_1\cdot v_0
  +u_0^4M_2\cdot v_0\ .
\end{eqnarray}
There is only one term at $\beta^{\frac{3}{2}}$ order, $\propto n(n\cdot v_0)$.
For this term to vanish, one should demand
\begin{equation}\label{non-standard-extra-1}
  n\cdot v_0=0\ .
\end{equation}
This equation has solution only if $v_0$ has more than one components.
Therefore, we expect this non-standard solution to exist only for
$p\geq 3$. This is compatible with the general structures of (\ref{small-roots})
and the explanations provided below this equation.
At the next order $\mathcal{O}(\beta^{\frac{9}{4}})$,
there are three terms which should cancel. After imposing
(\ref{non-standard-extra-1}), one obtains
\begin{equation}\label{non-standard-extra-2}
  n(n\cdot v_1)+u_0M_1\cdot v_0=0
\end{equation}
Here, $v_1$ can be decomposed to components parallel and orthogonal to $n$.
Let us write $v_1=v_{1\parallel}+v_{1\bot}$ and further define
$v_{1\parallel}=c_1 n$. Using
(\ref{non-standard-extra-1}), (\ref{non-standard-extra-2}) can be
written as
\begin{equation}
  0=Dc_1 n+\frac{u_0}{3}n\sum_{l\in{\rm odd}}l^2\cdot (v_0)_l\ \rightarrow\
  c_1=-\frac{u_0}{3D}\sum_{l\in{\rm odd}}l^2(v_0)_l\ ,
\end{equation}
where $D\equiv\lceil\frac{p}{2}\rceil$.
Finally, the terms at $\beta^3$ order demand the following equation,
\begin{eqnarray}
  0&=&2(v_0)_m+u_0^2n_m(n\cdot v_2)
  +2u_0u_1 n_m(n\cdot v_1)
  +u_0^3 (M_1\cdot v_1)_m+3u_0^2u_1 M_1\cdot v_0
  +u_0^4(M_2\cdot v_0)_m
  \nonumber\\
  &=&\left[2(v_0)_m+\frac{Du_0^3c_1}{3}m^2+\frac{u_0^4}{8}m^2
  \sum_{l\in{\rm odd}}l^2(v_0)_l
  \right]\\
  &&+n_m\left[u_0^2(n\cdot v_2)-u_0u_1(n\cdot v_1)
  +\frac{u_0^3}{3}\sum_{l\in{\rm odd}}\left(l^2(v_{1\bot})_l+c_1(l^2-2)
  +{\textstyle \frac{u_0}{8}}(l^4-8l^2)(v_0)_l\right)\right]\ .\nonumber
\end{eqnarray}
We explicitly decomposed the terms into those parallel to $n$
(third line) and those containing orthogonal components to $n$
(second line). The parallel components on the third line can be canceled by
tuning $v_2$. The orthogonal component extracted from the second line
determines $v_0$. This equation is given by
\begin{equation}
  0=2(v_0)_m +\frac{u_0^4}{72}\sum_{l\in{\rm odd}}q_ml^2(v_0)_l
\end{equation}
where we defined
\begin{equation}
  q_m\equiv m^2-\frac{1}{D}\sum_{l\in{\rm odd}}l^2
  =m^2-\frac{4D^2-1}{3}
\end{equation}
which satisfies $n\cdot q=0$. Therefore, the eigenvector equation
to be satisfied by $v_0$ is given by
\begin{equation}
  \mathcal{M}_{ml}(v_0)_l
  \equiv\sum_{l\in{\rm odd}}(m^2-{\textstyle \frac{4D^2-1}{3}})
  l^2(v_0)_l=-\frac{144}{u_0^4}(v_0)_m\ .
\end{equation}
The only nonzero eigenvector (unnormalized yet) satisfying this equation is
given by
\begin{equation}
  (v_0)_m\propto q_m=m^2-\frac{4D^2-1}{3}\ ,
\end{equation}
with the eigenvalue
\begin{equation}
  -\frac{144}{u_0^4}=\sum_{l\in{\rm odd}}l^2\left(l^2-\frac{4D^2-1}{3}\right)
  =\frac{16}{45}D(D^2-1)(4D^2-1)\ .
\end{equation}
All other eigenvectors have zero eigenvalues.
Recalling that $(v_{\rm odd})_m\approx(v_0)_m=(-1)^mm^2a_m\rho_m
\approx-m^2\rho_m$ for odd $m$ at the leading order, one obtains
\begin{equation}
  \rho_m\approx-\frac{v_m}{m^2}\propto 1-\frac{4D^2-1}{3m^2}\ .
\end{equation}
The normalized $\rho_m$ can be computed from the condition
$A\cdot\rho=1$. Noting that $A_m\approx 2$ for odd $m$'s, one obtains
\begin{equation}
  \rho_{2m-1}=\frac{1-\frac{4\lceil\frac{p}{2}\rceil^2-1}{3(2m-1)^2}}
  {2\sum_{l=1}^{\lceil\frac{p}{2}\rceil}
  \left(1-\frac{4\lceil\frac{p}{2}\rceil^2-1}{3(2l-1)^2}\right)}\ .
\end{equation}
At large $p$, one obtains
\begin{equation}
  \rho_{2m-1}=\frac{1-\frac{4D^2-1}{3(2m-1)^2}}{2D-\frac{\pi^2(4D^2-1)}{12}}
  \stackrel{D\rightarrow\infty}\longrightarrow
  \frac{4}{\pi^2(2m-1)^2}\ .
\end{equation}
Thus, although the distribution in this non-standard branch is different
from the standard one at finite $p$,
the large $p$ limit is precisely the same as the standard solution.
The free energy $\log Z$ and entropy are also same in the $p\rightarrow\infty$
limit.

Although the calculus is  more
involved in the non-standard branch at finite $p$, the large $p$ limit only
uses basic structures. The large $p$ analysis can thus be easily
generalized to other non-standard solutions.
We discuss the cases with scaling $u=u_0\beta^{\frac{3}{2n}}$, where
$n$ is a finite positive integer. That is, either $p$ is finite, or $p$ is large but
$n$ does not scale in large $p$. In this case, we need to expand
(\ref{expand-functions}) up to $u^{2n}$ order. Here, we only need to know
that fact that the coefficient of $u^d$ in (\ref{eigen-eqn})
is $ml^2$ times a degree $d-2$ polynomial in $m^2$ and $l^2$.
The last statement is true for finite $p$, and also true at large
$p$ if $d\ll p$. If $d\propto p\gg 1$, the coefficients will contain factorials
rather than polynomials, in which case our simple procedures
below will not hold. One can show that the condition to be met by the leading
$\mathcal{O}(\beta^0)$ part $v_0$ of $v_{\rm odd}$ are
\begin{equation}
  \sum_{l\in{\rm odd}}l^{2k}(v_0)_l=0\ \ \
  \textrm{for}\ \ k=0,1,\cdots,n-2\ .
\end{equation}
The matrix $\mathcal{M}$ appearing in the eigenvector equation
$\mathcal{M}\cdot v_0=u_0^{-2n}v_0$ is proportional to
\begin{equation}
  (\mathcal{M})_{ml}\propto\left(m^{2(n-1)}
  +a_{n-2}m^{2(n-2)}+\cdots a_1 m^{2}+a_0\right)l^{2(n-1)}\ ,
\end{equation}
where $a_0,\cdots,a_{n-2}$ are chosen to satisfy
\begin{equation}
  \sum_{m\in{\rm odd}}m^{2k}\left(m^{2(n-1)}
  +a_{n-2}m^{2(n-2)}+\cdots a_1 m^{2}+a_0\right)=0\ \ \
  \textrm{for}\ \ k=0,1,\cdots,n-2\ .
\end{equation}
$v_0$ satisfying this eigenvector equation is proportional to
\begin{equation}
  (v_0)_m\propto
  m^{2(n-1)}
  +a_{n-2}m^{2(n-2)}+\cdots a_1 m^{2}+a_0\ .
\end{equation}
In the large $D$ limit, note that the coefficients are proportional to
$a_0\propto D^{2(n-1)}$, $a_1\propto D^{2(n-2)},\cdots$, $a_{n-2}\propto D^2$.
So in this limit, $v_0$ is determined by the last
term proportional to $a_0$, implying that $v_0\propto n$. (We have checked
that the coefficients of $D^{2(n-1)}$ in $a_0$ are nonzero with increasing absolute
values, till $n\leq 5$.) This leads to
\begin{equation}
  \rho_{2m-1}\stackrel{D\rightarrow\infty}{\longrightarrow}
  \frac{4}{\pi^2(2m-1)^2}
\end{equation}
for the non-standard solution $u\propto \beta^{\frac{3}{2n}}$ at
finite $n$ which does not scale in $p\rightarrow\infty$.
Therefore, for infinitely many branches labeled by finite
$n=1,2,\cdots$ which do not scale with large $p$, we obtain precisely
the same eigenvalue distribution and the free energy,
$\log Z\approx-\frac{4N^2\beta^3}{\pi^2}$. We find an infinite
degeneracy of small black hole saddle points. In general,
$n$ can grow until $n\leq D = \left\lceil \textstyle{\frac{p}{2}} \right\rceil$ for the class (A). The computation at $p\rightarrow\infty$
and fixed nonzero $\frac{n}{p}$ is currently beyond our scope.

A possible scenario at large $p$ and nonzero $\frac{n}{p}$ is that the free
energies $\log Z$ may exhibit a `dense spectrum,' depending on an
effectively continuous parameter $\frac{n}{p}$. It would be very interesting
to check if this scenario is true, because in this case the extra continuous
parameter might be identified as that of the small hairy black holes
in AdS$_5\times S^5$ \cite{Markeviciute:2018yal,Markeviciute:2018cqs}.
If this is true, the reason why we got the same large $p$ free energies at
finite $n$ is because the effective continuous parameter $\frac{n}{p}$ is
all at the same value $\frac{n}{p}\rightarrow 0$. Physically, this would
be because the graviton hair outside the event horizon carries much smaller
charges than the black hole, so that their effects to the thermodynamics
are negligible.\footnote{It would be also interesting if some of these saddles are related to the fully localized 10d black holes in AdS$_5 \times S^5$ \cite{Hanada:2016pwv,Berenstein:2018lrm,Hanada:2018zxn,Hanada:2019czd}.}

\subsection{Small black holes with extra spin}

We introduce one more fugacity conjugate to $J_1-J_2$ and extend
the small black hole analysis of the previous subsection.
For simplicity, we only consider the standard branch. The small black hole
limits will correspond to the spinning BMPV black holes \cite{Breckenridge:1996is}
in flat spacetime. This apparently trivial extension exhibits significantly new
physics. There appear `entropic instabilities' at $J_1\neq J_2$,
which are very similar to the super-radiant instabilities of spinning
non-BPS black holes in AdS. This demands a special consideration to properly
define and compute the physical quantities of the BPS black holes from QFT.
We explain the situations in some detail before our microscopic studies.

In non-BPS cases, the instability of spinning black holes in AdS is
a long-standing question. A classic problem is the instability of Kerr-AdS
black holes. Due to the super-radiance
and the reflection by the AdS wall, over-spinning Kerr-AdS black holes exhibit
both thermodynamic instabilities \cite{Hawking:1999dp} and dynamical
instabilities of the quasi-normal modes
\cite{Cardoso:2004hs,Kunduri:2006qa,Cardoso:2006wa,Kodama:2009rq,Murata:2008xr}.
The two instabilities are related, in that the time evolution of
the tachyonic unstable modes obeys the second law of black hole
thermodynamics. The thermodynamic instability happens
due to the divergence of the thermal partition function of the radially
quantized dual CFT,
\begin{equation}\label{partition-rotating}
  Z(\beta,\Omega_i)={\rm Tr}\left[e^{-\beta(H-\sum_i\Omega_i J_i)}\right]\ ,
\end{equation}
where $i$ runs over all possible angular momenta.
The trace diverges when $\Omega_i^2>1$ for some $i$, since then
a derivative acquires the fugacity factor
$e^{-\beta(1-|\Omega_i|)}$ greater than $1$. Namely, let us call $\partial$
a derivative weighted by a fugacity greater than $1$.
If a local operator $\hat{\mathcal{O}}$ contributes to this partition function, then
all its conformal descendants taking the form of $\partial^n\hat{\mathcal{O}}$
will also contribute. The fugacities carried by
this infinite tower of operators can be indefinitely large at large $n$,
making the trace ill-defined.\footnote{The conformal descendant viewpoint of the
Kerr-AdS instability, as well as the related novel features
explained around (\ref{large-spin}), were all explained to us by Shiraz Minwalla.
We thank him for sharing the insights.}
If the fugacity factor
of an operator becomes $1$, this means that this operator can assume a nonzero
expectation value, implying a Bose-Einstein condensation. This is a signal
of the formation of hairy black holes in AdS. See
\cite{Basu:2010uz,Bhattacharyya:2010yg} and references therein, for instance.
An odd aspect of the unstable Kerr-AdS black holes is that infinitely many
operators want to condense at the same time.

In the BPS sector, dynamical instabilities due to tachyonic quasi-normal
modes are absent. But there can be thermodynamic instabilities
of BPS black holes at fixed charges, which are very similar to the Kerr-AdS
instability. This instability is simply the entropic subdominance of the black hole
in the ensemble sum. Thermodynamic instabilities in a similar sense were
studied in \cite{Bena:2011zw} for the BMPV black holes.
Since we realize the BMPV black holes as small black holes in AdS,
there appear more thermodynamic instabilities than \cite{Bena:2011zw}.
As a familiar non-BPS analogue, Kerr black
holes may be unstable only in AdS. It happens
even for the small Kerr black holes in AdS because the large AdS plays
the role of a reflecting wall for the
super-radiance of an over-spinning black hole, which causes the instability.
This means that, even if we can regard large AdS as an infrared regulator
of the asymptotically flat gravity, stability issues
can depend on the presence of AdS. The
thermodynamic stability of BPS black holes also depends on the
AdS embedding. Since we expect this to be a generic phenomenon of
AdS embedding, we elaborate on both the QFT and
gravity aspects of the instabilities.

Recall from section 2 the definition of the 2-parameter index,
\begin{equation}\label{index-refined-def}
  Z(\beta,\gamma)={\rm Tr}\left[(-1)^F (-e^{-3\beta})^{(2R+J_1+J_2)}
  e^{-\gamma(J_1-J_2)}\right]\ ,
\end{equation}
where we took $x=-e^{-\beta}$, $y=e^{-\gamma}$.
The BPS black holes carrying the extra spin $J_1-J_2$
are known from \cite{Chong:2005da,Kunduri:2006ek}.
The entropy of such black holes can be understood by
Legendre transforming the following `large $N$ free energy' \cite{Hosseini:2017mds},
\begin{equation}
  \log Z=\frac{N^2}{2}\frac{\Delta^3}{\omega_1\omega_2}=
  -\frac{N^2}{2}\frac{8\beta^3}{(\pi+3i\beta)^2+\gamma^2}\ ,
\end{equation}
where $\omega_1=-\pi i+3\beta+\gamma$, $\omega_2=-\pi i+3\beta-\gamma$
and $\Delta=2\beta$. The small black hole limit corresponds to keeping the leading
term at $|\beta|\ll 1$, at finite $\gamma$. This yields
\begin{equation}\label{BMPV-free}
  \log Z\ \rightarrow\ -\frac{4N^2\beta^3}{\pi^2+\gamma^2}\ .
\end{equation}
We make the Legendre transformation  at fixed charges
$q\equiv 3(2R+J_1+J_2)$, $j\equiv J_1-J_2$,
\begin{equation}
  S(\beta,\gamma;q,j)=\log Z+\beta q+\gamma j\ \rightarrow\
  q=\frac{12N^2\beta^2}{\pi^2+\gamma^2}\ ,\ \
  j=-\frac{8N^2\beta^3\gamma}{(\pi^2+\gamma^2)^2}\ .
\end{equation}
The solution for $\beta,\gamma$ is given by
\begin{equation}\label{legendre-sol}
  \beta=\frac{\pi q^2}{6N^2\sqrt{\frac{q^3}{27N^2}-j^2}}\ ,\ \
  \gamma=-\frac{\pi j}{\sqrt{\frac{q^3}{27N^2}-j^2}}\ ,
\end{equation}
and the entropy is given by
\begin{equation}
  S(q,j)=\pi\sqrt{\frac{q^3}{27N^2}-j^2}\ .
\end{equation}
This is a familiar expression for the BMPV black hole entropy
\cite{Breckenridge:1996is}, except that $q$ assumes a different
normalization from the more canonical one.
$\gamma$ should be finite and real for typical BMPV black holes.
For instance, the second term
$-j^2$ inside the square root of the entropy formula is comparable to the first
term $\frac{q^3}{27N^2}$ only when $\gamma$ is finite.
To reach the extreme case of vanishing entropy, or the closed time-like curve (CTC)
bound $j^2\rightarrow\frac{q^3}{27N^2}$, one should take
$\gamma\rightarrow\pm\infty$.\footnote{From (\ref{legendre-sol}), $\beta$
would diverge if the limit $j^2\rightarrow \frac{q^3}{27N^2}$ is applied literally,
violating the small black hole setup. So this limit should be understood as
$\frac{q^4}{N^4}\ll \frac{q^3}{27N^2}-j^2\ll\frac{q^3}{27N^2}$, where
the first inequality ensures $\beta\ll 1$.}
So the general BMPV black holes are realized as
small AdS black holes with $|\beta|\ll 1$ and $\gamma\in(-\infty,\infty)$.
See Appendix B for taking this limit on the AdS black hole solution.

In the BMPV limit of the previous paragraph, it is easy to see that
the index (\ref{index-refined-def}) has a thermodynamic instability
similar to that of (\ref{partition-rotating}). This is because if we take
$|\beta|\ll 1$, keeping $\gamma$ finite at a nonzero
real value will make the trace to diverge. To be definite,
let us take $\gamma<0$. We also define the two complex coordinates of
the spacetime $\mathbb{C}^2\sim\mathbb{R}^4$ to be $z_1,z_2$ so that
the two BPS derivatives $\partial_{z_1}$, $\partial_{z_2}$ carry spins
$(J_1,J_2)=(1,0), (0,1)$ respectively. Then consider
the conformal descendants of any gauge invariant
BPS operator $\hat{\mathcal{O}}$ contributing to the index,
given by $(\partial_{z_1})^n\hat{\mathcal{O}}$ with $n=1,2,\cdots$.
The extra fugacity factors carried by these new operators are
$e^{n(|\gamma|-3\beta)}$, which in the BMPV limit become
$e^{n|\gamma|}>1$. This factor can grow indefinitely large at large $n$,
making the trace ill-defined. The arguments are in complete parallel to
those for the over-rotating Kerr-AdS black holes at $|\Omega_i|>1$.
Let us take $|\gamma|=-\gamma$ to be a very small positive number,
still satisfying $|\gamma|\gg\beta$, going slightly
beyond the onset of the instability. If $\hat{\mathcal{O}}$ is weighted
by a fugacity $e^{-\mu}<1$, the net fugacity for the $n$'th descendant is
given by $e^{n|\gamma|-\mu}$. The trace diverges due to the sequence
of operators at very large spin,
\begin{equation}\label{large-spin}
  n>\frac{\mu}{|\gamma|}\gg 1\ .
\end{equation}
This is a characteristic feature
of the instabilities of over-spinning AdS black holes: modes with large
angular momenta start to cause the instability at the onset point.
Precisely the same feature is found with Kerr-AdS black holes.

We explain this instability
from the gravity side, with the BMPV small black holes in AdS.
An unstable black hole can increase its entropy by
`emitting' some of its charges as gravitons outside its event horizon.
For BPS black holes, the word `emission' should be simply understood as
moving to a different configuration in the ensemble
with graviton hairs outside the black holes.
After emitting charges at the same order as the black hole charge,
it is in general difficult to construct the full solution in which
the black hole and the hair back-react to each other.
However, for small black holes in AdS,
\cite{Basu:2010uz,Bhattacharyya:2010yg} established
a simple way to construct hairy black holes.
The nontrivial part of the metric of
the small black hole is contained in a small region, say of radial size $r$
which is much smaller than the AdS radius $\ell$. On the other hand,
the wavefunction of the emitted graviton hair carries small charges,
which we take to be the same order as the small black hole charges.
This wavefunction extends over the AdS scale $\ell$, so their condensate density
will be suppressed by $\frac{r}{\ell}$. From this, one can argue
\cite{Basu:2010uz,Bhattacharyya:2010yg} that the back-reaction of
the graviton wavefunction and the small black hole metric to each other is
negligible, to the leading order in the small charge parameter
$\frac{r}{\ell}\ll 1$. This means that one can superpose the two solutions
at the leading order.

The arguments of \cite{Basu:2010uz,Bhattacharyya:2010yg} can be applied
to our BMPV small black holes in AdS. In the microcanonical ensemble at fixed net
charges $q$ and $j$, we consider thermodynamically competing configurations in which
BPS gravitons are superposed with
the small black holes. The entropy carried by the gravitons will be negligible
compared to the entropy change of the small black hole which scales like $N^2$.
This is true even if the portions of charges carried by the gravitons scale
like $N^2$ \cite{Kinney:2005ej}.
So if the entropy of the black hole can
increase by losing the charges, this will be a channel
of the thermodynamic instability. We will
show that such hairy small black holes exist precisely for $|\gamma|>3\beta$.
Again we take $\gamma<0$, meaning $j\equiv J_1-J_2>0$.

Let us start by defining a function related to the BMPV entropy,
\begin{equation}
  F(q,j)\equiv\left(\frac{S}{\pi N^2}\right)^2=
  \frac{1}{27}\left(\frac{q}{N^2}\right)^3-\left(\frac{j}{N^2}\right)^2
  \equiv\frac{\hat{q}^3}{27}-\hat{j}^2\ .
\end{equation}
The normalized charges $\hat{q}=\frac{q}{N^2}$,
$\hat{j}=\frac{j}{N^2}$ are much smaller than
$1$ for small black holes. Now consider the hairy BPS black hole
which contains the gravitons at charges $\Delta R$, $\Delta J_1$, $\Delta J_2$.
Our interest is simply finding a hairy configuration with larger entropy,
rather than constructing the configuration with maximal entropy at given $q,j$.
We seek for such configurations when the graviton charges are
much smaller than $q,j$.
Since the graviton hair is also BPS, their charges should satisfy certain
positivity bounds. The general conditions are
$R_I+R_J\geq 0$, $R_I+J_i\geq 0$, $J_1+J_2\geq 0$ for the pair sums of
distinct charges \cite{Kinney:2005ej,Choi:2018hmj}.
These demand the following conditions for the graviton charges:
\begin{equation}\label{graviton-charge-range}
  \Delta R\geq 0\ ,\ \ \Delta J_1+\Delta J_2\geq 0\ ,\ \
  \Delta R+\Delta J_i\geq 0\ .
\end{equation}
The charges carried by the new black hole core are given by
\begin{equation}
  q^\prime=q-3(2\Delta R+\Delta J_1+\Delta J_2)\ ,\ \
  j^\prime=j-(\Delta J_1-\Delta J_2)\ .
\end{equation}
So the entropy change by going to the hairy black hole is
\begin{equation}\label{entropy-change}
  \Delta F\equiv F(q^\prime,j^\prime)-F(q,j)\approx
  2\hat{j}(\Delta J_1-\Delta J_2)
  -\frac{\hat{q}^2}{3}(2\Delta R+\Delta J_1+\Delta J_2)\ .
\end{equation}
At the last step, we used the fact that $\Delta R$, $\Delta J_i$ are
small and made a linearized approximation. If the graviton charges
can be chosen to
meet $\Delta F>0$, the hairy black hole has larger entropy.
$\hat{q},\hat{j}$ for the black hole should first meet the CTC bound
$\hat{j}<\frac{\hat{q}^{\frac{3}{2}}}{3\sqrt{3}}$. For the black hole in
the regime $|\gamma|>3\beta$, one finds $\hat{j}>\frac{\hat{q}^2}{6}$
by using (\ref{legendre-sol}). So we study whether $\Delta F$ can be
positive or not in the range
\begin{equation}
  \frac{\hat{q}^2}{6}<\hat{j}<\frac{\hat{q}^{\frac{3}{2}}}{3\sqrt{3}}\ .
\end{equation}
It suffices to consider the first inequality only,
to study the thermodynamic instability.

We first consider the region $|\gamma|< 3\beta$, in which we do not expect
any thermodynamic instability. In this case, applying
$0<\hat{j}<\frac{\hat{q}^2}{6}$, one obtains
\begin{equation}
  \Delta F<-4\hat{j}\left[\Delta R+\Delta J_2\right]\leq 0
\end{equation}
from (\ref{entropy-change}) and (\ref{graviton-charge-range}).
Therefore, one indeed finds no instabilities at $|\gamma|< 3\beta$. On the other
hand, let us consider the region $|\gamma|>3\beta$ by setting
$\frac{\hat{q}^2}{6}=\hat{j}(1-\epsilon)$ with $\epsilon\in(0,1)$ not too close
to $1$. In this case one obtains
\begin{equation}
  \Delta F=2\hat{j}\left[\epsilon\Delta J_1
  -(2-2\epsilon)\Delta R-(2-\epsilon)\Delta J_2\right]\ .
\end{equation}
At any $\epsilon$, one can take the graviton system at
large enough $\Delta J_1$ so that the quantity inside the square bracket
is positive. Taking $\Delta J_1$ arbitrarily large is possible because
the graviton wavefunctions in AdS exist at arbitrary large angular momentum.
(In the CFT dual language, they can be constructed from the conformal
descendants with sufficiently many derivatives.) So for $|\gamma|>3\beta$, one can always construct small
hairy black holes at larger entropies. For small $\epsilon$, one goes slightly
beyond the stability bound. In this case,
the value of $\Delta J_1>\frac{2}{\epsilon}(\Delta R+\Delta J_2)$
required for the instability is very large. Therefore, at the onset of
instability, the modes with infinite angular momentum first cause
the instability. This is precisely in accordance with
the conformal descendant picture that we presented around (\ref{large-spin}).

Now we make the following microscopic studies.
Rather than searching for the entropically most dominant black holes,
we microscopically study the unstable BMPV small black hole from the matrix
model by going beyond the `legal regime' $|{\rm Re}(\gamma)|>3{\rm Re}(\beta)$.
Since we clearly
expect there to be interesting large $N$ phenomena of BMPV black holes,
one should be able to address them from the index (\ref{index-2-para}).
So how can we compute the free energy at such unstable saddle points?
Clearly, the trace definition of the partition function
is ill-defined at $|\gamma|>3\beta$. Also, the matrix model expressions
involving the infinite sum over $n$ like (\ref{index-matrix-model}) do not make
sense because a fugacity factor is larger than $1$.
We can get hints from how the over-spinning BMPV black hole solutions at
$\frac{\hat{q}^2}{6}<\hat{j}<\frac{\hat{q}^{\frac{3}{2}}}{3\sqrt{3}}$ can be
understood from the classical gravity dual. Consider the solutions at $\hat{j}<\frac{\hat{q}^2}{6}$. From these solutions without
any instability, one can simply extend the parameters carried by the solutions
beyond the bound. No pathologies or singularities arise at the level of
semi-classical saddle point solutions. From QFT, we can similarly
construct the large $N$ semi-classical saddle points in a stable region and
then extend the large $N$ semi-classical results to
the unstable region by changing $\gamma$.
More concretely, if one wishes to compute within the context of
small black holes at $|\beta|\ll 1$, it will be more
convenient to first take $\gamma\equiv i\xi$ to be pure imaginary
for a while. The sum over the conformal descendants
associated with the fugacity factor $e^{in\xi}$ will not diverge.
The infinite series in (\ref{index-matrix-model})
is well defined, and one can also consider the truncations
at finite $p$. In this setting, we repeat the analysis of
section 5.1. After all the calculations,
the free energy can be continued to real $\gamma$.

As in section 5.1, we first consider the $p$'th truncated model.
We take the gap parameter $t=\sin^2\frac{\theta_0}{2}$
to assume the standard small $\beta$ scaling, $t=1+u\beta^{\frac{3}{2}}+\cdots$.
One obtains
\begin{equation}
  R_{ml}-\delta_{ml}=(-1)^{m+l+1}ml^2\left[
  -\frac{8(-e^{i\xi})^m}{\left(1-(-e^{i\xi})^m\right)^2}\delta_{ml}
  +u_0^2\right]\beta^3+\mathcal{O}(\beta^{\frac{9}{2}})\ .
\end{equation}
From this, the determinant of $R-{\bf 1}_{p\times p}$ is given by
\begin{equation}\label{det-BMPV}
  \det(R-{\bf 1})=\#\left(
  \prod_{n=1}^p\frac{-8(-e^{i\xi})^n}{(1-(-e^{i\xi})^n)^2}\right)
  \left(1-\frac{u_0^2}{8}\sum_{m=1}^p
  \frac{(1-(-e^{i\xi})^m)^2}{(-e^{i\xi})^m}\right)\beta^{3p}+
  \textrm{higher orders in }\beta
\end{equation}
where $\#$ is a number independent of $u_0,\beta$.
So ${\rm det}(R-{\bf 1})=0$ demands
\begin{equation}
  u_0^2=-\frac{8}{\sum_{m=1}^p\left(2-(-e^{i\xi})^m-(-e^{-i\xi})^m\right)}\ .
\end{equation}
To compute $\rho_{n\leq p}$, the equation
$(R-{\bf 1})\rho=0$ is given at the leading order by
\begin{equation}
  0=\sum_{l=1}^p (-1)^ll^2\left(u_0^2
  -8\frac{(-e^{i\xi})^m}{(1-(-e^{i\xi})^m)^2}\delta_{ml}\right)
  \rho_l\ .
\end{equation}
We want this to hold at $u_0$ for which (\ref{det-BMPV}) vanishes at
the leading order. The two equations will take the same form if the $l$ dependence
of $\rho_l$ is chosen to be
$\frac{(-1)^l}{l^2}\frac{(1-(-e^{i\xi})^l)^2}{(-e^{i\xi})^l}$. Normalizing it
to satisfy $A\cdot \rho=1$, one obtains
\begin{equation}
  \rho_n=-\frac{(-1)^n}{2}\frac{\frac{2-(-e^{i\xi})^n-(-e^{-i\xi})^n}{n^2}}
  {\sum_{m=1}^p\frac{2-(-e^{i\xi})^m-(-e^{-i\xi})^m}{m^2}}+
  \mathcal{O}(\beta^{\frac{3}{2}})
\end{equation}
for $n=1,\cdots,p$.

We would now like to consider the Yang-Mills matrix model at $p\rightarrow\infty$.
One finds
\begin{equation}\label{rho-BMPV-1}
  \rho_n=-\frac{(-1)^n}{2}
  \frac{\frac{2-(-e^{i\xi})^n-(-e^{-i\xi})^n}{n^2}}
  {\frac{\pi^2}{3}-{\rm Li}_2(-e^{i\xi})-{\rm Li_2(-e^{-i\xi})}}\ ,
\end{equation}
where we used the series definition of the dilogarithm function
\begin{equation}
  {\rm Li}_2(x)\equiv\sum_{n=1}^\infty\frac{x^n}{n^2}
\end{equation}
which converges for $|x|\leq 1$.
Here, one can use the following identity
\begin{equation}\label{dilog-id}
  {\rm Li}_2(-e^{i\xi})+{\rm Li}_2(-e^{-i\xi})=
  -\frac{(2\pi i)^2}{2!}B_2\left({\textstyle \frac{\xi+\pi}{2\pi}}\right)
  =\frac{\xi^2}{2}-\frac{\pi^2}{6}
\end{equation}
for $-\pi<\xi<\pi$, where $B_2(x)\equiv x^2-x+\frac{1}{6}$. Plugging this into
(\ref{rho-BMPV-1}), one obtains
\begin{equation}\label{rho-BMPV-2}
  \rho_n=\frac{(-1)^{n-1}}{\pi^2-\xi^2}\cdot
  \frac{2-(-e^{i\xi})^n-(-e^{-i\xi})^n}{n^2}\ .
\end{equation}
At real $\xi$, the real $\rho_n$'s in the range $-\pi<\xi<\pi$
yields a real non-negative function $\rho(\theta)$ for real $\theta$.
In particular, the sums over $n$ can be done explicitly to yield
\begin{eqnarray}\label{rho-theta-BMPV}
  \rho(\theta)&=&\frac{1}{2\pi}\left(1+2\sum_{n=1}^\infty\rho_n\cos(n\theta)\right)\\
  &=&\frac{1}{2\pi}\left[1+\frac{1}{\pi^2-\xi^2}
  \sum_\pm\left({\rm Li}_2(e^{i(\xi\pm\theta)})
  +{\rm Li}_2(e^{-i(\xi\pm\theta)})-2{\rm Li}_2(-e^{\pm i\theta})\right)\right]
  \ .\nonumber
\end{eqnarray}
For real $-\pi<\theta<\pi$, one can use (\ref{dilog-id}) to simplify
the last term $\sum_\pm {\rm Li}_2(-e^{\pm i\theta})$. Also,
the other ${\rm Li}_2$ functions can be simplified
at real $-\pi<\xi<\pi$ by using
\begin{equation}
  {\rm Li}_2(e^{ix})+{\rm Li}_2(e^{-ix})=2\pi^2B_2\left(
  {\textstyle \frac{x}{2\pi}}-n\right)
\end{equation}
for $2\pi n<x<2\pi (n+1)$, $n\in\mathbb{Z}$. From the
identities at $n=0$ and $n=-1$,
\begin{equation}\label{dilog-id-2}
  {\rm Li}_2(e^{ix})+{\rm Li}_2(e^{-ix})=
  \left\{\begin{array}{ll}
    2\pi^2\left[\left(\frac{x}{2\pi}\right)^2-\frac{x}{2\pi}+\frac{1}{6}\right]&
    \textrm{for }0<x<2\pi\\
    2\pi^2\left[\left(\frac{x}{2\pi}\right)^2+\frac{x}{2\pi}+\frac{1}{6}\right]
    &\textrm{for }-2\pi<x<0
  \end{array}\right.\ ,
\end{equation}
one obtains
\begin{equation}
  |x|=\frac{x^2}{2\pi}+\frac{\pi}{3}
  -\frac{{\rm Li}_2(e^{ix})+{\rm Li}_2(e^{-ix})}{\pi}
\end{equation}
for $-2\pi<x<2\pi$. Applying all these identities,
one obtains
\begin{equation}\label{rho-theta-BMPV-real}
  \rho(\theta)=
  \frac{1}{\pi^2-\xi^2}\left(\pi-\frac{|\theta+\xi|+|\theta-\xi|}{2}\right)
\end{equation}
for real $-\pi<\xi<\pi$ and real $-\pi<\theta<\pi$.

The free energy $\log Z$ can be computed from
\begin{equation}\label{free-fourier-2}
  \log Z=N^2\sum_{n=1}^\infty\frac{a_n-1}{n}\rho_n^2\ .
\end{equation}
The function $a_n-1$ in the BMPV limit is given by
\begin{equation}
  a_n(\beta,\gamma)-1=-\frac{(1-e^{-2n\beta})^3}
  {(1-(-1)^ne^{-n(3\beta+i\xi)})(1-(-1)^ne^{-n(3\beta-i\xi)})}
  \ \rightarrow\ -\frac{8n^3\beta^3}{2-(-e^{i\xi})^n-(-e^{-i\xi})^n}\ .
\end{equation}
Plugging this in (\ref{free-fourier-2}) and again using (\ref{dilog-id}),
one obtains
\begin{equation}\label{BMPV-free-real}
  \log Z=-\frac{4N^2\beta^3}{\pi^2-\xi^2}\ .
\end{equation}
This finishes our calculations at real $\xi$. If one naturally
assumes the analyticity of the free energy in complex $\gamma=i\xi$,
one obtains the free energy
\begin{equation}\label{BMPV-free-complex}
  \log Z=-\frac{4N^2\beta^3}{\pi^2+\gamma^2}
\end{equation}
which precisely accounts for the BMPV black holes embedded in large AdS.

One would ultimately want to understand the complex eigenvalue distributions at
complex or real $\gamma$. This problem is very tricky, and presumably impossible
within our computational framework here. Let us just outline the subtleties of
the problem.

Knowing all $\rho_n$'s, one may think that computing $\rho(\theta)$
at real $\xi,\theta$ and then analytically continuing it
would yield $\rho(\theta)$ in the complex $\theta$ plane.
Then it may naively look that this would determine the complex cut, following
the procedures of section 2. This is subtler than it looks, as we explain now.
The distribution function that we computed for real $\xi,\theta$
can be complexified by going back to the last expression of (\ref{rho-theta-BMPV}).
At real $\xi,\theta$, we applied the identities (\ref{dilog-id}),
(\ref{dilog-id-2}) to derive (\ref{rho-theta-BMPV-real}).
For complex $\xi,\theta$, one can apply similar identities
after replacing the ranges of $x$ by the ranges of ${\rm Re}(x)$ for
complex $x$. Repeating the calculus, one obtains
\begin{equation}
  \rho(\theta)=\frac{1}{\pi^2-\xi^2}\left[\pi-
  \frac{{\rm sgn}({\rm Re}(\theta+\xi))(\theta+\xi)+
  {\rm sgn}({\rm Re}(\theta-\xi))(\theta-\xi)}{2}\right]
\end{equation}
for $-\pi<{\rm Re}(\theta)<\pi$ and $-\pi<{\rm Re}(\xi)<\pi$.
The expression has branch cuts at ${\rm Re}(\theta)=\pm{\rm Re}(\xi)$,
with the branch points $\theta=\pm\xi$.
From this, the function $s(\theta)=\int d\theta\rho(\theta)$ is given
by
\begin{equation}\label{s-BMPV}
  s(\theta)=\left\{
  \begin{array}{lll}
    s_1(\theta)&\equiv\frac{(\theta+\pi)^2}{2(\pi^2-\xi^2)}-\frac{1}{2}&
    \textrm{ for }-\pi<{\rm Re}(\theta)<-{\rm Re}(\xi)\\
    s_2(\theta)&\equiv\frac{\theta}{\pi+\xi}&
    \textrm{ for }-{\rm Re(\xi)}<{\rm Re}(\theta)<{\rm Re}(\xi)\\
    s_3(\theta)&\equiv-\frac{(\theta-\pi)^2}{2(\pi^2-\xi^2)}+\frac{1}{2}&
    \textrm{ for }\ {\rm Re}(\xi)<{\rm Re}(\theta)<\pi
  \end{array}
  \right.
\end{equation}
where the integration constants are chosen to
meet the requirements $s(\pm\pi)=\pm\frac{1}{2}$, $s(0)=0$.
The ${\rm Im}[s(\theta)]=0$ lines passing through either
$\theta=-\pi,0,\pi$ are locally straight lines. They generally do not
meet at the branch cuts, not forming a continuous cut. So blindly making
an analytic continuation of $\rho(\theta)$ with the data given, one generally
finds a piecewise continuous cut consisting of three straight lines.
Furthermore, one finds that the total probability obtained by
integrating $\rho(\theta)$ along these straight lines is always less than
$1$, unless ${\rm Im}(\xi)=0$. This means that there are missing
eigenvalues which are not captured by this calculus.
This phenomenon happens because $s(\theta)$ of (\ref{s-BMPV}) has
discontinuities across the branch cut ${\rm Re}(\theta)=\pm {\rm Re}(\xi)$.
This is a singularity which occurs by
taking the $\beta\rightarrow 0^+$ limit at complex $\xi$. Presumably
at nonzero small $\beta$, the singular function is resolved
into a better behaved function with the resolution size given by $\beta$.
Probably, the missing eigenvalues are hidden in this singular region
which is beyond the scope of our computational framework.
We leave this tricky question unsolved in this paper.
Anyway, we emphasize again that
having computed the free energy (\ref{BMPV-free-real}) and naturally assuming
the analyticity in $\xi=-i\gamma$, one can convincingly compute the free energy
(\ref{BMPV-free-complex}) for the BMPV black holes at real $\gamma$.

In section 5.3, we extend all the analysis above to unequal electric
charges $R_I$, leading to the free energy
$\log Z=-\frac{4N^2\beta_1\beta_2\beta_3}{\pi^2+\gamma^2}$ and the entropy
$S=\pi\sqrt{\frac{8(R_1+J_+)(R_2+J_+)(R_3+J_+)}{N^2}-(J_1-J_2)^2}$.

We finish this subsection by comparing our results to that of
\cite{Bena:2011zw}, which also discussed thermodynamic instabilities of
BMPV black holes. \cite{Bena:2011zw} considered two possible configurations
which can compete with the BMPV black holes, all in strictly asymptotically
flat background. One is the black ring, and another
is the BMPV black hole surrounded by the smooth solutions called `supertubes.'
The latter configuration may be regarded as a sort of graviton hair outside
the black hole event horizon. However, these hairs are different from those
we studied in this subsection in two qualitative manners.
Firstly, our graviton hairs are extended in the AdS box of size $\ell$
much larger than the size $r$ of the black hole. Therefore, the existence of
our hairs depends on putting AdS as a large IR regulator.
The hairy black holes of \cite{Bena:2011zw} became more dominant than the
BMPV black holes very near the CTC bound, which is
$j\approx\frac{q^{\frac{3}{2}}}{3\sqrt{3}N}$ in our setting. On the other hand,
our hairs extended in AdS cause instabilities at much smaller values of spin, $j=\frac{q^2}{6N^2}$. The stability
of a black hole depends on the presence/absence of the AdS box, even if it is
much larger than the black hole. Secondly, from our AdS embedding viewpoint,
the hairs discussed in \cite{Bena:2011zw} are localized in a small spatial region
of size much smaller than  $\ell$. To construct
their solutions, the interaction of the hair and the black hole is important.
This type of hairs is excluded in our considerations by assumption, since we
relied on the non-interacting picture of \cite{Basu:2010uz,Bhattacharyya:2010yg}.
The results of \cite{Bena:2011zw} imply that, even for small AdS black holes,
there could be subtler localized graviton hairs for which interactions are important.

\subsection{Three electric charges and extra spin}

\def\up{\Upsilon}
\def\bE{\mathbb{E}}
\def\qe{\mathfrak{q}}
\def\kq{\mathfrak{q}}
\def\fq{\mathfrak{q}}
\def\fs{\mathfrak{s}}
\def\fc{\mathfrak{c}}
\def\fD{\mathfrak{D}}
\def\fA{\mathfrak{A}}
\def\rx{\mathrm{x}}
\def\ii{\mathrm{i}}
\def\ri{\mathrm{i}}
\def\rj{\mathrm{j}}
\def\bi{\mathbf{i}}
\def\bj{\mathbf{j}}
\def\bz{\mathbf{z}}
\def\by{\mathbf{y}}
\def\bx{\mathbf{x}}
\def\ba{\mathbf{a}}
\def\BC{\mathbb{C}}
\def\BR{\mathbb{R}}
\def\BE{\mathbb{E}}
\def\BZ{\mathbb{Z}}
\def\BI{\mathbb{I}}
\def\CalN{\mathcal{N}}
\def\CalA{\mathcal{A}}
\def\CalV{\mathcal{V}}
\def\CalP{\mathcal{P}}
\def\CalR{\mathcal{R}}
\def\CalZ{\mathcal{Z}}
\def\CalH{\mathcal{H}}
\def\CalW{\mathcal{W}}
\def\CalU{\mathcal{U}}
\def\CalO{\mathcal{O}}
\def\CalM{\mathcal{M}}
\def\CalS{\mathcal{S}}
\def\CalT{\mathcal{T}}
\def\qe{\mathfrak{q}}
\def\CalL{\mathcal{L}}
\def\CalX{\mathcal{X}}
\def\Tr{{\rm Tr}}
\def\sE{\mathscr{E}}
\def\sM{\mathscr{M}}
\def\ve{{\varepsilon}}
\def\bbZ{\mathbb{Z}}
\def\ta{\mathtt{a}}
\def\tb{\mathtt{b}}
\def\tc{\mathtt{c}}
\def\td{\mathtt{d}}
\def\tm{\mathtt{m}}
\def\tn{\mathtt{n}}
\def\tP{\mathtt{P}}
\def\fz{\mathfrak{z}}
\def\sY{\mathsf{Y}}
\def\sQ{\mathsf{Q}}
\def\sP{\mathsf{P}}
\def\sI{\mathsf{I}}
\def\sK{\mathsf{K}}
\def\sJ{\mathsf{J}}
\def\sI{\mathsf{I}}
\def\sS{\mathsf{S}}
\def\sF{\mathsf{F}}
\def\sL{\mathsf{L}}
\def\sw{\mathsf{w}}
\def\pa{\partial}

 \def\p{\partial}
 \def\pb{\bar{\partial}}
 \def\a{\alpha}
 \def\b{\beta}
 \def\g{\gamma}
 \def\d{\delta}
 \def\eps{\epsilon}

 \def\th{\theta}
 \def\vt{\th}
 \def\k{\kappa}
 \def\l{\lambda}
 \def\m{\mu}
 \def\n{\nu}
 \def\x{\xi}
 \def\r{\rho}
 \def\u{\upsilon}
 \def\vr{\varrho}
 \def\s{\sigma}
 \def\t{\tau}
 \def\th{\theta}
 \def\z{\zeta }
 \def\vp{\varphi}
 \def\G{\Gamma}
 \def\D{\Delta}
 \def\T{\theta}
 \def\X{\Xi}
 \def\P{\Pi}
 \def\S{\Sigma}
 \def\L{\Lambda}
 \def\O{\Omega}
 \def\o{\omega }
 \def\U{\Upsilon}

\def\beq{\begin{equation}}
\def\eeq{\end{equation}}

We consider small black hole saddle points of the matrix model
at three independent R-charges $R_I$, $I=1,2,3$. This will account for
the small black holes with three independent electric charges.
Let us write $x_I ^2 \equiv e^{-\Delta_I} $, $x_1 x_2 x_2 e^{\g} = e^{-\o_1}$, and $x_1 x_2 x_3 e^{-\g} = e^{-\o_2}$ so that
\begin{align}
    a_n = 1- \frac{\prod_{I=1} ^3 (1- e^{-n \Delta_I})}{(1-e^{-n \o_1})(1-e^{-n \o_2})} = 1- \frac{\prod_{I=1}^3 1- x _I ^{2n} }{(1- x_1 ^n x_2 ^n x_3 ^n e^{n\g})(1- x_1 ^n x_2 ^n x_3 ^n e^{-n\g})}.
\end{align}
In the small black hole limit, we set $x_I = - e^{-\b_I}$ where $\b_I$ goes to zero in the same order. Then note that
\begin{align} \label{eq:an}
    a_n = 1 + \frac{8 n^3 (-1)^n e^{n \g}
   }{((-1)^{n+1} + e^{
    n \g})^2} \b_1 \b_2 \b_3 + \mathcal{O}(\b^5).
\end{align}
In particular, the first correction term of order $\b^3$ is proportional to $\b_1 \b_2 \b_3$, while more complicated terms appear in higher orders.

Recall from section 2 that $\r_n$'s are determined by the equations
\begin{equation} \label{eq:consts}
    (R-{\bf 1})\r=0,\quad A\cdot \r=1,
\end{equation}
where
\begin{equation} \label{eq:rmat}
R_{ml} = a_l \sum_{k=1}^l \left( B^{m+k-\frac{1}{2}} (t) +B^{\vert m-k +\frac{1}{2} \vert} (t) \right) P_{l-k} (1-2t)\ ,\
    A_m = a_m \left(P_{m-1} (1-2t)- P_m (1-2t)\right)\ .
\end{equation}
The existence of nontrivial solution implies $\det (R-{\bf 1}) = 0$, from which we can determine $t$ perturbatively by expanding it around $t=1$. Since only the combination $\b_1 \b_2 \b_3$ appears at the lowest order correction to $a_n$, the correction to $t$ at the lowest orders also only depends on $\b_1\b_2\b_3$ as far as they are not affected by the $\mathcal{O}(\b^5)$ correction to $a_n$.
The general solution can be written in the form of $t= 1+t_1 (\b_1 \b_2 \b_2)^{\frac 1 2} + t_2 \b_1\b_2 \b_3 + \mathcal{O}(\b^{\frac 7 2}) $, where the higher order terms involve more complicated combinations of $\b_I$'s.
More explicitly, we expand
\begin{align}
    P_{l} (1-2t) = (-1)^l \left[ 1+ l(l+1)\left( t_1 (\b_1 \b_2 \b_3)^{\frac 1 2} + t_2 \b_1 \b_2 \b_3 \right) +\frac{l(l^2-1) (l+2)}{4} t_1 ^2 \b_1 \b_2 \b_3  \right] + \mathcal{O}(\b^{\frac 7 2}),
\end{align}
and
\begin{align}
    B^{l+\frac 1 2} (t) = \d_{l,0} + (-1)^l \left[ t_1 (\b_1 \b_2 \b_3)^{\frac 1 2} + t_2 \b_1 \b_2 \b_3 +\frac{l(l+1)}{2} t_1 ^2 \b_1 \b_2 \b_3 \right] + \mathcal{O}(\b^{\frac 7 2}).
\end{align}
We can argue $t=1+t_1 (\b_1\b_2\b_3)^{\frac 1 2} + t_2 \b_1 \b_2 \b_3 + \mathcal{O}(\b^{\frac 7 2})$ as follows. Let us start from more general ansatz $t=1+t_1 \b^{\frac 3 2} + t_2 \b^3 + t_3 \b^{\frac 7 2} + t_4 \b^4 +t_5 \b^{\frac 9 2}$. Recall that $a_n = 1+ a^{(1)}_n \b_1 \b_2 \b_3 + \cdots$. We compute $R_{ml}$ up to the order of $\b^{\frac 9 2}$. If we choose $1$ from $a_n$, $t_3$, $t_4$, and $t_5$ in $B$ and $P_l$ could contribute in principle. But an explicit computation shows that this is not the case and their contributions actually vanish up to the order of $R_{ml} \sim \mathcal{O}(\b^{\frac 9 2})$. Then only $t_1$ and $t_2$ contribute, so that they should be accompanied with $(\b_1 \b_2 \b_3)^{\frac 1 2}$ and $\b_1 \b_2 \b_3$ since only $\b_1\b_2\b_3$ appear from $a_n$ at this order.

By substituting these expressions into \eqref{eq:rmat}, we obtain
\begin{align}
    R_{ml} - \d_{ml} =  (-1)^{m+l+1} m l^2 \left[ \frac{ - 8\left(- e^{ \g} \right)^m}{\left( 1- (- e^{ \g})^m \right) ^2} \d_{ml} + t_1 ^2  \right] \b_1 \b_2 \b_3 + \mathcal{O}(\b^{\frac 9 2})\ .
\end{align}
The determinant takes the form of
\begin{eqnarray}
    \det (R- {\bf 1})&=&\# \left(  \prod_{n=1} ^p \frac{-8 (- e^{\g} )^n}{(1-(-e^\g)^n)^2} \right) \left( 1 - \frac{t_1 ^2}{8} \sum_{m=1} ^p \frac{(1-(-e^\g )^m )^2}{(-e^\g )^m} \right) (\b_1 \b_2 \b_3)^{p} + \text{higher orders in $\b$} \nonumber\\
    &=&\# \left(  \prod_{n=1} ^p \frac{-8 (- e^{\g} )^n}{(1-(-e^\g)^n)^2} \right) \left( 1 + \frac{t_1 ^2}{8} \sum_{m=1}^p \left( 2-(-e^\g)^m -(-e^\g)^{-m} \right) \right) (\b_1 \b_2 \b_3)^{p}\nonumber\\
    &&+ \text{higher orders in $\b$}\ .
\end{eqnarray}
Hence, we obtain
\begin{align}
    t_1 ^2 = -\left( \frac{\sum_{m=1}^p (2-(-e^\g)^m -(-e^\g)^{-m}}{8} \right)^{-1}.
\end{align}

Next, we solve the equations \eqref{eq:consts} for $\rho_n$.
Note that these equations are linear, and thus only has to be confirmed by direct substitution once the solution is given. Indeed, it can be checked that the solution is
\begin{align}
    \r_n = - \frac{(-1)^n}{2} \frac{\frac{2-(-e^\g)^n -(-e^\g)^{-n}}{n^2}}{\sum_{l=1} ^p \frac{2-(-e^\g)^l -(-e^\g)^l)}{l^2}} + \mathcal{O}({\b^\frac 3 2})
, \quad n=1,\cdots, p,
\end{align}
by direct substitution.
In the strict $\b = 0$ limit, we have $t=1$ and the gap closes. Then the moments $\r_n$'s get identified with the Fourier coefficients of the distribution. Thus we obtain the distribution by the Fourier expansion formula, along with the limit $p\to \infty$ as
\begin{align}
\begin{split}
    \r(\th) &= \frac{1}{2\pi} \left( 1+ 2 \sum_{n=1} ^\infty \r_n \cos n\th \right) = \frac{1}{2\pi} \left( 1 -  \frac{\sum_{n=1} ^\infty   \frac{2-(-e^\g)^n -(-e^\g)^{-n}}{n^2}(-1)^n \cos n\th }{\sum_{n=1} ^\infty \frac{2-(-e^\g)^n -(-e^\g)^n)}{n^2}} \right) .
\end{split}
\end{align}
The distribution is precisely the same as that studied in section 5.2. This
can be summed to yield the same closed-form expression.

Finally, we can easily compute the leading contribution to the free energy by
\begin{equation}
    \frac{\log Z}{N^2} = -\sum_{n=1} ^\infty \frac{1-a_n}{n} \r_n ^2 = - \frac{2\b_1 \b_2 \b_3 }{\sum_{n=1} ^\infty \frac{2- (-e^\g)^n -(-e^\g)^{-n}}{n^2}}
    = -\frac{4\b_1 \b_2 \b_3}{\pi^2 + \g^2}\ .
\end{equation}
We extremize the function
\begin{align}
    S( \b_I,\g;q_I,j )\equiv -\frac{4 N^2 \b_1 \b_2 \b_3}{\pi^2 + \g^2} +  \sum_{I=1} ^3 \b_I q_I + \g j,
\end{align}
to obtain the entropy, where $q_I = 2R_I + J_1 + J_2$ and $j= J_1 - J_2$ are
integral quantized charges. The solution for $\b_I$ and $\g$ is
\begin{align}
    \b_I = \frac{\pi}{2N^2} \frac{q_1 q_2 q_3}{q_I} \frac{1}{\sqrt{\frac{q_1 q_2 q_3}{4N^2}-  j^2}}, \quad \g = - \frac{\pi j}{\sqrt{\frac{q_1 q_2 q_3}{N^2} -j^2}},
\end{align}
and the entropy is given by
\begin{align}
    S(q_I,j) = \pi\sqrt{\frac{q_1 q_2 q_3}{N^2} - j^2}.
\end{align}

\section{Conclusion}

In this paper, we studied the large $N$ saddle points of the matrix model
for the index of 4d maximal super-Yang-Mills theory on $S^3\times \mathbb{R}$,
and investigated the physics of the holographically dual black holes.
The study was made in two closely related directions.

Firstly, we studied the large $N$ saddle points of the truncated matrix models.
These truncations were investigated in \cite{Aharony:2003sx,Copetti:2020dil} prior
to our studies. The truncated models admit various numerical and analytic approaches
which would have been much more difficult in the full Yang-Mills matrix model.
We explored various saddle points of these matrix models.
We found the numerical saddles which microscopically account for
the known AdS black holes in semi-quantitative manners.
In particular, we found that multiple branches of saddle points have
to be patched to describe the known black holes.
It would be interesting to know how many branches participate
at larger values of $p$.

Secondly, we analytically constructed the exact saddle points for the small black
holes, at charges much smaller than $N^2$. The solutions
are found for the whole infinite sequence of the truncated models labeled by
$p\geq 1$, so one obtains the exact saddle points
of the full Yang-Mills theory by sending $p\rightarrow\infty$.
These saddles perfectly account for the thermodynamics of
small AdS black holes of \cite{Gutowski:2004ez,Chong:2005da,Kunduri:2006ek}.
Small AdS$_5$ black holes are related to the 5d
asymptotically flat black holes. Thus we have provided a first-principle
microscopic account for the black holes of
\cite{Strominger:1996sh,Breckenridge:1996is}. For the BMPV black holes
\cite{Breckenridge:1996is}, we found their thermodynamic instabilities
when embedded into AdS, both from the gravity and QFT sides.
This is in close parallel to the instability
of the over-spinning Kerr-AdS black holes \cite{Hawking:1999dp}.

There are many directions to be further explored.
Firstly, we found infinitely many small black hole saddles which exhibit
the same large $N$ thermodynamics. We speculated a scenario in which these
degenerate saddles are part of a continuous spectrum of saddles, motivated
by the small hairy black holes. However, further computations have to be done
to confirm or rule out this scenario, as explained at the end of section 5.1.

We would also like to better understand the possibilities of
new black hole like saddle points away from the tachyonic region of $\rho_1$.
The key motivation of studying the $\rho_1$ tachyon region is that it hosts
the black holes which cause the Hawking-Page transition, determining the
dominant AdS thermodynamics in the grand canonical ensemble. However,
in the microcanonical ensemble, other saddle points could be meaningful
as independent black hole solutions. In the last viewpoint, there is no
particular reason to focus only on the $\rho_1$ tachyon region.
We are particularly interested in analytically constructing new
small black hole solutions away from the $\rho_1$ tachyon region.
One motivation for this study is that small black holes are likely to be
related to asymptotically flat black holes in 5d Minkowski background,
and many results are known for them.
As briefly commented in section 3.2, we found some numerical evidences
that such small black holes could exist around the tachyonic regions of $\rho_2$.
They were obtained by taking limits of the single-cut saddles,
but we can also relax our ansatz to the multi-cut saddles. Appendix A
provides the technical backgrounds for such extensions.
Presumably, constructing analytic saddle points for small black holes
will be possible, along the line of section 5. It will be interesting to see
if some of them quantitatively account for the physics of more nontrivial
asymptotically flat black holes, such as multi-centered black holes
or black rings.

As a related matter, we also want to study various large black hole limits
analytically. Although large black holes enjoy certain degrees of universality,
we strongly feel that there could be more nontrivial large black hole saddles
in the multi-cut sectors.
We also expect that their large black hole limits can be easily solvable
analytically at $p\rightarrow\infty$, by slightly generalizing the calculus of
our section 4 or that of \cite{Choi:2018hmj}.  It will also be interesting to
numerically check how these new large black hole branches connect to the new small
black hole branches.

We eagerly hope to come back in the near future
with solutions to these questions.

\vskip 0.5cm

\hspace*{-0.8cm} {\bf\large Acknowledgements}
\vskip 0.2cm

\hspace*{-0.75cm} We thank Dongmin Gang, Eunwoo Lee, June Nahmgoong, Jun Nian
and especially Shiraz Minwalla for helpful discussions related to this project.
This work is supported in part by the National
Research Foundation (NRF) of Korea Grant 2018R1A2B6004914 (SC, SK),
NRF-2017-Global Ph.D. Fellowship Program (SC), the US Department of
Energy under grant DE-SC0010008 (SJ), the NRF grant
2021R1A2C2012350 (SK), a KIAS Individual Grant PG081602 at Korea Institute for Advanced Study (SC), and CERN and CKC fellowship (SJ).

\appendix

\section{Matrix model analysis}

In this appendix, we study our matrix models at complex coupling constants.
Let us write the index in terms of
$z_a = e^{i\a_a}$, $a=1,\cdots, N$,
\begin{eqnarray}
    Z&=&
    \int \prod_{a=1} ^N dz_a \exp \left[ - \sum_{a=1} ^N \log z_a + \sum_{a\neq b} \log\left( 1-\frac{z_b}{z_a} \right) + \sum_{n=1} ^\infty \frac{a_n}{n} \sum_{a,b=1}^N z_a ^n z_b ^{-n} \right]\nonumber\\
    &=& \int \prod_{a=1} ^N dz_a\, e^{-N^2 \left[ \frac{1}{N} \sum_{a=1} ^N \log z_a -\frac{1}{2N^2} \sum_{a\neq b} \log(z_a-z_b)^2 -\frac{1}{N^2} \sum_{n=1} ^\infty \frac{a_n}{n} \sum_{a,b=1}^N z_a ^n z_b ^{-n} \right]}\ .
\end{eqnarray}
In the large $N$ continuum limit, the eigenvalues $z_a$ accumulate on a curve $\g$,
with a certain distribution $\r(z)$. In terms of $\rho(z)$,
the saddle point equation reads
\begin{eqnarray}
    0&=&\log z -\int_\gamma dz' \log(z-z')^2 \r(z')
    - \sum_{n=1}^\infty \frac{a_n}{n} \int_\g dz' (z^n z'^{-n} +z^{-n} z'^n)  \r(z') \nonumber\\
    & =& \log z - \int_\g dz' \log(z-z')^2 \r(z') - \sum_{n=1}^\infty \frac{a_n}{n} (z^n \r_{-n} + z^{-n} \r_n),
\end{eqnarray}
where $\log(z-z')^2 $ should be understood as the principal value
\begin{equation}
    \log(z-z')^2 = \log (z_+ -z') + \log (z_- -z'), \quad z,z'\in \g,
\end{equation}
which gets rid of the singularity at $z=z'$. Here, $f (z_\pm)$ is defined as the limit of $f (z')$ where $z'$ tends to $z \in \g$ from the left and the right of $z$, with respect to the orientation of $\g$. We also defined the moments of distribution $\r_n \equiv \int_\g dz\, z^n \r(z)$, $n\in \mathbb{Z}$.

The solutions to the saddle point equation are called equilibrium distributions or equilibrium densities. Now, as a formal technique to solve this saddle point equation, it is convenient to treat the moments as independent variables temporarily. In other words, the above saddle point equation can be understood as the continuum limit of the saddle point equation of the holomorphic matrix model with the potential
\begin{align}
    W(z) = \log z -\sum_{n=1} ^\infty \frac{a_n}{n} (z^n \r_{-n} +z^{-n} \r_n).
\end{align}
This provides a systematic way of truncating the model up to a given number $p$, by restricting the infinite summation up to $p$. From now on, we will consider this truncated holomorphic matrix model, and find the equilibrium distribution from it. Then, we will reconnect to the $\mathcal{N}=4$ index by solving further $\r_n = \int_\g dz\, z^n \r(z) $.

In our application, we will assume the equilibrium distribution is even,
which means the invariance of $z\r(z)$ under the transformation $z \to z^{-1}$. This implies $\r_n = \r_{-n}$ for $n\in \mathbb{Z}$, so in this case the potential becomes
\begin{align} \label{eq:poten}
    W(z) = \log z - \sum_{n=1} ^p \frac{\l_n}{n} (z^n + z^{-n}),
\end{align}
where we defined the new couplings $\l_n = a_n \r_n$. Note that $\r_n = \int_\g dz\, z^n \r(z) = \int_\g dz \frac{z^n+z^{-n}}{2} \r(z) $ for even distributions.

Before we move on to find the equilibrium density by the matrix model analysis,
we describe how one can obtain the leading contribution to the free energy once the equilibrium distribution is given. We simply evaluate the action at the given equilibrium density $\r(z)$,
\begin{eqnarray} \label{eq:freeori}
    -\frac{\log Z}{N^2}
    &=&\int_\g dz\, \r(z) \log z -\frac{1}{2} \int _{\g^{\times 2}} dz\,dz' \,\log(z-z')^2 \r(z)\r(z') -\sum_{n=1} ^p \frac{a_n}{n} \int_{\g^{\times 2}} dz \,dz'\, z^n z'^{-n} \r(z)\r(z') \nonumber\\
    &=& \int_{\g ^{\times 2}} dz \,dz ' \sum_{n=1}^\infty \frac{1}{n} \left(  \frac{z'}{z} \right)^n \r(z) \r(z') - \sum_{n=1} ^p \frac{a_n}{n} \int_{\g^{\times 2}} dz\,dz' \, z^n z'^{-n} \r(z) \r(z') \nonumber\\
    &=& \sum_{n=1} ^\infty \frac{1}{n} \r_n \r_{-n}-\sum_{n=1 } ^p \frac{a_n}{n} \r_n \r_{-n} ,
\end{eqnarray}
where we expanded $\log(z-z')^2$ in $\vert z \vert> \vert z' \vert$ in the second line. Since the integrand is symmetric in $z$ and $z'$, there is no loss of generality. Note that we have taken the truncation number $p$. By taking the limit $p\to \infty$ to this expression, we finally obtain the leading contribution to the free energy from the given equilibrium density.

Let us consider the holomorphic matrix model
\begin{equation} \label{eq:part}
    Z = \prod_{a=1} ^N \int_\Gamma dz_a \prod_{i<j}
    (z_a -z_b )^2 \exp \left( -N W(z_a) \right),
\end{equation}
where the $W(z)$ possibly contains complex-valued coupling constants, and $\G$ is not necessarily unit circle but the $N$-product of a contour $\g$ in the complexified $z$-plane $\mathbb{C}$, properly chosen to make the above partition function convergent (We directly adopt the eigenvalue representation without further explanation. For more detail on holomorphic matrix models and their complex saddles, see \cite{LAZ03,EKR15, AAM13, AAM16} for instance). This partition function can be written as
\begin{equation}
    Z = \prod_{a=1} ^N \int_\G dz_a \, e^{-N^2 S_N},
\end{equation}
with the action
\begin{equation} \label{eq:action}
    S_N \equiv \frac{1}{N} \sum_a W(z_a)
    -\frac{1}{2N^2} \sum_{a\neq b} \log (z_a -z_b )^2.
\end{equation}
The saddle point equation reads
\begin{equation} \label{eq:saddleeq}
    N W'(z_a) + \sum_{b\neq a} \frac{2}{z_b -z_a} =0 , \quad a =1, \cdots, N.
\end{equation}
Let us give a parametrization $f : [0,1] \longrightarrow \g$ of the curve $\g$. Then we can denote the eigenvalue density by
\begin{equation}
    \r(s) = \frac{1}{N} \sum_{a=1} ^N \d(s-s_a),\quad s\in [0,1],
\end{equation}
where $f(s_a) = z_a \in \g$ is the position of the $a$'th eigenvalue. Also with the solutions $z_a$ to the saddle point equation, let us define the discrete resolvent
\begin{equation} \label{eq:resolvent}
    \o(z) = \frac{1}{N} \sum_a \frac{1}{z-z_a}.
\end{equation}
Note that we have
\begin{equation} \label{eq:rel}
    \o(z_+) - \o (z_-) = - \frac{2\pi i}{N} \sum_{a=1} ^N \d(s-s_a)
    = -2\pi i \r(s), \quad f(s) = z \in \g,
\end{equation}
due to the property of the delta function. Now the saddle point equation \eqref{eq:saddleeq} implies the Riccati equation
\begin{equation} \label{eq:riccati}
    \frac{1}{N} \o' (z) + \o (z)^2 -  W'(z) \o (z) = \frac{1}{ N}
    \sum_a \frac{W'(z) - W'(z_a)}{z-z_a}.
\end{equation}

We study the continuum limit $N \to \infty$ of the model \eqref{eq:part}. In the large $N$ limit, the saddle points $z_a$'s accumulate on the curve $\g = \g_1 \cup \cdots \cup \g_s$, where $s$ is the number of cuts. Hence we can define a normalized positive eigenvalue density $\r(z)$ supported on $\g$
\begin{equation}
    \frac{1}{N} \sum_{a=1} ^N \d(s-s_a ) \longrightarrow \r(s) \, ds = \r(z) \,dz , \quad \int_\g \rho(z) \, dz  =1.
\end{equation}
Now the action can be written by taking the limit to \eqref{eq:action}
\begin{equation} \label{eq:actcont}
    S[\r] =  \int_\g W(z) \r(z)  dz  - \frac{1}{2}
    \int_{\g^{\times 2}} dz\,  dz'  \,  \log (z-z')^2 \r(z) \r(z').
\end{equation}
Again, the integral on the second term
is the principal value which removes the singularity of $\log(z-z')^2$ at $z=z'$.

Let us define the continuum limit of the resolvent \eqref{eq:resolvent},
\begin{equation}
    \o(z) = \int_\g P\frac{\r(z') dz' }{z-z'},
\end{equation}
where $P$ denotes the principal value. Note that $\o(z)$ is the Cauchy transform of $\r(z)$ on $\g$, so that $\o(z)$ is an analytic function in $\mathbb{C} \setminus \g$. Then the Sokhotski-Plemelj theorem states that
\begin{align} \label{eq:spt}
\begin{split}
    &\o(z_+) = \o(z) -\pi i \r(z) \\
    &\o(z_-) = \o(z) + \pi i \r(z) \end{split}\;\;, \quad\quad z \in \g.
\end{align}
In particular, we have
\begin{equation} \label{eq:orho}
   \o(z_+)-\o(z_-) = - 2\pi i \r(z), \quad z \in \g.
\end{equation}
This is just the large $N$ limit of \eqref{eq:rel}. Also, the continuum limit of the Riccati equation \eqref{eq:riccati}, called the Dyson-Schwinger equation, is obtained as
\begin{equation} \label{eq:dyson}
    \o(z) ^2 - W'(z) \o (z) = - \int_\g \frac{W'(z) -W'(z')}{z-z'} \r(z')  dz'  \equiv -P(z).
\end{equation}
Solving for the resolvent $\o(z)$, we get
\begin{equation} \label{eq:osol}
    \o(z) = \frac{1}{2}\left( W'(z) \pm \sqrt{W'(z)^2 - 4P(z)} \right).
\end{equation}
Recall that the eigenvalues are found at the singularities of the resolvent, which in large $N$ limit accumulate on the curve $\g$. This is precisely the square-root branch cut appearing in this expression. Therefore, by using \eqref{eq:orho} we find
\begin{equation} \label{eq:rho}
    \r(z) = \frac{1}{2\pi} \sqrt{4 P(z) - W'(z)^2}, \quad z \in \g.
\end{equation}
Hence, the equilibrium density can be achieved by computing $P(z)$ and substituting this into this equation. It is more convenient to use the variable $y(z)$ which we define by
\begin{equation} \label{eq:ydef}
    y(z) \equiv W'(z) - 2\o(z).
\end{equation}
It is straightforward from \eqref{eq:osol} and \eqref{eq:rho} that
\begin{equation}
    \r(z) = \pm \frac{y(z_\pm)}{2\pi i} , \quad z \in \g.
\end{equation}
In terms of $y(z)$, the Dyson-Schwinger equation \eqref{eq:dyson} becomes
\begin{equation} \label{eq:DSeq}
    y(z) ^2 = \left( W'(z)\right)^2 -4\int_{\g} \frac{W'(z)-W'(z')}{z-z'} \r(z') dz' .
\end{equation}
We solve this equation to obtain $y(z)$, and therefore $\r(z)$. The detail of the procedure largely depends on the problem we would like to address, determined by the potential $W(z)$. We present this procedure shortly, for our
truncated models.

Before proceeding, we describe how to obtain the support $\g$ of the equilibrium
distribution. First, the notion of probability should be well-defined from equilibrium density. In particular, the integral of the density along segments on the cut should be positive real number. Hence for each zero $a$ of $y(z)^2$, we consider a function
\begin{align}
    G(z) \equiv \int_a ^z y(z') dz',
\end{align}
and the Stokes lines defined by
\begin{equation}
    0 = \text{Re}\, G(z).
\end{equation}
A Stokes line could end either at an endpoint, i.e., a zero of $y(z)^2$, or at infinity. Due to the constraint $\int_\g dz\,\r(z)=1$, the equilibrium distribution cannot have a non-compact support, and we should only consider compact Stokes lines connecting endpoints. Moreover, due to the positive-definiteness of the eigenvalue density, not all the compact Stokes lines support the eigenvalue density. More specifically, the points in the neighborhood of a cut satisfy $\text{Re}\, G(z) <0$, while other Stokes lines separates the region where $\text{Re}\,G(z) <0$ and the region where $\text{Re}\, G(z)>0$. This can be seen as follows. For a given point $z\in \g$ on the cut, let us consider the points $z_+ + i dz$ and $z_- - i dz$ near $z$ to the left and right of the cut respectively, where $dz$ is the line segment along the cut. Then we have
\begin{align}
\begin{split}
   &\text{Re}\,G(z_+ +i dz) = \text{Re}\,G(z) - \text{Im} \left( G'(z_+) dz \right) = -\text{Im} \left(y(z_+) dz \right)\\
   &\text{Re}\,G(z_- -i dz) = \text{Re}\,G(z) + \text{Im} \left( G'(z_-) dz \right) = \text{Im} \left(y(z_-) dz \right)
\end{split}
\end{align}
where we have used $\text{Re}\, G(z)=0$ for $z\in \g$ and the definition of $G(z)$ in the second equality. Since $\r(z) = \pm \frac{y(z_\pm)}{2\pi i}$, we obtain
\begin{align}
    &\text{Re}\,G(z_+ +i dz) = \text{Re}\,G(z_- -i dz)=  - 2\pi\, \text{Re} \left(\r(z) dz\right) <0
\end{align}
due to the positive-definiteness of the density. Therefore, we conclude that the equilibrium density is supported on the union of compact Stokes lines which satisfy $\text{Re}\,G(z) <0$ on their neighborhoods.

Now we specialize to our truncated model, i.e. the model with the potential
\begin{align}
    W(z) = \log z -  \sum_{n=1} ^p \frac{\l_n}{n} (z^n + z^{-n}),
\end{align}
where $\l_n \in \mathbb{C}$, $n=1,\cdots, p$ are complex-valued couplings (as we mentioned earlier, they will be identified as $\l_n = a_n \r_n$ in our application to the $\mathcal{N}=4$ index).
A direct computation shows that
\begin{eqnarray}
    \int_\g \frac{W'(z)-W'(z')}{z-z'} \r(z')  dz
    &=& \int_\g \left[ -\frac{1}{zz'} +\sum_{n=1} ^p \l_n
    \left(\!-\!\!\sum_{i+j=n-2}\! z ^i z'^j - z^{-n-1} z'^{-n-1} \sum_{i+j=n} z^i z'^j \right) \right] \rho (z') dz'
    \nonumber\\
    &\equiv& -\sum_{i=-p-1} ^{p-2}  \frac{c_i}{4} z^i,
\end{eqnarray}
where in the last line we defined $c_i$ as the coefficient of $z^i$. Hence the Dyson-Schwinger equation \eqref{eq:DSeq} yields the following equation
\begin{align} \label{eq:sc}
    y(z)^2 = \left( W'(z) \right)^2 + \sum_{i=-p-1} ^{p-2} c_i z^i.
\end{align}
Note that the resolvent has the following asymptotic behavior at large $z$,
\begin{equation}
        \o(z) = \frac{1}{z} + \sum_{n=1} ^{p-1} \frac{1}{z^{n+1}} \int_\g z'^n \r(z')  dz' + \mathcal{O}(z^{-p-1})\ .
\end{equation}
Then by the definition \eqref{eq:ydef}, $y(z)$ has the following asymptotics as $z\to \infty$
\begin{equation}
    y(z) = -\frac{1}{z} - \sum_{n=1} ^p \l_n z^{n-1} + \sum_{n=1} ^{p-1} \l_n z^{-n-1} -2 \sum_{n=1} ^{p-1}  \r_n z^{-n-1} + \mathcal{O}(z^{-p-1}),
\end{equation}
where we remind $\r_n \equiv \int_\g z^n \r(z)  dz$ for $n\in \mathbb{Z}$ is the $n$'th moment of the distribution. By matching this with \eqref{eq:sc} up to the order of $z^{-1}$, we get
\begin{align}
    c_{i} = 4 \l_{i+2} + 4 \sum_{j=1} ^{p-i-2} \r_j \l_{i+j+2}, \quad i =-1,\cdots, p-2.
\end{align}
Also, the remaining $c_i$'s can be computed as
\begin{align}
    c_{-i-3} = 4 \sum_{j=1} ^{p-i-1} \l_{i+1+j} \r_{-j}, \quad i=-1, \cdots, p-2.
\end{align}
All in all, this leads to an algebraic curve parametrized by $z$ and $y$:
\begin{equation} \label{eq:scfin}
    y(z)^2 = \left( \frac{1}{z}\!-
    \!\sum_{i=1} ^p \l_i (z^{i-1}\! -\!z^{-i-1}) \right)^2
    \!+4 \sum_{i=1} ^{p} \l_i z^{i-2}
    + 4 \sum_{i=1} ^{p-1} \sum_{j=1} ^{p-i} \l_{i+j} \left( z^{i-2}    \r_j + z^{-i-2}  \r_{-j} \right) + 4 z^{-2} \sum_{j=1} ^p \l_{j} \r_{-j} .
\end{equation}
This curve is referred to as the spectral curve. Note that the spectral curve is
a double covering of the $z$-plane $\mathbb{C}$, upon which $y(z)$ is single-valued. The two sheets of the covering are connected through the square-root branch cut, and the equilibrium density is supported precisely on this cut. Let us write the spectral curve into the factorized form,
\begin{align} \label{eq:scnormal}
    y(z)^2 = \l_p ^2 \frac{\prod_{i=1} ^{2p} (z-a_i ^-) (z-a_i ^+) }{z^{2p+2}},
\end{align}
where the endpoints $a_i ^\pm$ are determined by $\l_i$ and $\r_i$, from equating \eqref{eq:scfin} and \eqref{eq:scnormal}. In practice, we first fix the endpoints depending on the phase that we are interested in, and then solve them and $\r_i$ in terms of $\l_i$ by equating the two expressions.

We focus on the one-cut gapped phase. We also require the equilibrium density to be even, meaning that $z y(z)$ is invariant under the transformation $z \mapsto z^{-1}$. First, note that only two of the $4p$ roots $a^\pm_i$ above should be single roots while all the other roots have to be double roots, since we want only one square-root branch cut connecting two endpoints to be created. Also, we study the most generic case of such, requiring the $2p-1$ double roots are all distinct. To meet the requirement for even distribution, a root should be accompanied with its reciprocal. In particular, we have two single roots $a$ and $a^{-1}$, $2p-2$ double roots $d_i$ and $d_i ^{-1}$, $i=1,\cdots,p-1$, and one remaining double root. This last double root should be $-1$ due to the even property. To sum up, we set
\begin{eqnarray}
    a_i ^\pm = d_i&&\textrm{ for } \quad i= 1,\cdots, p-1,\nonumber\\
    a_{i+p-1} ^\pm = d_i^{-1}&&\textrm{ for } \quad i=1,\cdots,p-1\nonumber\\
    a_{2p-1} ^\pm = -1&&\textrm{ and } \quad a_{2p}^\pm=a^{\pm 1}\ .
\end{eqnarray}
Then the spectral curve simplifies into
\begin{align} \label{eq:specfac}
     y(z)^2 = \l_p ^2 \frac{  (z-a)(z-a^{-1}) (z+1)^2  \prod_{i=1} ^{p-1} (z-d_i)^2 (z-d_i ^{-1})^2}{z^{2p+2}}.
\end{align}
By equating the coefficients of $z^m$, $m=0, \cdots, 4p$, of the numerator, we get $4p+1$ equations. Applying the even property to \eqref{eq:scfin}, we get the equality $\r_i = \r_{-i}$, $i=1, \cdots p-1$. The even property reduces the number of nontrivial equations to $2p+1$, one of which is only a trivial equation. In total, there are $2p$ nontrivial equations. Meanwhile, there are $2p$ undetermined variables, $\r_{-i}$, $i=1,\cdots, p$, $d_i$, $i=1,\cdots, p-1$, and $a$. Accordingly, the $2p$ undetermined variables are fixed in terms of the couplings $\l_i$ by the $2p$ equations.

It is rather difficult to solve them in the current form.
We rewrite the polynomial appearing in \eqref{eq:specfac} as
\begin{align}
    (z+1) \prod_{i=1} ^{p-1} (z-d_i) (z-d_i ^{-1}) \equiv z^{p-\frac{1}{2}} \left( z^{p-\frac{1}{2}} + z^{-p+\frac{1}{2}} + \sum_{i=1} ^{p-1} \frac{Q_i}{2 \l_p} \left( z^{i-\frac{1}{2}} + z^{-i+\frac{1}{2}} \right) \right),
\end{align}
where we just changed $p-1$ unknown variables from $d_i$ to  $Q_i$, $i=1,\cdots, p-1$. Let us also define $Q_p = 2\l_p$ for notational convenience. Finally, let us change the unknown variable $a$ to $A\equiv \frac{a+ a^{-1}}{2}$. Then we have to solve
\begin{align}
\begin{split}
    & z^{-\frac{1}{2}} \sqrt{z^2 +1 -2 A z} \sum_{i=1} ^p \frac{Q_i}{2} \left( z^{i-\frac{1}{2}} + z^{-i+\frac{1}{2}} \right) \\
    &=\sqrt{\left( 1+\sum_{i=1}^p \l_i (z^i + z^{-i}) \right)^2 +4 \sum_{i=1} ^{p-1} \sum_{j=1} ^{p-i} \l_{i+j} \r_{j} (z^i + z^{-i}) +4 \sum_{j=1} ^p \l_j \r_{j} }.
\end{split}
\end{align}
Note that $(1+z^2 -2A z)^{-\frac{1}{2}} = \sum_{l=0} ^\infty P_l (A) z^l$, i.e., the generating function of Legendre polynomials. Hence this equation can be rewritten as
\begin{align} \label{eq:match}
\begin{split}
    &\sum_{i=1} ^p \frac{Q_i}{2} \left( z^{i-\frac 1 2} +z^{-i + \frac 1 2}\right) \\
    &=\sum_{l=0} ^{\infty} z^{l+\frac 1 2} P_l (A)  \sqrt{\left( 1+\sum_{i=1}^p \l_i (z^i + z^{-i}) \right)^2 +4 \sum_{i=1} ^{p-1} \sum_{j=1} ^{p-i} \l_{i+j} \r_{j} (z^i + z^{-i}) +4 \sum_{j=1} ^p \l_j \r_{j} }.
\end{split}
\end{align}
Now we can solve for $A$ and $Q_i$ by expanding in $\vert z \vert <1 $ and matching the $p+1$ coefficients of $z^n$, for $n=-p+\frac 1 2 , -p+\frac 3 2, \cdots, \frac 1 2$. The square root on the right hand side can be expanded up to the order of $z^{\frac 1 2}$ as
\begin{align}
    z^{-p} \sqrt{\left( z^p \sum_{i=1} ^p \l_i z^{-i} \right)^2 +2 \l_p z^p + \mathcal{O}(z^{p+1})} = \sum_{i=1} ^p \l_i z^{-i} + 1 + \mathcal{O}(z).
\end{align}
Therefore the right hand side of \eqref{eq:match} is expanded as
\begin{align}
\begin{split}
    &\sum_{i=1} ^p \sum_{l=0} ^{i} \l_i P_l (A) z^{l-i + \frac 1 2} +  z^{\frac 1 2} + \mathcal{O} (z) =\sum_{i=1} ^p \sum_{l=0} ^{p-i} \l_{l+i} P_l (A) z^{-i + \frac 1 2} +  z^{\frac 1 2} + \mathcal{O}(z).
\end{split}
\end{align}
By comparing this with the left hand side of \eqref{eq:match}, we get
\begin{align}\label{Q-appendix}
    Q_i = 2 \sum_{l=0} ^{p-i} \l_{l+i} P_l (A),\quad i = 1, \cdots p,
\end{align}
from the coefficients of $z^{-i+\frac 1 2}$, $i=1, \cdots p$, and
\begin{align} \label{eq:width}
    Q_1 = 2\sum_{i=1} ^p \l_i P_i (A) + 2,
\end{align}
from the coefficient of $z^{\frac 1 2}$. In particular, there are $p+1$ equations for $p+1$ variables, the width of the cut $A\left( \equiv \cos \th_0  \right)$ and $Q_i$, $i=1,\cdots, p$, so that they are completely fixed in terms of the couplings $\l_i$ by these equations. From them, we finally achieve the following expression for the equilibrium distribution,
\begin{align} \label{eq:equil}
    \r(\th) = \frac{1}{\pi} \sqrt{\sin^2 \frac{\th_0}{2} -\sin^2 \frac{\th}{2}} \sum_{n=1} ^p Q_n \cos \left( n- \frac 1 2 \right) \th, \quad\quad \begin{split}& z= e^{i \th} \\ &a= e^{i \th_0}. \end{split}
\end{align}
(\ref{eq:equil}), (\ref{Q-appendix}), (\ref{eq:width}) derive the equations
(\ref{gap-solution}), (\ref{Q-definition}), (\ref{normalization}) in section 2.

We can also compute the moments $\r_n$, $n=1, \cdots p$, by either expanding the right hand side of \eqref{eq:match} to even higher orders of $z$ or directly computing them from \eqref{eq:equil}. In our application of the truncated matrix model to the $\mathcal{N}=4$ index, we have to further relate $\l_n = a_n \r_n$, and solve $\r_n = \int_\g dz\, z^n \r(z)$. Namely, even if we had solved the moments $\r_i$ in terms of the couplings $\l_i$, the couplings are now given by the moments so that we have to solve $\r_n = \int_\g dz\, z^n \r(z)$ along with \eqref{eq:width} by substituting the equilibrium density \eqref{eq:equil}. This is precisely the equations
\begin{eqnarray}
    (R -{\bf 1})\r = 0, \quad A \cdot \r =1,
\end{eqnarray}
which we explained and solved in sections 2 $\sim$ 5.

\section{BPS black holes in AdS}

We summarize the properties of the BPS black holes of
\cite{Kunduri:2006ek} and \cite{Chong:2005da}. We also explain
the small black hole limit of these solutions, identifying the charged
asymptotically flat 5d black holes of \cite{Strominger:1996sh} and
the spinning BMPV black holes \cite{Breckenridge:1996is}.

We consider BPS black holes in AdS$_5$ with one $U(1)_R$ electric charge
$R$ and two angular momenta $J_1,J_2$. (In the notation of \cite{Kunduri:2006ek},
we set $\mu_1=\mu_2=\mu_3\equiv\mu$.) We first present the solutions in the notation
of \cite{Kunduri:2006ek}, and then relate it to the convention of \cite{Chong:2005da}
later. The black hole metric is given by
\begin{eqnarray}\label{metric}
  ds^2\!&=&\!-f^2(dt+\omega_\psi d\psi+\omega_\phi d\phi)^2
  +f^{-1}h_{mn}dx^mdx^n\\
  h_{mn}dx^mdx^n\!&=&\!r^2\left[\frac{dr^2}{\Delta_r}
  +\frac{d\theta^2}{\Delta_\theta}\right]
  +\mathcal{M}_{ij}d\phi^i d\phi^j\ ,\ \
  \textrm{where }i,j=1,2\ ,\ \phi^i=(\phi^1,\phi^2)=(\psi,\phi)\nonumber\\
  \mathcal{M}\!&=&\!r^2\left(\!\begin{array}{cc}
    \frac{c_\theta^2}{\Xi_b^2}
    \left(\Xi_b+c_\theta^2(\rho^2g^2\!+\!2(1\!+\!bg)(a\!+\!b)g)\right)&
    \frac{s_\theta^2 c_\theta^2}{\Xi_a\Xi_b}\left(\rho^2 g^2
    \!+\!2(a\!+\!b)g\!+\!(a\!+\!b)^2g^2\right)\\
    \frac{s_\theta^2 c_\theta^2}{\Xi_a\Xi_b}\left(\rho^2 g^2
    \!+\!2(a\!+\!b)g\!+\!(a\!+\!b)^2g^2\right)
    &\frac{s_\theta^2}{\Xi_a^2}
    \left(\Xi_a+s_\theta^2(\rho^2g^2\!+\!2(1\!+\!ag)(a\!+\!b)g)\right)
  \end{array}\!\!\right)
  \nonumber\\
  \Delta_r\!&=&\!r^2\left(g^2r^2+(1\!+\!ag\!+\!bg)^2\right)\ \ ,
  \ \ \ \Delta_\theta=1-a^2g^2\cos^2\theta-b^2g^2\sin^2\theta\nonumber\\
  \Xi_a\!&=&\!1-a^2g^2\ \ ,\ \ \ \Xi_b=1-b^2g^2\ \ ,\ \ \
  \rho^2=r^2+a^2\cos^2\theta+b^2\sin^2\theta\nonumber
\end{eqnarray}
where
\begin{eqnarray}
  f^{-1}&=&1+\frac{\sqrt{\Xi_a\Xi_b}(1+g^2\mu)-\Xi_a\cos^2\theta
  -\Xi_b\sin^2\theta}{g^2r^2}\\
  \omega_\psi&=&-\frac{g\cos^2\theta}{r^2\Xi_b}
  \left[\rho^4+(2r_m^2+b^2)\rho^2+\frac{1}{2}
  \left(\beta_2-a^2b^2+g^{-2}(a^2-b^2)\right)\right]\nonumber\\
  \omega_\phi&=&-\frac{g\sin^2\theta}{r^2\Xi_a}
  \left[\rho^4+(2r_m^2+a^2)\rho^2+\frac{1}{2}
  \left(\beta_2-a^2b^2+g^{-2}(b^2-a^2)\right)\right]\nonumber
\end{eqnarray}
and
\begin{eqnarray}
  r_m^2&=&\frac{a+b}{g}+ab\ \ ,\ \ \
  \mu=\frac{1}{3\sqrt{\Xi_a\Xi_b}}\left[2r_m^2+3g^{-2}
  \left(1-\sqrt{\Xi_a\Xi_b}\right)\right]\\
  \beta_2&=&3\Xi_a\Xi_b\mu^2-\frac{6\sqrt{\Xi_a\Xi_b}(1-\sqrt{\Xi_a\Xi_b})}{g^2}\mu
  +\frac{3(1-\sqrt{\Xi_a\Xi_b})^2}{g^4}\ .\nonumber
\end{eqnarray}
We defined $(c_\theta,s_\theta)\equiv(\cos\theta,\sin\theta)$,
and $g\equiv\ell^{-1}$ is the inverse-radius of AdS$_5$ and $S^5$.
The $U(1)_R\subset U(1)^3\subset SO(6)$ vector potential is given by
\begin{equation}\label{vector-potential}
  A=(f-1)dt+f(\omega_\psi d\psi+\omega_\phi d\phi)+U_\psi d\psi+U_\phi d\phi
\end{equation}
where
\begin{eqnarray}
  U_\psi&=&\frac{g\cos^2\theta}{\Xi_b}\left[\rho^2+2r_m^2+b^2
  -\sqrt{\Xi_a\Xi_b}\mu+g^{-2}\left(1-\sqrt{\Xi_a\Xi_b}\right)\right]\\
  U_\phi&=&\frac{g\sin^2\theta}{\Xi_a}\left[\rho^2+2r_m^2+a^2
  -\sqrt{\Xi_a\Xi_b}\mu+g^{-2}\left(1-\sqrt{\Xi_a\Xi_b}\right)\right]\ .\nonumber
\end{eqnarray}
Here we shifted the definition of $A$ in \cite{Kunduri:2006ek} by a pure
gauge $-dt$, which yields vanishing $A$ at the spatial infinity
of AdS$_5$. This solution is expressed in co-rotating coordinates.
To get to the canonical coordinates of asymptotic AdS, one should
use the coordinates $\tilde{t},\tilde\psi,\tilde\phi$ defined by
$t=\tilde{t}$, $\psi=\tilde\psi-g\tilde{t}$, $\phi=\tilde\phi-g\tilde{t}$.
Upon the coordinate transformation, the definition of energy changes as
$E\rightarrow\tilde{E}=E+g(J_1+J_2)$, where $J_1$ and $J_2$ are angular
momenta conjugate to $\phi^1=\psi$, $\phi^2=\phi$, respectively.
The energy (mass) of the black hole is given by the BPS relation
$\tilde{E}=g(R_1+R_2+R_3+J_1+J_2)$, where $R_1+R_2+R_3\equiv 3R$ is the
$U(1)_R$ charge.

The charges $R$, $J_1\equiv J_\psi$, $J_2\equiv J_\phi$
and the Bekenstein-Hawking entropy $S$ are given by
\begin{eqnarray}\label{BH-charges}
  R&=&\frac{\pi}{4G g}\left[\mu+\frac{\mu^2g^2}{2}\right]\\
  J_1&=&\frac{\pi}{4G}\left[\frac{3g\mu^2}{2}+g^3\mu^3
  +g^{-3}\left(\sqrt{{\textstyle\frac{\Xi_a}{\Xi_b}}}-1\right)
  (1+g^2\mu)^3\right]\nonumber\\
  J_2&=&\frac{\pi}{4G}\left[\frac{3g\mu^2}{2}+g^3\mu^3
  +g^{-3}\left({\textstyle \sqrt{\frac{\Xi_b}{\Xi_a}}}-1\right)
  (1+g^2\mu)^3\right]\nonumber\\
  S&=&\frac{\pi^2}{2G}\sqrt{(1+3g^2\mu)\mu^3-\frac{9g^2\mu^4}{4}
  -\frac{(\sqrt{\Xi_a}-\sqrt{\Xi_b})^2}{g^6\sqrt{\Xi_a\Xi_b}}(1+g^2\mu)^3}\ .
  \nonumber
\end{eqnarray}
In the normalization of AdS$_5\times S^5$,
$N^2=\frac{\pi}{2g^3G}$. We exchanged the definitions of $J_1,J_2$ relative to
\cite{Choi:2018hmj}. We also multiplied $g^{-1}$ to $R$ relative to the definition
of $Q_I$ in \cite{Kunduri:2006ek}.
Since these black holes carry three charges depending on $2$ parameters
$a,b$, they satisfy a charge relation. The following two
expressions for the entropy $S$ in terms of dependent
charges $R,J_1,J_2$ are often very useful \cite{Kim:2006he,Choi:2018hmj}:
\begin{equation}
  S=2\pi\sqrt{3R^2-\frac{N^2}{2}(J_1+J_2)}
  =2\pi\sqrt{\frac{R^3+\frac{N^2}{2}J_1J_2}{\frac{N^2}{2}+3R}}\ .
\end{equation}
The equivalence of the two expressions is the charge relation.

\begin{figure}[!t]
\centering
\includegraphics[width=0.5\textwidth]{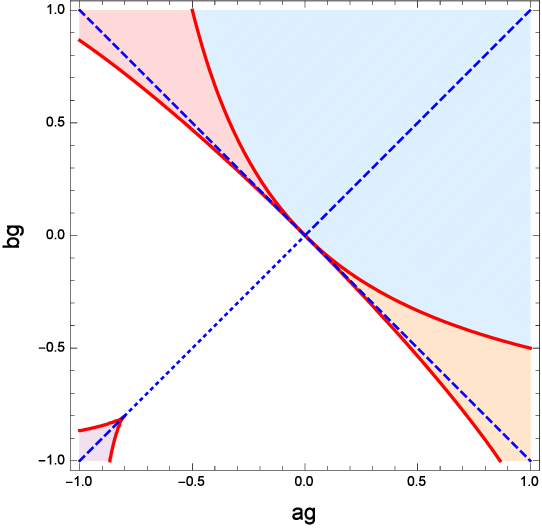}
\caption{Four equivalent fundamental domains in the parameter space
for the CTC-free black holes are shown in blue, red, orange and purple.
The black hole entropy vanishes at the red lines. The dashed lines
at $b=a$ and $b=-a$ both describe black holes at $J_1=J_2$.}
\label{bh-regions}
\end{figure}
\cite{Kunduri:2006ek} assumed $a,b\geq 0$, but this condition is unnecessary.
The only condition to be imposed on $a,b$ in the range
$-g^{-1}<a,b<g^{-1}$ is the absence of closed timelike
curves (CTC), which requires that the expression inside the square root of
$S$ in (\ref{BH-charges}) is positive. It will be important to understand that
negative $a$ or $b$ is allowed. They provide solutions inequivalent
to those with $a,b>0$. In particular, it will be shown that BMPV
black holes in the small black hole limit can have negative $a$ or $b$. To clearly
see the fundamental domain of the parameters $(a,b)$, it is helpful to
compare them with the two parameters appearing in the solutions
of \cite{Chong:2005da}. The two parameters of \cite{Chong:2005da} are also
called $a,b$. Let us call them $\tilde{a},\tilde{b}$
in order to distinguish them with $a,b$ of \cite{Kunduri:2006ek}.
As explained in section 2.3 of \cite{Kunduri:2006ek}, $\tilde{a},\tilde{b}$ and
$a,b$ are related by
\begin{eqnarray}
  \frac{g^{-2}(1+ag+bg)^2}{1-a^2g^2}&=&
  \frac{g^{-1}(\tilde{a}+\tilde{b}+\tilde{a}\tilde{b}g)
  +g^{-2}(1+\tilde{a}g+\tilde{b}g)^2}{1-\tilde{a}^2g^2}\\
  \frac{g^{-2}(1+ag+bg)^2}{1-b^2g^2}&=&
  \frac{g^{-1}(\tilde{a}+\tilde{b}+\tilde{a}\tilde{b}g)
  +g^{-2}(1+\tilde{a}g+\tilde{b}g)^2}{1-\tilde{b}^2g^2}\ .\nonumber
\end{eqnarray}
These equations can be solved for $\tilde{a}$, $\tilde{b}$ in terms of
$a,b$, yielding the following unique solution:
\begin{eqnarray}\label{ab-cclp}
  \tilde{a}&=&
  \frac{13a^2 + 8ab - 5b^2 + (12a^3  + 20a^2 b  + 8ab^2)g + (12a^3b + 13a^2b^2)g^2}
  {12 a+12 b+(13 a^2 +20 ab+13b^2)g + (8a^2b  + 8ab^2)g^2 - 5 a^2b^2g^3}\\
  \tilde{b}&=&
  \frac{-5a^2 + 8ab + 13b^2 + (8a^2b + 20ab^2 + 12b^3)g + (13a^2b^2 + 12ab^3)g^2}
 {12a+ 12b + (13a^2 + 20ab + 13b^2)g+ (8a^2b + 8ab^2)g^2 - 5 a^2b^2g^3}
  \ .\nonumber
\end{eqnarray}
The CTC-free condition reads $\tilde{a}+\tilde{b}+\tilde{a}\tilde{b}g>0$
\cite{Chong:2005da}. This in terms of $(\hat{a},\hat{b})\equiv(ag,bg)$
is given by
\begin{equation}\label{CTC}
  \frac{(\hat{a}\!+\!\hat{b}\!+\!\hat{a}\hat{b})^2
  (32\hat{a}^3(1\!+ \!\hat{b})+\hat{b}(32 \!+ \!61\hat{b}\!+\!32\hat{b}^2)
  +\hat{a}^2(61\!+\!118\hat{b}\!+\!61\hat{b}^2)+2\hat{a}
  (16\!+\!59\hat{b}\!+\!59\hat{b}^2\!+\!16\hat{b}^3))}
   {(\hat{b} (12 + 13 \hat{b}) +\hat{a}^2 (13 + 8 \hat{b} - 5 \hat{b}^2)
   + 4 \hat{a} (3 + 5 \hat{b} + 2 \hat{b}^2))^2}>0\ .
\end{equation}
$(\tilde{a},\tilde{b})$ satisfying (\ref{CTC}) makes a 1-to-1 map
to the black hole solutions. However, the parameters
$(a,b)$ of \cite{Kunduri:2006ek} make a 4-to-1 map to
$(\tilde{a},\tilde{b})$. So the CTC-free region $-g^{-1}<a,b<g^{-1}$
can be divided into four equivalent fundamental domains. These four
regions are shown in Fig. \ref{bh-regions}. One interesting point is
about the black holes at equal rotations $J_1=J_2$. From an obvious
exchange symmetry, such solutions are obtained at $a=b$. This also
yields $\tilde{a}=\tilde{b}$ from \ref{ab-cclp}. However,
one can easily check that $a=-b$ also maps to $\tilde{a}=\tilde{b}$.
Since the solutions are obviously same if $\tilde{a},\tilde{b}$ are,
$a=-b$ also maps to black holes at $J_1=J_2$.
On these two lines, the maps between $\tilde{a}$ and $a$ are different:
\begin{equation}
  \tilde{a}(a)=\left\{\begin{array}{ll}
    \frac{a(4+5a)}{6+4a-a^2}&\textrm{on the branch }a=b\\
    \frac{a^2}{6-5a^2}&\textrm{on the branch }a=-b
  \end{array}\right.\ .
\end{equation}
There are four branches in the four regions, shown as dashed blue lines in
Fig. \ref{bh-regions}: (1) the $a=b$ branch for
$a>0$ in the blue region of Fig. \ref{bh-regions}, (2) the $a=-b$ branch
for $a<0$ in the red region, (3) the $a=-b$ branch for $a>0$ in the orange
region, and (4) the $a=b$ branch for $a<-\frac{4}{5}$ in the purple region.
In all these four branches, as one increases $|a|$ from $0$ to $1$
or from $\frac{4}{5}$ to $1$, $\tilde{a}$ increases from $0$ to $1$.
Below, we shall discuss the black hole solutions in the fundamental domain
shown by the blue color in Fig. \ref{bh-regions}. This domain includes
all solutions at $a,b>0$, but it also
includes some regions with $a>0,b<0$ or $a<0,b>0$.

Let us explain the small black hole limit in the blue domain of Fig. \ref{bh-regions}.
There are many equivalent ways of describing this limit. With our motivation to
study the asymptotically flat black holes from small AdS black holes, it is
most convenient to view it as sending the AdS size to infinity.
Namely, we send the inverse-radius $g$ of AdS to zero, while scaling
$a,b$ suitably. The small black hole limit is given by
\begin{equation}\label{BMPV-scaling}
  g\rightarrow 0\ ,\ \ a_+\equiv\frac{a+b}{g}\rightarrow\textrm{finite}
  \ ,\ \ a_-\equiv a-b\rightarrow\textrm{finite}\ .
\end{equation}
If the radial coordinate $r$ is much smaller than $g^{-1}$, i.e.
$r\sim\mathcal{O}(g^0)$, the black hole solution reduces to
\begin{eqnarray}
  ds^2&=&-f^2(dt+\omega)^2+f^{-1}\left[dr^2
  +r^2\left(d\theta^2+\cos^2\theta d\psi^2+\sin^2\theta d\phi^2\right)\right]\\
  A&=&(f-1)dt+f\omega\ \ ,\ \ \
  f^{-1}=1+\frac{a_-^2+8a_+}{12r^2}\ \ ,\ \ \
  \omega=-\frac{a_+a_-}{2r^2}
  \left(\cos^2\theta d\psi-\sin^2\theta d\phi\right)\ .\nonumber
\end{eqnarray}
This is the BMPV black hole solution \cite{Breckenridge:1996is} written
in the form of \cite{Gauntlett:2002nw}, except that we again shifted
$A$ by a pure gauge $-dt$.  Therefore, the solution in the scaling limit
is approximately given by the asymptotically flat BPS black hole solution
if one keeps $r$ to be much smaller than large $g^{-1}$.
Note that the scaling limit (\ref{BMPV-scaling}) typically has
one of $a,b$ to be negative, unless $J_1-J_2$ is fine-tuned to be very small.
The scaling limit amounts to approaching the red curve of
Fig. \ref{bh-regions} from the blue domain,
close to the origin due to the condition $a+b\sim\mathcal{O}(g)$.
If one further takes $a_-=0$ at nonzero $a_+>0$, one recovers the
non-rotating 5d charged black holes of \cite{Strominger:1996sh}.
(In fact the same black holes are obtained by setting $a_+=0$
at nonzero $a_-$, in the red or orange regions of Fig. \ref{bh-regions}.)

\end{document}